\documentclass[aps,prd,preprintnumbers,superscriptaddress,nofootinbib,showkeys,floatfix]{revtex4}
\pdfoutput=1
\usepackage[latin9]{inputenc}
\usepackage{amsmath}
\usepackage{amssymb}
\usepackage{graphicx}
\usepackage{hyperref}

\usepackage{xcolor}

\makeatletter

\providecommand{\tabularnewline}{\\}

\@ifundefined{textcolor}{}
{%
 \definecolor{BLACK}{gray}{0}
 \definecolor{WHITE}{gray}{1}
 \definecolor{RED}{rgb}{1,0,0}
 \definecolor{GREEN}{rgb}{0,1,0}
 \definecolor{BLUE}{rgb}{0,0,1}
 \definecolor{CYAN}{cmyk}{1,0,0,0}
 \definecolor{MAGENTA}{cmyk}{0,1,0,0}
 \definecolor{YELLOW}{cmyk}{0,0,1,0}
}

\PassOptionsToPackage{english}{babel}

\usepackage{float}
\usepackage{mathrsfs}
\usepackage{amsfonts}
\usepackage{array}
\usepackage{epsfig}

\def\beq{\begin{equation}}
\def\b0{\beta_0}

\def\eeq{\end{equation}}
\def\beeq{\begin{eqnarray}}
\def\eeeq{\end{eqnarray}}

\def\to{\rightarrow}

\makeatother

\begin{document}

\preprint{SMU-HEP-18-04}

\title{Mapping the sensitivity of hadronic experiments to nucleon structure}

\author{Bo-Ting Wang}
\email{botingw@mail.smu.edu}

\affiliation{Department of Physics, Southern Methodist University,\\
 Dallas, TX 75275-0181, U.S.A. }

\author{T. J. Hobbs}
\email{tjhobbs@smu.edu}

\affiliation{Department of Physics, Southern Methodist University,\\
 Dallas, TX 75275-0181, U.S.A. }

\affiliation{Jefferson Lab, EIC Center,
 Newport News, VA 23606, U.S.A. }

\author{Sean Doyle}

\affiliation{Department of Physics, Southern Methodist University,\\
 Dallas, TX 75275-0181, U.S.A. }

\author{Jun Gao}

\affiliation{INPAC, Shanghai Key Laboratory for Particle Physics and Cosmology,\\
 School of Physics and Astronomy,\\
 Shanghai Jiao-Tong University, Shanghai 200240, China}

\author{Tie-Jiun Hou}

\affiliation{School of Physics Science and Technology, Xinjiang University,\\
 Urumqi, Xinjiang 830046 China}

\author{Pavel~M. Nadolsky}
\email{nadolsky@physics.smu.edu}

\author{Fredrick I. Olness}

\affiliation{Department of Physics, Southern Methodist University,\\
 Dallas, TX 75275-0181, U.S.A. }
\begin{abstract}
	Determinations of the proton's collinear parton distribution functions
	(PDFs) are emerging with growing precision due to increased experimental
	activity at facilities like the Large Hadron Collider. While this
	copious information is valuable, the speed at which it is released makes it difficult to quickly assess its impact on the PDFs, short of performing
	computationally expensive global fits. As an alternative, we explore new methods for quantifying
	the potential impact of experimental data on the extraction of
	proton PDFs. Our approach relies crucially on the Hessian correlation
	between theory-data residuals and the PDFs themselves, as well as
	on a newly defined quantity --- the {\it sensitivity}
	--- which represents an extension of the correlation and
	reflects both PDF-driven and experimental uncertainties. This approach
	is realized in a new, publicly available analysis package \textsc{PDFSense},
	which operates with these statistical measures to identify particularly
	sensitive experiments, weigh their relative or potential impact on
	PDFs, and visualize their detailed distributions in a space of the parton
	momentum fraction $x$ and factorization scale $\mu$. This tool offers
	a new means of understanding the influence of individual measurements
	in existing fits, as well as a predictive device for directing future
	fits toward the highest impact data and assumptions. Along the way,
	many new physics insights can be gained or reinforced. As one of many
	examples, \textsc{PDFSense} is employed to rank the projected impact
	of new LHC measurements in jet, vector boson, and $t\bar{t}$ production
	and leads us to the conclusion that inclusive jet production at the LHC
	has a potential for playing an indispensable role in future PDF fits.
	These conclusions are independently verified by preliminarily fitting
	this experimental information and investigating the constraints they
	supply using the Lagrange multiplier technique.
\end{abstract}


\maketitle
\newpage{}\tableofcontents{}\newpage{} 

\section{Introduction \label{sec:Introduction}}

The determination of collinear parton distribution functions (PDFs)
of the nucleon is becoming an increasingly precise discipline with
the advent of high-luminosity experiments at both colliders and fixed-target
facilities. Several research groups are involved in the rich research
domain of the modern PDF analysis \cite{Dulat:2015mca,Harland-Lang:2014zoa,Ball:2017nwa,Alekhin:2017kpj,Accardi:2016qay,Abramowicz:2015mha,Alekhin:2014irh}.
By quantifying the distribution of a parent hadron's longitudinal
momentum among its constituent quarks and gluons, PDFs offer both
a description of the hadronic structure and an essential ingredient
of perturbative QCD computations. PDFs enjoy a symbiotic relationship
with high-energy experimental data, in the sense that they are crucial
for understanding hadronic collisions in the Standard Model (SM) and
beyond, while reciprocally benefiting from a wealth of high-energy
data that constrain the PDFs. In fact, since the start of the Large
Hadron Collider Run II (LHC Run II), the volume of experimental data
pertinent to the PDFs is growing with such speed that keeping pace
with the rapidly expanding datasets and isolating measurements of
greatest impact presents a significant challenge for PDF fitters.
This paper intends to meet this challenge by presenting a method for
identifying high-value experiments which constrain the PDFs and the
resulting SM predictions that depend on them.

That such expansive datasets can constrain the PDFs is a consequence
of the latter's universality \textemdash{} a feature which relies
upon QCD factorization theorems to separate the inherently nonperturbative
PDFs (at long distances) from process-dependent, short-distance matrix
elements. For instance, the cross section for inclusive single-particle
hadroproduction (of, \textit{e.g.}, a weak gauge boson $W/Z$) in
proton-proton collisions at the LHC is directly sensitive to the nucleon
PDFs via an expression of the form 
\begin{align}
 & \sigma(AB\to W/Z\!+\!X)\ =\ \sum_{n}\,\alpha_{s}^{n}(\mu_{R}^{2})\,\sum_{a,b}\int dx_{a}dx_{b}\,\label{eq:fact}\\
 & \times\,f_{a/A}(x_{a},\mu^{2})\,\hat{\sigma}_{ab\to W/Z+X}^{(n)}\big(\hat{s},\,\mu^{2},\mu_{R}^{2}\big)\,f_{b/B}(x_{b},\mu^{2})\ ,\nonumber 
\end{align}
in which $f_{a/A}(x_{a},\mu^{2})$ represents the PDF for a parton
of flavor $f_{a}$ carrying a fraction $x_{a}$ of the 4-momentum
of proton $p_{A}$ at a factorization scale $\mu$; the $n^{\mathit{th}}$-order
hard matrix element is denoted by $\hat{\sigma}_{ab\to W/Z+X}^{(n)}\big(\hat{s},\,\mu^{2},\mu_{R}^{2}\big)$
and is dependent upon the partonic center-of-mass energy $\hat{s}=x_{a}x_{b}s$,
in which $s$ in the center-of-mass energy of the initial hadronic
system; and $\mu_{R}$ is the renormalization scale in the QCD coupling
strength $\alpha_{s}(\mu_{R})$. In Eq.~(\ref{eq:fact}), subleading
corrections $\sim\!\!\Lambda^{2}/M_{W/Z}^{4}$ have been omitted,
and we emphasize that factorization theorems like Eq.~(\ref{eq:fact})
have been proved to arbitrary order in $\alpha_{s}$ for essential
observables in the global PDF analysis, such as the inclusive cross
sections in DIS and Drell-Yan processes. For compactness and generality,
we shall refer henceforth to a PDF for the parton of flavor $f$ simply
as $f(x,\mu)$.

Given this formalism, one is confronted with the problem of finding
those experiments that provide reliable new information about the
PDF behavior. With the proliferation of potentially informative new
data, incorporating them all into a global QCD fit inevitably incurs
significant cost both in terms of computational resources and required
fitting time. Indeed, tremendous progress in the precision of PDFs
and robustness of SM predictions is driven by the technology for performing
global analysis that has vastly grown in complexity and sophistication.
Nowadays, the state-of-the-art in perturbative QCD (pQCD) treatments
are done at NNLO (and increasingly even N$^{3}$LO), and advanced
statistical techniques are commonly employed in PDF error estimation.
The magnitude of this subject is vast, and we refer the interested
reader to Refs.~\cite{Gao:2017yyd,Butterworth:2015oua} for comprehensive
reviews. The tradeoff of this progress is that the impact of an experiment
on the ultimate PDF uncertainty is often hard to foresee without doing
a complicated fit. Various publications claim sensitivity of
new experiments to the PDFs. In this paper, we look into these claims
using statistical techniques that bypass doing the fits,
and with an eye on theoretical, experimental, and methodological components
relevant at the NNLO precision.

\begin{figure*}
\includegraphics[width=1\textwidth]{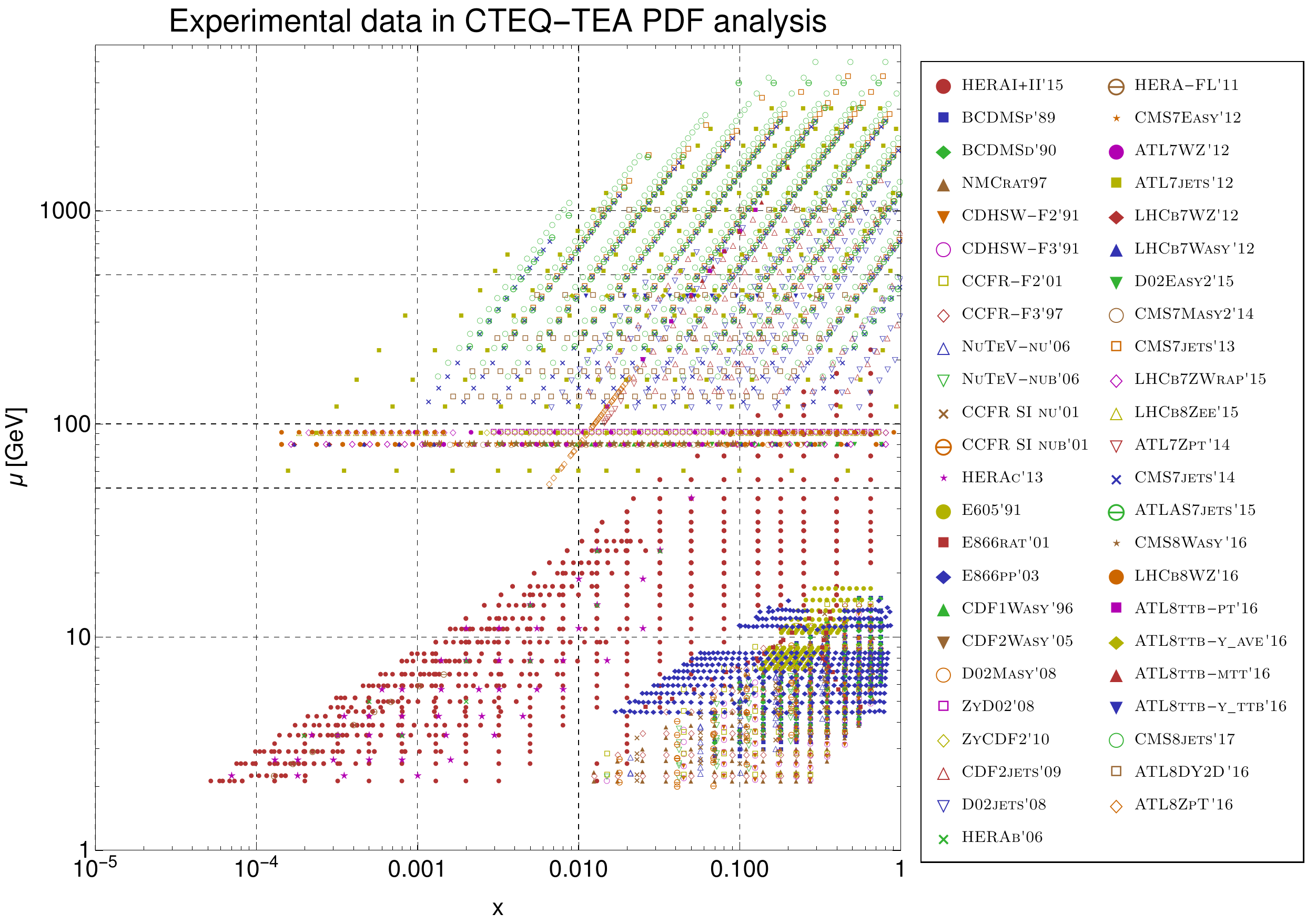}
\caption{A graphical representation of the space of $\{x,\mu\}$ points probed
by the full dataset treated in the present analysis, designated as
``CTEQ-TEA''. It represents an expansion to include newer LHC data of the CT14HERA2 dataset
\cite{Hou:2016nqm} fitted in the most recent CT14 framework \cite{Dulat:2015mca},
which involved measurements from Run II of HERA \cite{Abramowicz:2015mha}.
Details of the datasets corresponding to the short-hand names given in the legend
may be found in Tables~\ref{tab:EXP_1}--\ref{tab:EXP_3}.
}
\label{fig:data} 
\end{figure*}

The potential cost is steepened by the large size of the global datasets
usually involved. This point can be seen in Fig.~\ref{fig:data},
which plots the default dataset considered in the present analysis
in a space of partonic momentum fraction $x$ and factorization scale
$\mu$. We label these data as the ``CTEQ-TEA set,'' given that
it is an extension of the 3287 raw data points (given by the sum over
$N_{\mathit{pt}}$ in Tables~\ref{tab:EXP_1} and \ref{tab:EXP_2}) treated
in the NNLO CT14HERA2 analysis of Ref.~\cite{Hou:2016nqm}, now augmented
by the inclusion of 734 raw data points (given by the sum over $N_{\mathit{pt}}$
in Table~\ref{tab:EXP_3}) from more recent LHC data. These raw measurements
can ultimately be mapped to 5227 typical $\{x,\mu\}$ values
in Fig.~\ref{fig:data}, such that each symbol corresponds to a data point from
an experiment shown in the legend, at the approximate $x$ and $\mu$
values characterizing the data point as described in Appendix~\ref{sec:supp}.
The experiments are labeled by a short-hand name which includes the
year of final publication ({\it e.g.}, ``HERAI+II'15'' --- corresponding to the
2015 combined HERA Run I and Run II data),
following the translation key also given in Tables~\ref{tab:EXP_1}\textendash \ref{tab:EXP_3}
of App. \ref{sec:Tables}. The experiments included in the CT14HERA2
analysis are listed in the left column and upper part of the right column of the legend, while
the newer LHC data considered for the upcoming CTEQ-TEA analysis are the last 14
entries of the right column.

The growing complexity of PDF fitting stimulates development of
less computationally involved approaches to estimate the
impact of new experimental data on full global fits, such as Hessian
profiling techniques \cite{Camarda:2015zba} and Bayesian reweighting
\cite{Ball:2010gb,Ball:2011gg} of PDFs. Although these approaches
do simulate the expansion of a particular global fit
by including theretofore absent dataset(s), they are also limited in
that the interpretation of their outcomes is married to the specific
PDF parametrization and definition of PDF errors. For example, conclusions
obtained by PDF reweighting regarding the importance of a given data
set strongly depend on the assumed statistical tolerance or the choice
of reweighting factors \cite{Sato:2013ika,Paukkunen:2014zia}.

Parallel to these efforts, the notion of using correlations between
the PDF uncertainties of two physical observables was proposed in
Refs.~\cite{Pumplin:2001ct,Nadolsky:2001yg} as a means of quantifying
the degree to which these quantities were related based upon their
underlying PDFs. The PDF-mediated correlation $C_{f}$ in this case,
which we define in Sec.~\ref{sec:Correlations}, embodies
the Pearson correlation coefficient computed by a generalization of
the ``master formula'' \cite{Pumplin:2002vw} for the Hessian PDF
uncertainty. The Hessian correlation was deployed extensively in Ref.~\cite{Nadolsky:2008zw}
to explore implications of the CTEQ6.6 PDFs for envisioned LHC observables.
It proved to be instrumental for identifying the specific PDF flavors
and $x$ ranges most correlated with the PDF uncertainties for $W,$
$Z,$ $H$, and $t\bar{t}$ production cross sections as well as other
processes. The Pearson correlation coefficient has also
proven to be informative in the approach based on Monte-Carlo PDF replicas,
see, e.g., Refs.~\cite{Ball:2008by,Carrazza:2016htc}.
However, the PDF-mediated correlation with a theoretical
cross section is only partly indicative of the sensitivity of the
experiment. The constraining power of the experiment also depends
on the size of experimental errors that were not normally considered
in correlation studies, as well as on correlated systematic effects
that are increasingly important.

As a remedy to these limitations, we introduce a new format for the
output of CTEQ-TEA fits and a natural extension of the correlation
technique to quantify the sensitivity of any given experimental data
point to a PDF-dependent observable of the user's choice. In this
approach, we work with \emph{statistical residuals} quantifying the
goodness-of-fit to individual data points. We demonstrate that the
complete set of residuals computed for Hessian PDF sets characterizes
the CTEQ-TEA fit well enough to permit a means of gauging the influence
of empirical information on PDFs in a fashion that does not require
complete refits.

A generalization of the PDF-mediated correlations called
the \textit{sensitivity $S_{f}$} \textemdash{} to be characterized
in detail in Sec.~\ref{sec:Sensitivities} \textemdash{}
better identifies those experimental data points that tightly
constrain PDFs both by merit of their inherent precision and their
ability to discriminate among PDF error fluctuations. Such an approach
aids in identifying regions of $\{x,\mu\}$ for which PDFs are
particularly constrained by physical observables.

\begin{figure*}
\includegraphics[width=0.47\textwidth]{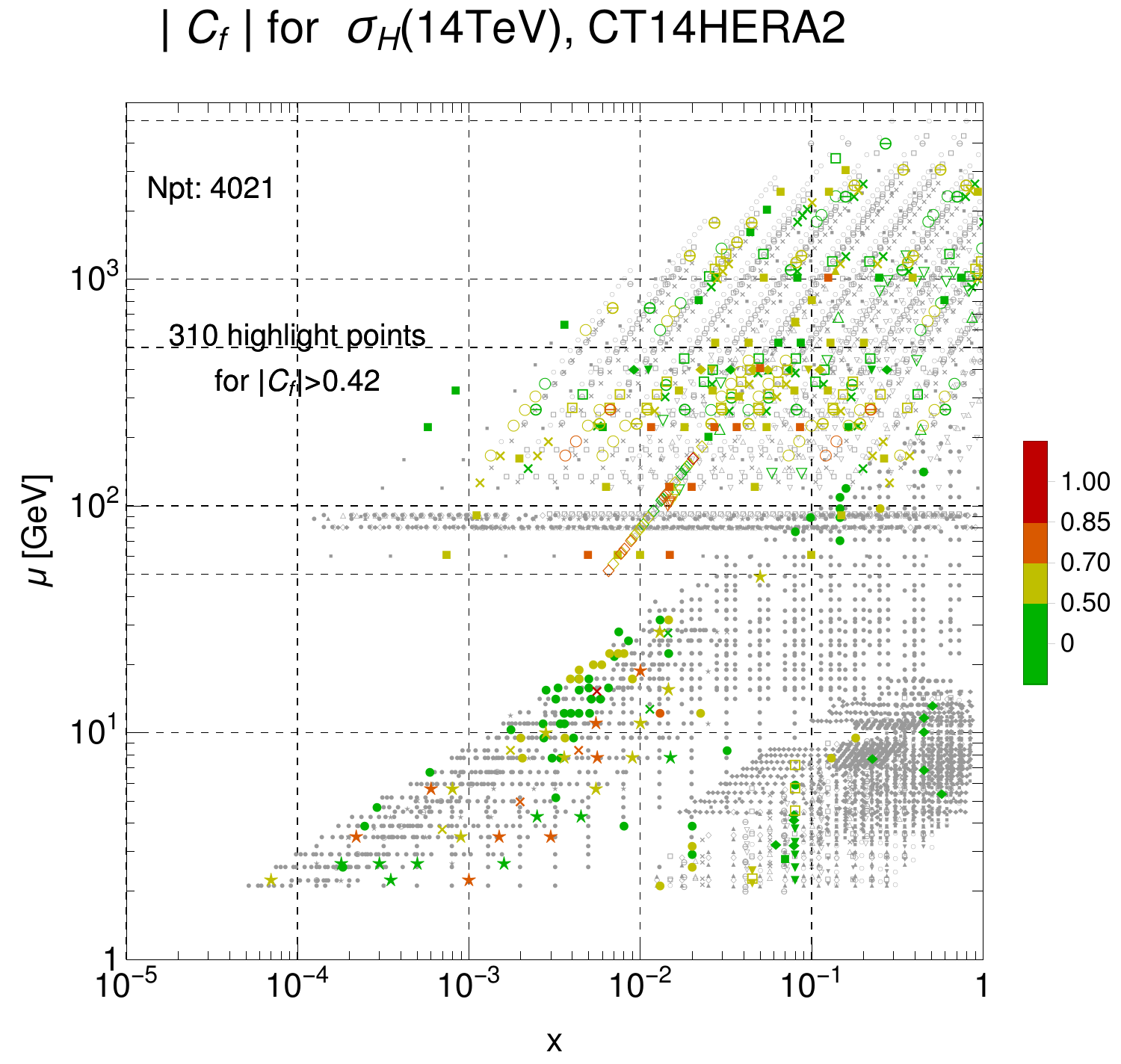}\ \ \
\includegraphics[width=0.47\textwidth]{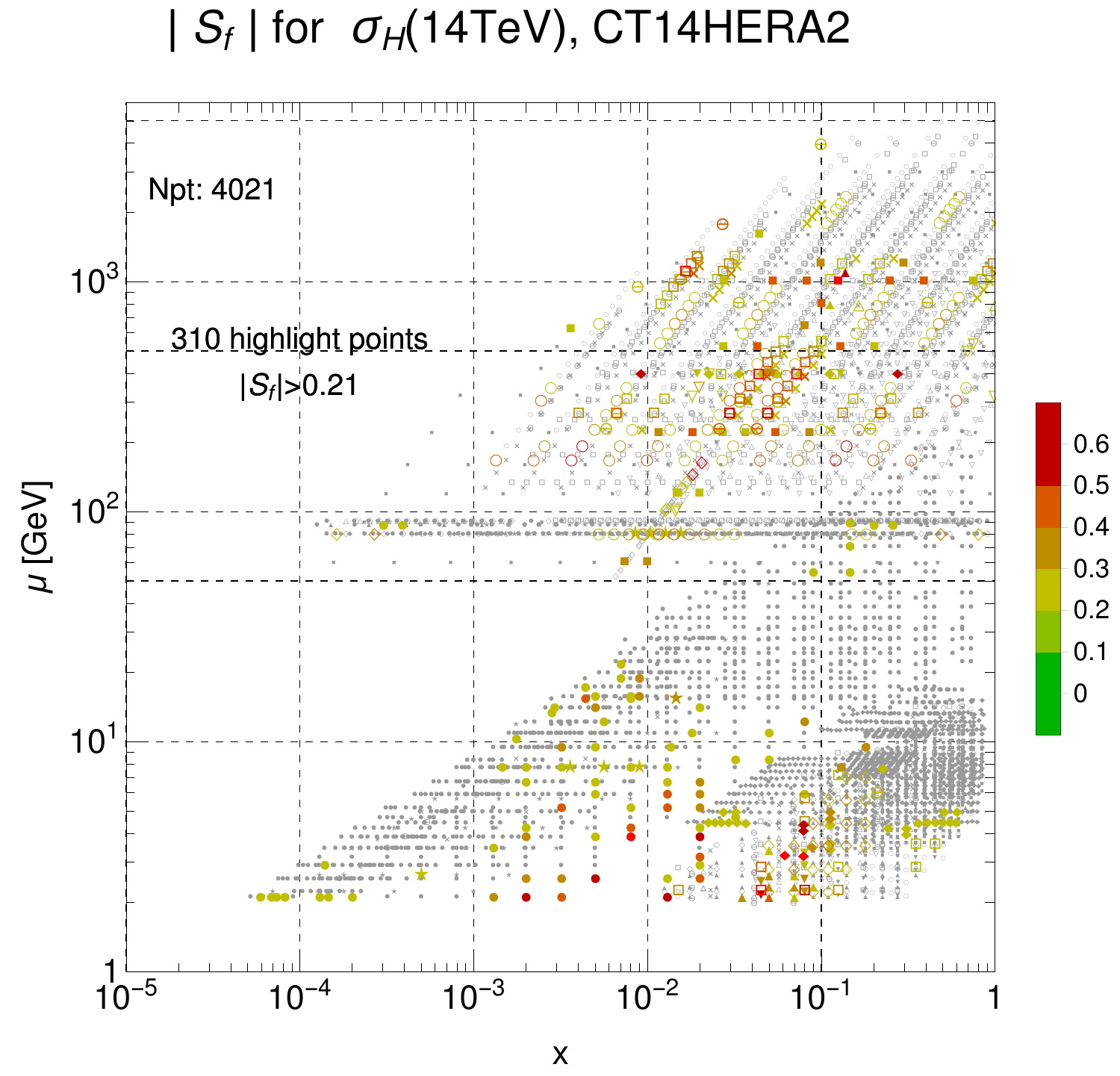}
\caption{
  For the full CTEQ-TEA dataset of Fig.~\ref{fig:data}, we show the
absolute correlation $|C_{f}|$ and sensitivity $|S_{f}|$ associated
with the 14 TeV Higgs production cross section $\sigma_{H^{0}}(14\,\mathrm{TeV})$.
310 input data points with most significant magnitudes of $|C_{f}|$
and $|S_{f}|$ are highlighted with color.
When only the $|C_{f}|$ plot is considered, only a very small
subpopulation of jet production data (diagonal open circles and closed squares
with $\mu\gtrsim100$ GeV) exhibits significant correlations with
$|C_f|>0.7$ (orange and red colors),
as well as some HERA DIS, high-$p_T$ $Z$ boson, and $t\bar{t}$ production data points.
Our novel definition for the sensitivity in the right panel, on the other
hand, reveals more points that have comparable potency for constraining
the Higgs cross section. In this case, a larger fraction of the jet production
points is important (especially CMS measurements of CMS8jets'17 and CMS7jets'14),
as well as a number of other processes at smaller $\mu$, particularly
DIS data from HERA, BCDMS, NMC, CDHSW, and CCFR (experiments HERAI+II'15, BCDMSd'90,
NMCrat'97, CDHSW-F2'91, CCFR-F2'01, CCFR-F3'97). Although its cumulative impact
is comparatively modest, ATLAS $t\bar{t}$
production data (ATL8ttb-pt'16, ATL8ttb-y\_ave'16, ATL8ttb-mtt'16, ATL8ttb-y\_ttb'16)
register significant per-point sensitivities, as do E866 $pp$ Drell-Yan pair production
(E866pp'03), LHCb $W, $Z production (LHCb7ZWrap'15, LHCb8WZ'16), and charge lepton
asymmetries at D0 and CMS (D02Masy'08, CMS7Masy2'14, CMS7Easy'12). Similarly,
some of the high-$p_T$ $Z$ production information (ATL7ZpT'14, ATL8ZpT'16)
from ATLAS provide modest constraints.
\label{fig:CorrSensH14}}
\end{figure*}

In fact, in the numerical approach presented in the forthcoming sections,
the user can quantify the sensitivity of data not only to individual
PDF flavors, but even to specific physical observables, including
the modifications due to correlated systematic uncertainties in every
experiment of the CT14HERA2 analysis. For example, for Higgs boson
production via gluon fusion ($gg\to H$) at the LHC 14 TeV, the short-distance
cross sections are known up to N$^{3}$LO with a scale uncertainty
of about 3\% \cite{Anastasiou:2016cez}. It has been suggested that
$t\bar{t}$ production and high-$p_{T}$ $Z$ boson production on their own
constrain the gluon PDF in the $x$ region
sensitive to the LHC Higgs production, and that these are comparable to the
constraints from LHC and Tevatron data \cite{Czakon:2016olj,Boughezal:2017nla}.
Verifying the degree to which this hypothesis is true has been difficult
without actually including all these data in a fit.

As an alternative to doing a full global fit, we can critically assess
this supposition in the context of the entire global dataset
of Fig.~\ref{fig:data} using the Hessian correlations and sensitivities,
$|C_{f}|$ and $|S_{f}|$.
The detailed procedure is explained in Secs.~\ref{sec:Correlations} and \ref{sec:Sensitivities}.
In the example at hand, we could rely on the established wisdom that
the theoretical cross sections that have an especially large correlation
with $\sigma_{H^{0}}$ may constrain the PDF dependence of
$\sigma_{H^{0}}$; say, when $|C_f|\gtrsim 0.7$ \cite{Nadolsky:2008zw}. 
Along this reasoning, the left frame in Fig.~\ref{fig:CorrSensH14}
illustrates 310 experimental data points in $\{x,\mu\}$ space 
that have the highest absolute correlation, $|C_{f}|$, between the point's
statistical residual defined in
Sec.~\ref{sec:Correlations} and the cross section
$\sigma_{H^{0}}$ at 14 TeV via  the CT14HERA2 NNLO PDFs.
To locate such points in the figure,
we highlighted them with color according to
the convention shown on the color scale to the right.
The respective $|C_f|$ for the highlighted data points ranges between
0.42 and 1. The rest of the data points have smaller correlations and are
shown in gray.

We find that the 310 data points with the highest correlation
for $\sigma_{H^{0}}$ belong to 20 experiments. Nearly all of them
are contributed by HERA Neutral Current (NC) DIS,
LHC and Tevatron jet production, and HERA charm production.
Some of the data points with highest $|C_f|$ come from high-$p_T$ $Z$ boson
and even $t\bar t$ production. 

The correlations $C_f$, however, do not reflect the experimental
uncertainties, which vary widely across the experiments.  
In the left panel of Fig.~\ref{fig:CorrSensH14},
fewer than 30 points have a strong correlation of $0.7$ or more; but
more data points impose relevant constraints in the global fit. 
To
include the information about the uncorrelated and correlated
experimental errors, in the right panel
of Fig.~\ref{fig:CorrSensH14}, we plot the distributions of 310 data
points with the highest sensitivity parameter $S_f$, which more faithfully
reproduces the actual constraints during the fitting.
In general, we find substantial differences between the
 $C_f$ and $S_f$ distributions. Even the most significant correlations,
of order $|C_{f}| \sim 0.7$ and above, do not guarantee a significant
contribution of the experimental point to the log-likelihood
$\chi^2$ if the errors are large. On the other hand,
we argue that $|S_f|$ is closely related to a
contribution of the data point to $\chi^2$. According to the
distribution in the right figure, the 310 data points with the highest
sensitivity $|S_f|$ to $\sigma_{H^0}(\mbox{14 TeV})$ arise from 27
experiments. Among these data points,  only some have
a large correlation $|C_f|$ with
$\sigma_{H^0}(\mbox{14 TeV})$. Nonetheless, they have medium-to-large
sensitivity, $|S_f| > 0.21$, according to the criterion developed
in Sec.~\ref{sec:Sensitivities}.
We stress that, while one might suggests plausible dynamical reasons why
certain experiments might be particularly sensitive to Higgs production via
the gluon PDF, ({\it e.g.}, via the leading-order $qg$ and $gg$
hard cross sections in jet production and DGLAP scaling violations in
inclusive DIS), this reasoning alone does not predict the actual
sensitivity revealed by $S_f$ in the presence of multiple experimental
constraints.  

As one noticeable difference from the $|C_f|$ figure,
while inclusive DIS at HERA continues to
contribute a large number of data points  (about 80) with a high
$|S_f|$, also the fixed-target DIS experiments (BCDMS, NMC, CDHSW,
CCFR) contribute about the same number of sensitive points in the
right panel that were not identified by large correlations. Other sensitive
points belong to the jet production
data sets from ATLAS and CMS and some vector boson production
experiments (muon charge asymmetries at D0, CMS;
E866 low-energy Drell-Yan production; LHCb 7 TeV $W$ and $Z$ cross
sections).

On the other
hand, HERA charm production, ATLAS 7/8 TeV high-$p_T$ $Z$ production,
have suppressed sensitivities despite their large correlations,
reflecting the larger experimental uncertainties in these measurements. 
While the LHC $t\bar t$ production experiments have large per-point
sensitivities, they contribute relatively little to $\chi^2$ because
of their small total number of data points. From this comparison,
one finds, perhaps somewhat unexpectedly, that fixed-target DIS
experiments impose important constraints on $\sigma_{H^0}(\mbox{14
  TeV})$, thus complementing the HERA inclusive DIS data.
One would conclude that efforts to constrain PDF-based SM predictions
for Higgs production by relying only on a few points of $t\overline{t}$
data, but to the neglect of high-energy jet production points, would
be significantly handicapped by the absence of the latter. We will
return to this example in Sec.~\ref{sec:CaseCTEQ-TEA}.

The discriminating power of a sensitivity-based analysis 
therefore forms the primary motivation
for this work, and we present the attendant details below. To assess
information about the PDFs encapsulated in the residuals for large
collections of hadronic data implemented in the CTEQ-TEA global analysis,
we make available a new statistical package \textsc{PDFSense} to map
the regions of partonic momentum fractions $x$ and QCD factorization
scales $\mu$ where the experiments impose strong constraints on the
PDFs. In companion studies,
we have applied \textsc{PDFSense} to select new data sets
for the next generation of the
CTEQ-TEA global analysis, to quantitatively explore the physics
potential for constraining the PDFs at a future
Electron-Ion Collider (EIC) \cite{Accardi:2012qut,Boer:2011fh,Abeyratne:2012ah,Aschenauer:2014cki}
and Large Hadron-Electron Collider (LHeC) \cite{AbelleiraFernandez:2012cc}, and to investigate the
potential of high-energy data to inform lattice-calculable
quantities \cite{Lin:2017snn} like the Mellin moments of structure functions \cite{Gockeler:1995wg} and quark quasi-distributions \cite{Ji:2013dva}.
We reserve many instructive results for follow-up publications currently in preparation,
while presenting select calculations in this article to demonstrate the power of the method. We find that the
sensitivity technique generally agrees with the preliminary
CTEQ-TEA fits and Hessian reweighting realized in the \textsc{ePump} program
\cite{Schmidt:2018hvu}. However, assessing the sensitivity is much
simpler than doing the global fit. It does not require access to a fitting
program or the application of (potentially subtle) PDF reweighting techniques.  

The remainder of the article proceeds as follows. Pertinent aspects
of the PDFs and their standard determination via QCD global analyses
are summarized in Sec.~\ref{sec:PDF-preliminaries}. Then, 
we introduce {\it normalized residual variations} to extract,
visualize, and quantify statistical information about the global QCD
fit. In Sec.~\ref{sec:QuantifyingDistributionsOfResiduals}, we
construct a number of statistical quantities that
characterize the PDF constraints in the global analysis using the
residual variations. In Sec.\ \ref{sec:CaseCTEQ-TEA}, we apply
the thus constructed sensitivity parameter
to examine the impact of various CTEQ-TEA datasets
on extractions of the gluon PDF $g(x,\mu)$. In this section and
in the conclusion contained
in Sec.\ref{sec:Conclusions}, we emphasize a number of {\it physics insights}
that we obtained by applying our sensitivity analysis techniques. Additional
aspects of the technique and supplementary tables are reserved for
Apps.~\ref{sec:supp}, \ref{sec:Tables}, and \ref{sec:SM}.

\section{PDF preliminaries \label{sec:PDF-preliminaries}}

\subsection{Data residuals in a global QCD analysis \label{sec:Data-residuals}}

While various theoretical models exist for computing nucleon PDFs
\cite{Farrar:1975yb,Hobbs:2014lea,Hobbs:2013bia},
unambiguous evaluation of the PDFs entirely in terms of
QCD theory is not yet possible due to the fact that the PDFs can in
general receive substantial nonperturbative contributions at infrared
momenta. For this reason, precise PDF determination has proceeded
mainly through the technique of the QCD global analysis \textemdash{}
a method enabled by QCD factorization and PDF universality.

In this approach, a highly flexible parametric form is ascribed for
the various flavors in a given analysis at a relatively low scale
$Q_{0}^{2}$. For example, one might take the input PDF for a given
quark flavor $f$ to be a parametric form 
\begin{equation}
f(x,\mu^{2}=Q_{0}^{2})=A_{f,0}\,x^{A_{f,1}}(1-x)^{A_{f,2}}\,F(x;\,A_{f,3},\dots)\ ,\label{eq:fitform}
\end{equation}
in which $F(x;\:A_{f,3},\dots)$ can be a suitable polynomial function,
\textit{e.g.}, a Chebyshev or Bernstein polynomial, or replaced with
a feed-forward neural network $\mathrm{NN}_{f}(x)$ as in the NNPDF
approach. While the full statistical theory for PDF determination
and error quantification is beyond the intended range of this analysis,
roughly speaking, a best fit is found for a vector $\vec{A}$ of $N$
PDF parameters $A_{l}$ by minimizing a goodness-of-fit function $\chi^{2}$
describing agreement of the QCD data and physical observables computed
in terms of the PDFs. Based on the behavior of $\chi^{2}$ in the
neighborhood of the global minimum, it is then possible to construct
an ensemble of error PDFs to quantify uncertainties of PDFs at a predetermined
probability level.

There are various ways to evaluate uncertainties on PDFs,
\emph{e.g.}, the Hessian \cite{Pumplin:2001ct,Pumplin:2002vw},
the Monte Carlo \cite{Giele:1998gw,Giele:2001mr}, and the
Lagrange Multiplier approaches \cite{Stump:2001gu}. In this analysis
our default PDF input set is CT14HERA2, which uses the Hessian method
to estimate uncertainties and is therefore based on the quadratic
assumption for $\chi^{2}(\vec{A}$) in the vicinity of the global
minimum. In the Hessian method, an orthonormal basis of PDF parameters
$\vec{a}$ is derived from the input PDF parameters $\vec{A}$ by
the diagonalization of a Hessian matrix $H$, which encodes the second-order
derivatives of $\chi^{2}$ with respect to $A_{l}$. The eigenvector
PDF combinations $\vec{a}_{l}^{\pm}$ are found for two extreme variations
from the best-fit vector $\vec{a}_{0}$ along the direction of the
$l^{th}$ eigenvector of $H$ allowed at a given probability level.
The uncertainty on a QCD observable $X$ can then be estimated with
one of the available ``master formulas'' \cite{Pumplin:2002vw,Nadolsky:2001yg},
the ``symmetric'' variety of which is 
\begin{align}
\Delta X & =\frac{1}{2}\sqrt{\sum_{l=1}^{N}(X_{l}^{+}-X_{l}^{-})^{2}}\ .\label{DelX}
\end{align}

In the CTEQ-TEA global analysis, the $\chi^{2}$ function accounts
for multiple sources of experimental uncertainties, as well as for
some prior theoretical constraints on the $a_{l}$ parameters. Consequently,
the global $\chi^{2}$ function takes the form 
\begin{equation}
\text{\ensuremath{\chi}}_{global}^{2}=\sum_{E}\chi_{E}^{2}+\chi_{th}^{2}\ ,
\label{eq:chi2glob}
\end{equation}
where the sum runs over all experimental datasets $(E);$ and $\chi_{th}^{2}$
imposes theoretical constraints. The complete formulas for $\chi_{E}^{2}$
and $\chi_{th}^{2}$ can be found in Ref.~\cite{Gao:2013xoa}. For
the purposes of this paper, we express $\chi_{E}^{2}$ for each experiment
$E$ in a compact form as a sum of squared\emph{ shifted residuals}
$r_{i}^{2}(\vec{a})$, which are summed over $N_{\mathit{pt}}$ individual
data points $i$ in this experiment, as well as the contributions
of $N_{\lambda}$ best-fit nuisance parameters $\overline{\lambda}_{\alpha}$
associated with correlated systematic errors: 
\begin{align}
\chi_{E}^{2}(\vec{a}) & =\sum_{i=1}^{N_{\mathit{pt}}}\,r_{i}^{2}(\vec{a})+\sum_{\alpha=1}^{N_{\lambda}}\overline{\lambda}_{\alpha}^{2}(\vec{a})\ .\label{eq:chi2}
\end{align}
In turn, $r_{i}(\vec{a})$ for the $i^{th}$ data point is constructed
from the theoretical prediction $T_{i}(\vec{a})$ evaluated in terms
of PDFs, total uncorrelated uncertainty $s_{i}$, and the shifted
central data value $D_{i,sh}(\vec{a})$: 
\begin{align}
	r_{i}(\vec{a}) & =\frac{1}{s_{i}}\,\big(T_{i}(\vec{a})-D_{i,\mathit{sh}}(\vec{a})\big)\ .\label{eq:residual}
\end{align}
This representation arises in the Hessian formalism due to the presence
of correlated systematic errors in many experimental datasets, which
require $\chi_{E}^{2}$ to depend on nuisance parameters $\lambda_{\alpha}$.
This is in addition to the dependence of $\chi_{E}^{2}$ on the PDF
parameters $\vec{a}$ and theoretical parameters such as $\alpha_{s}(M_{Z})$
and particle masses. The $\lambda_{\alpha}$ parameters are optimized
for each $\vec{a}$ according to the analytic solution derived in
Appendix B of Ref.~\cite{Pumplin:2002vw}. Optimization effectively
shifts the central value $D_{i}$ of the data point by an amount determined
by the optimal nuisance parameters $\overline{\lambda}_{\alpha}(\vec{a})$
and the correlated systematic errors $\beta_{i\alpha}:$ 
\begin{equation}
	D_{i}\rightarrow D_{i,\mathit{sh}}(\vec{a})=D_{i}-\sum_{\alpha=1}^{N_{\lambda}}\beta_{i\alpha}\overline{\lambda}_{\alpha}(\vec{a})\ .
\end{equation}
It should be noted that the contribution of the squared best-fit nuisance
parameters to $\chi_{E}^{2}$ in Eq.~(\ref{eq:chi2}) is dominated
in general by the first term involving the shifted residuals, which
tends to be much larger \textemdash{} especially for more sizable
datasets.

We point out also that some alternative representations for $\chi^{2}$
include the correlated systematic errors via a covariance matrix $\left(\mbox{cov}\right)_{ij}$,
rather than the above mentioned CTEQ-preferred form that explicitly
operates with $\lambda_{\alpha}$. Various $\chi^{2}$ definitions
in use are reviewed in \cite{Ball:2012wy}, as well as in \cite{Alekhin:2014irh}.
Crucially, however, the representations based upon operating with
$\lambda_{\alpha}$ and $\left(\mbox{cov}\right)_{ij}$ are derivable
from each other \cite{Gao:2013xoa}. From an extension of the derivation
in Ref.~\cite{Pumplin:2002vw}, we may relate the shifted residual
to the covariance matrix at an $i^{th}$ point and optimal nuisance
parameters as 
\begin{align}
r_{i}(\vec{a})\  & =\ s_{i}\sum_{j=1}^{N_{\mathit{pt}}}(\mathrm{cov}^{-1})_{ij}\,\left(T_{j}(\vec{a})-D_{j}\right),\label{eq:res-cov}\\
\overline{\lambda}_{\alpha}(\vec{a}) & =\sum_{i,j=1}^{N_{\mathit{pt}}}(\mathrm{cov}^{-1})_{ij}\frac{\beta_{i\alpha}}{s_{i}}\frac{\left(T_{j}(\vec{a})-D_{j}\right)}{s_{j}},
\end{align}
where 
\begin{equation}
(\mathrm{cov}^{-1})_{ij}\ =\ \left[\frac{\delta_{ij}}{s_{i}^{2}}\,-\,\sum_{\alpha,\beta=1}^{N_{\lambda}}\frac{\beta_{i\alpha}}{s_{i}^{2}}A_{\alpha\beta}^{-1}\frac{\beta_{j\beta}}{s_{j}^{2}}\right]\ ,\label{eq:covmat}
\end{equation}
and 
\begin{equation}
A_{\alpha\beta}\ =\ \delta_{\alpha\beta}\,+\,\sum_{k=1}^{N_{\mathit{pt}}}\frac{\beta_{k\alpha}\beta_{k\beta}}{s_{k}^{2}}\ .
\end{equation}
Thus, even for those PDF analyses which operate with the covariance
matrix one is still able to determine the shifted residuals $r_{i}$
from $\left(\mbox{cov}^{-1}\right)_{ij}$ using Eq.~(\ref{eq:res-cov}).
In this article, we conveniently follow the CTEQ methodology and obtain
$r_{i}(\vec{a})$ directly from the CTEQ-TEA fitting program, together
with the optimal nuisance parameters $\overline{\lambda}_{\alpha}(\vec{a})$
and shifted central data values $D_{i,sh}(\vec{a}).$

\subsection{Visualization of the global fit with the help of residuals}

The shifted residuals $r_{i}$ draw our interest because, in consequence
of the definitions in Eqs.~(\ref{eq:chi2})-(\ref{eq:residual}),
they contain substantial low-level information about the agreement
of PDFs with every data point in the global QCD fit in the presence
of systematic shifts. The response of $r_{i}(\vec{a})$ to the variations
in PDFs depends on the experiment type and kinematic range associated
with the $i^{th}$ data point, and the totality of these responses
can be examined with modern data-analytical methods. The sum of squared
residuals over all points of the global dataset renders the bulk
of the log-likelihood, or experimental, component $\chi_{E}^{2}$
of the global $\chi^{2}$. In turn, the root-mean-squared residual
$\langle r_{0}\rangle_{E}$ for experiment $E$ and the central PDF
set $\vec{a}_{0}$ is tied to $\chi_{E}^{2}(\vec{a}_{0})/N_{\mathit{pt}},$
the standard measure of agreement with experiment $E$ at the best
fit: 
\begin{equation}
\langle r_{0}\rangle_{E}\equiv\sqrt{\frac{1}{N_{\mathit{pt}}}\sum_{i=1}^{N_{\mathit{pt}}}r_{i}^{2}(\vec{a}_{0})}=\sqrt{\frac{1}{N_{\mathit{pt}}}\left(\chi_{E}^{2}(\vec{a}_{0})-\sum_{\alpha=1}^{N_{\lambda}}\overline{\lambda}_{\alpha}^{2}(\vec{a_{0}})\right)}\approx\sqrt{\frac{\chi_{E}^{2}(\vec{a}_{0})}{N_{\mathit{pt}}}}.\label{r0E}
\end{equation}
Notice that $\langle r_0 \rangle_E \approx 1$ when the fit to the experimental data set $E$ is good.

We will now invoke the Hessian formalism to first organize the analysis
of the PDF dependence of individual residuals, and then introduce
a framework to evaluate sensitivity of individual data points to PDF-dependent
physical observables. To test the effectiveness of the proposed method,
we study constraints using CT14HERA2 parton distributions \cite{Hou:2016nqm}
fitted to datasets from DIS processes, $Z\rightarrow l^{+}l^{-}$,
$d\sigma/dy_{l}$, $W\rightarrow l\nu$, and jet production ($p_{1}p_{2}\rightarrow jjX)$.
We include both the experiments that were used to construct the CT14HERA2
dataset, as well as a number of LHC experiments that may be fitted
in the future. The experimental data sets are summarized in
Tables~\ref{tab:EXP_1}-\ref{tab:EXP_3}.

Given the urgency in improving
constraints on the gluon PDF for investigations of the Higgs sector,
we focus attention on several candidate experiments that may probe
$g(x,\mu)$: high-$p_T$ $Z$-boson
production (ATL8ZpT'16, ATL7ZpT'14), $t\bar{t}$ production (ATL8ttb-pt'16, ATL8ttb-y\_ave'16, ATL8ttb-mtt'16, ATL8ttb-y\_ttb'16),
as well as high-luminosity or alternative data sets for jet production, such as
the high-luminosity ATLAS 7 TeV jet data (ATLAS7jets'15) that is to replace the
counterpart low-luminosity set ATL7jets'12, or the CMS 7 TeV jet data set (CMS7jets'14) 
that extends to lower jet $p_T$ and higher rapidity,
$2.5<|y_{j}|<3$, than the previously fitted CMS 7 TeV
jet data set (CMS7jets'13).\footnote{As a result, a small number of data points
  that contributes to both the data sets CMS7jets'14 and CMS7jets'13 is double-counted
  in the histograms, without affecting the conclusions.
}
The dependence of such experiments on $g(x,\mu)$ is scrutinized in a
number of ways. We examine their statistical properties using both the
PDFs from the CT14HERA2 NNLO analysis, which already impose
significant constraints on the large-$x$ gluon using
the Tevatron inclusive jet data sets, CDF2jets'09 and D02jets'08; and in some comparisons using
a special version of the NNLO PDFs that are fitted to the same CT14HERA2
data set, except without including the above jet data sets. As yet
another aspect, we investigate a range of measurements of Drell-Yan
pair production cross sections and charge lepton asymmetries with the
goal to understand their sensitivity predominantly to the (anti)quark sector. 

To parametrize the response of a residual $\vec{r}_{i}$, we evaluate
it for every eigenvector PDF $\vec{a}_{l}^{\pm}$ of the CT14HERA2
PDF set with $N=28$ PDF parameters. Then, given the
{\it normalized residual variations}
\begin{equation}
\delta_{i,l}^{\pm}\equiv\left(r_{i}(\vec{a}_{l}^{\pm})-r_{i}(\vec{a}_{0})\right)/\langle r_{0}\rangle_{E}\label{deltapmil}
\end{equation}
between the residuals for the PDF eigenvectors $\vec{a}_{l}^{\pm}$
and for the CT14HERA2 central PDF $\vec{a}_{0}$, we construct a $2N$-dimensional
vector 
\begin{equation}
\vec{\delta}_{i}=\left\{ \delta_{i,1}^{+},\,\delta_{i,1}^{-},\,...,\delta_{i,N}^{+},\,\delta_{i,N}^{-}\right\} \label{deltail}
\end{equation}
for each data point of the global dataset.

The components of $\vec{\delta}_{i}$ parametrize responses of $r_{i}$
to PDF variations along the independent directions given by $\vec{a}_{l}^{\pm}$.
The differences are normalized to the central root-mean-square (r.m.s.)
residual $\langle r_{0}\rangle_{E}$ of experiment $E$ {[}see Eq.~(\ref{r0E}){]}
so that the normalized residual variations do not significantly depend on $\chi^{2}(\vec{a}_{0})/N_{\mathit{pt}},$
the quality of fit to experiment $E$. Recall that a substantial spread
over the fitted experiments is generally obtained for $\chi_{E}^{2}/N_{\mathit{pt}}$.
Moreover, it is reasonable to expect significantly larger values for
$\chi_{E}^{2}/N_{\mathit{pt}}$ for the experiments that have not been
yet fitted, but are included in the analysis of the residuals, \textit{e.g.},
the new LHC experiments shown in Fig.~\ref{fig:data}. With the definitions
in Eqs. (\ref{deltapmil}) and (\ref{deltail}), however, $\vec{\delta_{i}}$
is only weakly sensitive to $\chi_{E}^{2}/N_{\mathit{pt}}$. 

Thus, we represent the PDF-driven variations of the residuals of a
global dataset by a bundle of vectors $\vec{\delta}_{i}$ in a $2N$-dimensional
space.\footnote{In this section, we consider separate variations along $\vec{a}_{l}$
in the positive and negative directions. Alternatively, it is possible
to work with a vector of $N$ symmetric differences $\delta_{i,l}\equiv\left(r_{i}(\vec{a}_{l}^{+})-r(\vec{a}_{l}^{-})\right)/\left(2\langle r_{0}\rangle_{E}\right)$
and arrive at similar conclusions. Symmetric differences will be employed
to construct correlations and sensitivities in Sec.~\ref{sec:QuantifyingDistributionsOfResiduals}.
\label{fn:sym-deltail}} This mapping opens the door to applying various data-analytical methods
for classification of the data points and identifying the data points
of the utmost utility for PDF fits. As the length of $\vec{\delta}_{i}$
is equal to the PDF-induced fractional error on $r_{i}$ as compared
to the average residual at the best fit, it can be argued that important
PDF constraints arise 
from new data points that either have a large $|\vec{\delta}_{i}|$
or are otherwise distinct from the existing data points. Conversely,
new data points with a small $|\vec{\delta}_{i}|$, or the ones that
are embedded in the preexisting clusters of points, are not likely to
improve constraints on the PDFs.

\subsection{Manifold learning and dimensionality reduction \label{subsec:Manifold-learning}}

\subsubsection{\emph{PCA and t-SNE visualizations} \label{sec:embedding}}

We illustrate a possible analysis technique carried out with the help
of the TensorFlow Embedding Projector software for the visualization
of high-dimensional data \cite{EmbeddingProjector}. A table of 4021
vectors $\vec{\delta}_{i}$ for the CTEQ-TEA dataset (corresponding
to our total number of raw data points) is generated by our package
\textsc{PDFSense} and uploaded to the Embedding Projector website.
As variations along many eigenvector directions result only in small
changes to the PDFs, the 56-dimensional $\vec{\delta}_{i}$ vectors
can in fact be projected onto an effective manifold spanned by fewer
dimensions. Specifically, the Embedding Projector approximates the
56-dimensional manifold by a 10-dimensional manifold using principal
component analysis (PCA). In practice, this 10-dimensional manifold
is constructed out of the 10 components of greatest variance in the
effective space, such that the most variable combinations of $\delta_{i,l}$
are retained, while the remaining 46 components needed to fully reconstruct
the original 56-dimensional $\vec{\delta}_{i}$ are discarded. However,
because the 10 PCA-selected components describe the bulk of the variance
of $\delta_{i,l}$, the loss of these 46 components results in only
a minimal relinquishment of information, and in fact provides a more
efficient basis to study $\delta_{i,l}$ variations.

We encourage the reader to download the table of the normalized
residual variations $\vec\delta_i$ for 
CT14HERA2 NNLO from the \textsc{PDFSense}
website \cite{PDFSenseWebsite} and explore it for themselves using the Embedding
Projector \cite{EmbeddingProjector} or another program for multidimensional data visualization
such as a tour \cite{Cook:2018mvr}. These tools help to understand 
the detailed PDF dependence of individual data sets
{\it without doing the global fit}. 
Performing such task has been challenging for non-experts,
if not for the PDF fitters themselves. With the proposed method, we
can visually examine the PDF dependence of the residuals from the diverse data
sets before quantitatively characterizing these distributions using
the estimators developed in the next sections.
In the future, a computer algorithm
can be written to select the experimental data for PDF fits,
based on the residual variations, and with minimal involvement from humans.

\begin{figure*}
\centering{}\includegraphics[width=0.48\textwidth]{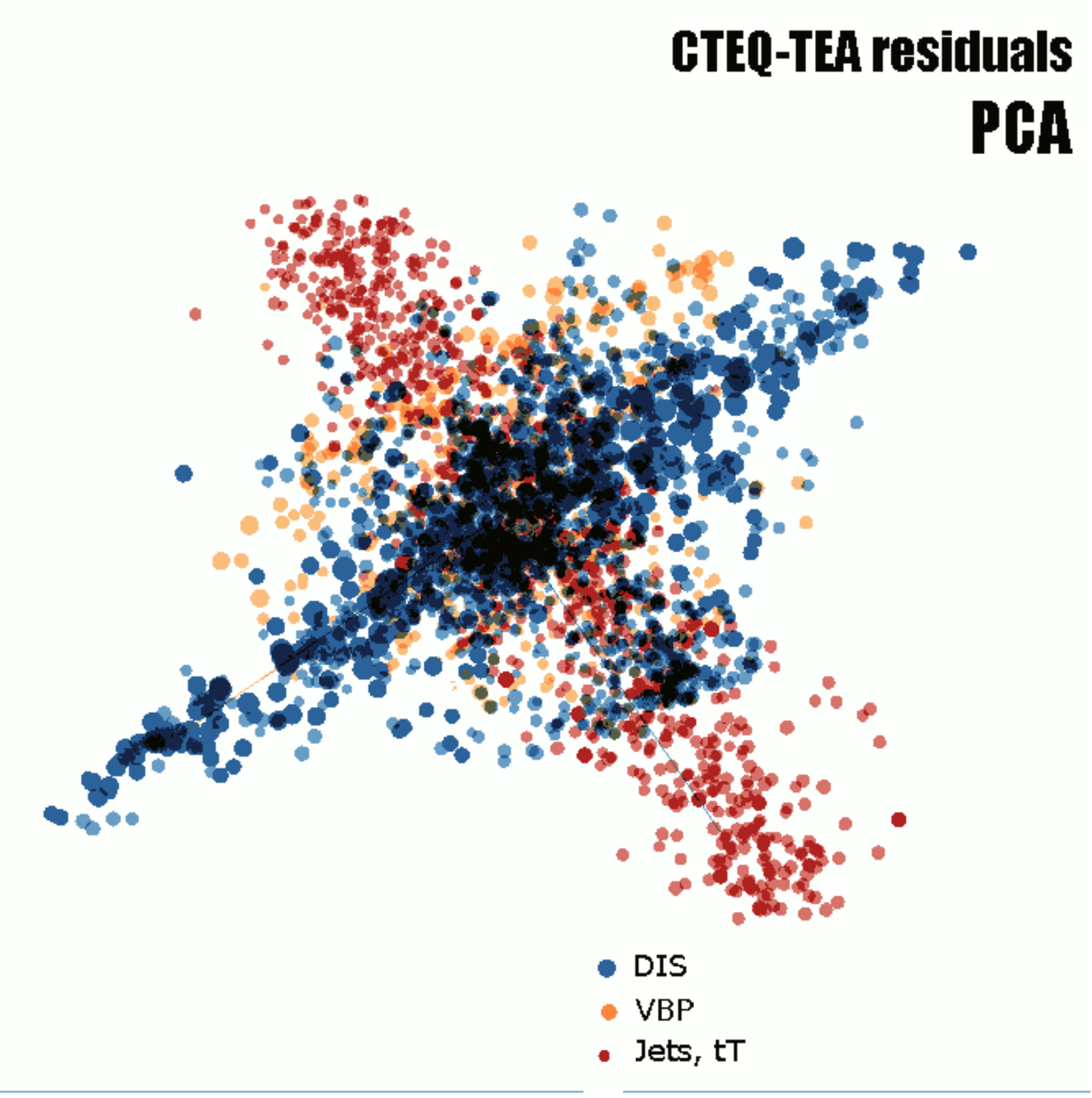}\ \ \includegraphics[width=0.48\textwidth]{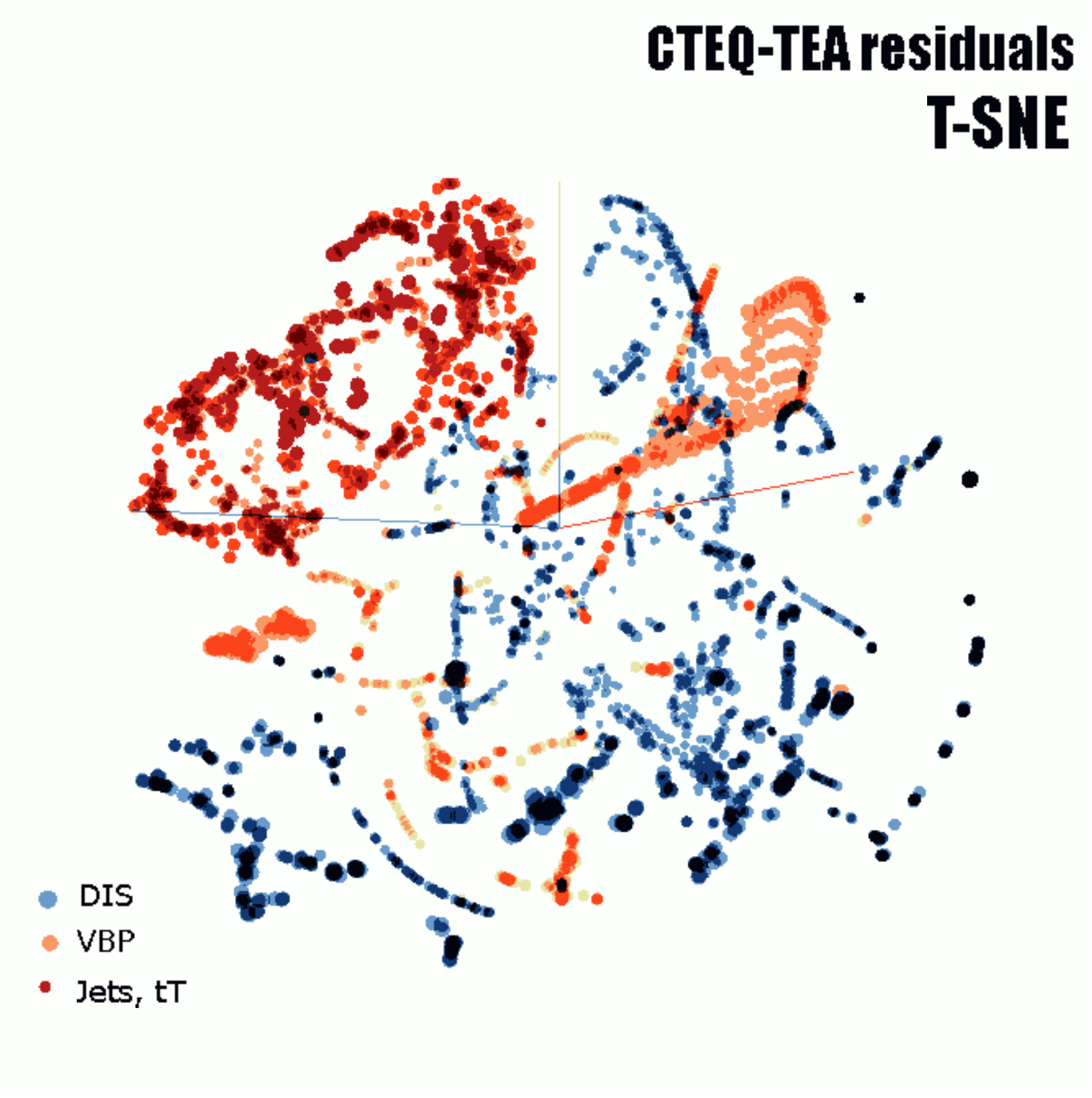}
\caption{Distributions of residual variations $\vec{\delta_{i}}$ from the
CTEQ-TEA analysis obtained by dimensionality reduction methods. Left:
a 3-dimensional projection of a 10-dimensional manifold constructed
by principal component analysis (PCA). Right: a distribution from
the 3-dimensional t-SNE clustering method. Blue, orange, and red colors
indicate data points from DIS, vector boson production, and jet/$t\bar{t}$
production processes. \label{fig:PCA-TSNE}}
\end{figure*}

To offer an illustration, while grasping the full PDF dependence of the 
data points in the original 56-parameter space is daunting, 
in the 10-dimensional representation obtained via PCA,
some directions result in efficient separation of the data points
of different types according to their residual variations. 
The left panel of Fig.~\ref{fig:PCA-TSNE} shows one such 3-dimensional
projection of $\vec{\delta}_{i}$ that separates clusters of residual
variations arising from data for DIS,
vector boson production, and jet/$t\bar{t}$ production.
In this example, the jet/$t\bar{t}$ cluster, shown in red, is roughly
orthogonal to the blue DIS cluster and intersects it. This separation
is quite remarkable, as it is based only on numerical properties of the
$\vec{\delta}_{i}$ vectors, and not on the meta-data about the types
of experiments that is entered only after the PCA is completed; in other
projections, the data types are not separated. The underlying
reasons for this separation, namely, dependence on independent PDF
combinations, will be quantified by the sensitivities in the next section.

As an alternative, the Embedding Projector can organize the $\vec{\delta}_{i}$
vectors into clusters according to their similarity using $t$-distributed
stochastic neighbor embedding (t-SNE) \cite{vanderMarten:2008}. A
representative 3-dimensional distribution of the vectors obtained
by t-SNE is displayed in the right panel of Fig.~\ref{fig:PCA-TSNE}.
In the figure, we show that the t-SNE method is able to identify and
separate the clusters of
data according to the experimental process (DIS, vector production, or
jet production). In fact, the reader can perform the t-SNE analysis
on the Embedding Projector website themselves and verify that it actually
sorts the $\vec\delta_i$ vectors into the clusters
according to their values of $x$ and
$\mu$, and even the experiment itself. This exercise demonstrates,
yet again, that the statistical residuals provided in \textsc{PDFSense}
reflect the key properties of the global fit.
Information can be extracted from them and examined  in a number of ways.  

\begin{figure*}[p]
\centering{}\includegraphics[width=0.9\textwidth,height=0.85\textheight]{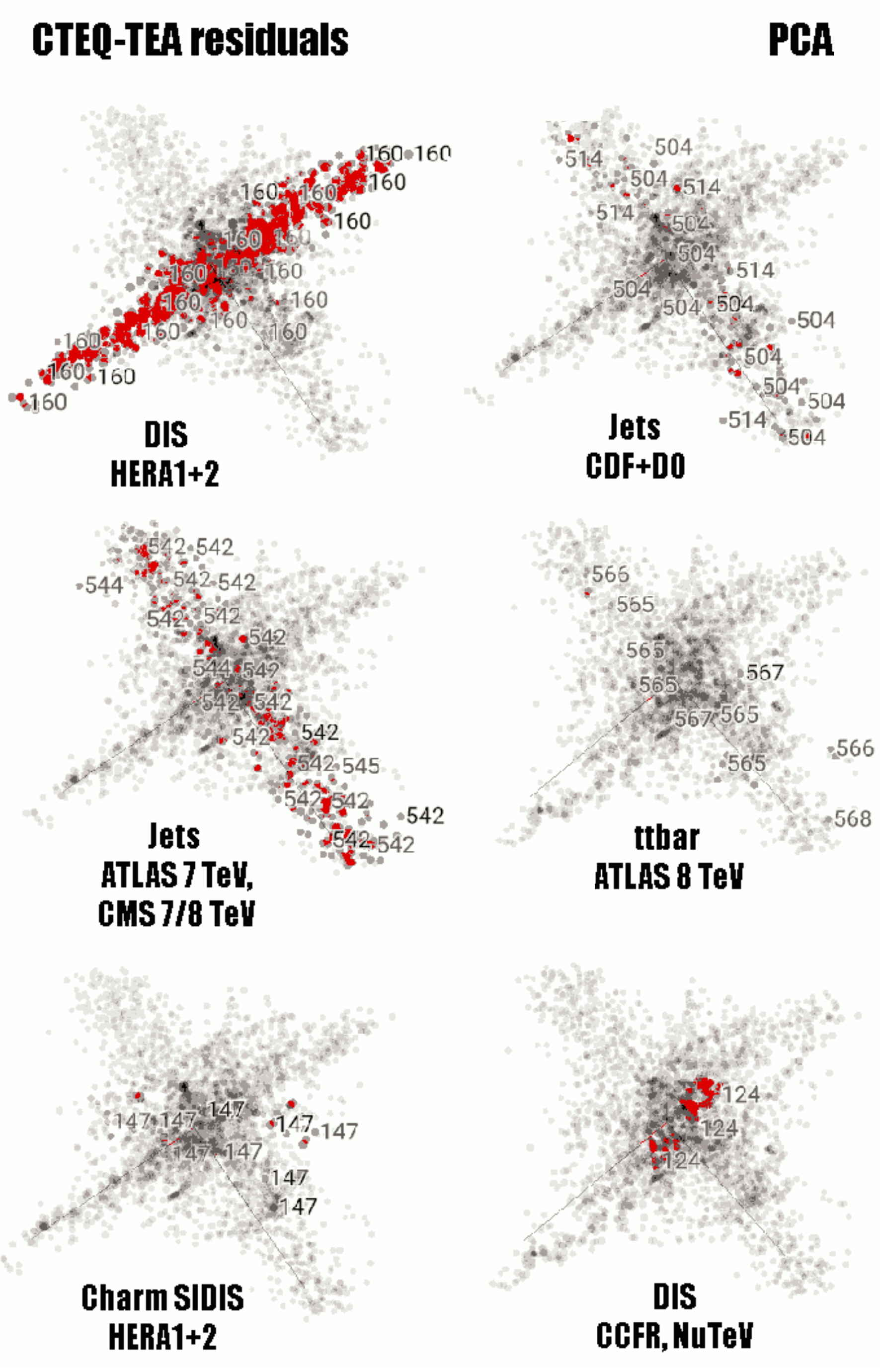}
\caption{The PCA distribution from Fig.~\ref{fig:PCA-TSNE}, indicating distributions
of points from classes of experiments. In the numbering scheme used here, points labeled
  1XX correspond to fixed-target measurements and 5XX to jet and $t\bar{t}$ production as given in Tables~\ref{tab:EXP_1}--\ref{tab:EXP_3}.
  The specific experiments are noted in the plots.
	}
\label{fig:PCA-CTExperiments}
\end{figure*}

The breakdown of the vectors over experiments in the PCA representation
is illustrated by Fig.~\ref{fig:PCA-CTExperiments}. Here, we see
that the bulk of the DIS cluster from the left Fig.~\ref{fig:PCA-TSNE}
originates with the combined HERA1+2 DIS data [HERAI+II'15]. 
The jet cluster in Fig.~\ref{fig:PCA-TSNE} will be dominated by
ATLAS and CMS inclusive jet datasets [CMS7jets'14, ATLAS7jets'15, and CMS8jets'17], 
which add dramatically more points across a wider kinematical range
on top of the CDF Run-2 and D0 Run-2 jet production datasets (CDF2jets'09)
and (D02jets'08). 

In contrast, although the $t\bar{t}$ production experiments (ATL8ttb-pt'16, ATL8ttb-y\_ave'16, ATL8ttb-mtt'16, ATL8ttb-y\_ttb'16) 
are generally characterized by large $\vec{\delta}_{i}$
vectors, they contribute only a few data points lying within the jet
cluster of Fig.~\ref{fig:PCA-CTExperiments} and, by themselves, will not make much difference in a global
fit. The same conclusion applies to data from high-$p_{T}$ $Z$ production,
which has too few points to stand out in a fit with significant inclusive
jet data samples. We return to this point in the discussion of reciprocated
distances below.

It is also interesting to note that semi-inclusive charm production
at HERA [HERAc'13] 
lies between, and partly overlaps with, the DIS and jet clusters. Finally,
CCFR/NuTeV dimuon semi-inclusive DIS [SIDIS] (CCFR-F2'01, CCFR-F3'97, NuTeV-nu'06, NuTeV-nub'06)
extends in an orthogonal direction, not well separated from the other datasets in the
selected three-dimensional projection.

\subsubsection{\emph{Reciprocated distances} \label{sec:Reciprocated-distances}}

\begin{figure*}
\centering{}\includegraphics[width=0.47\textwidth]{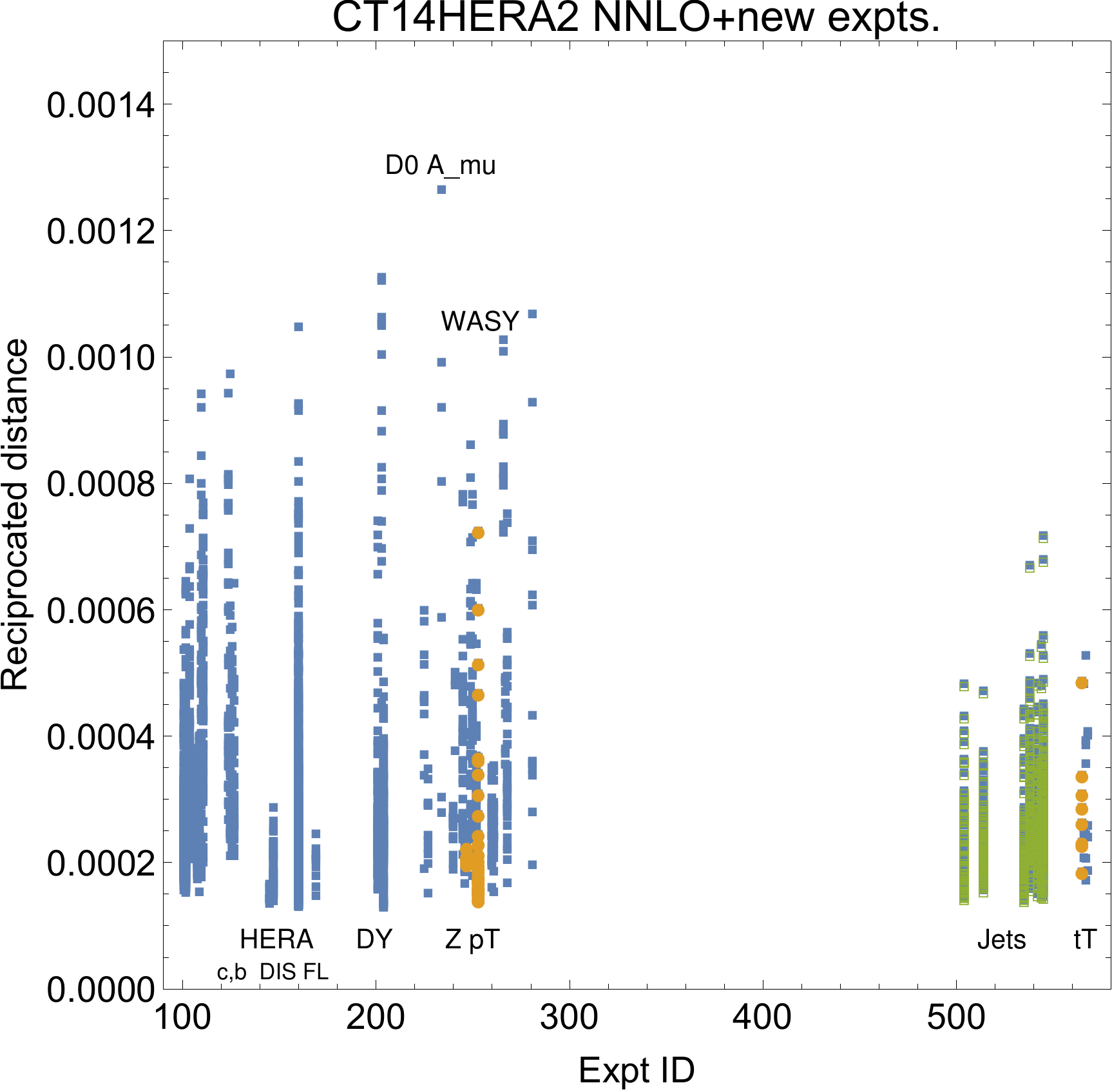}
\ \ \ \centering{}\includegraphics[width=0.47\textwidth]{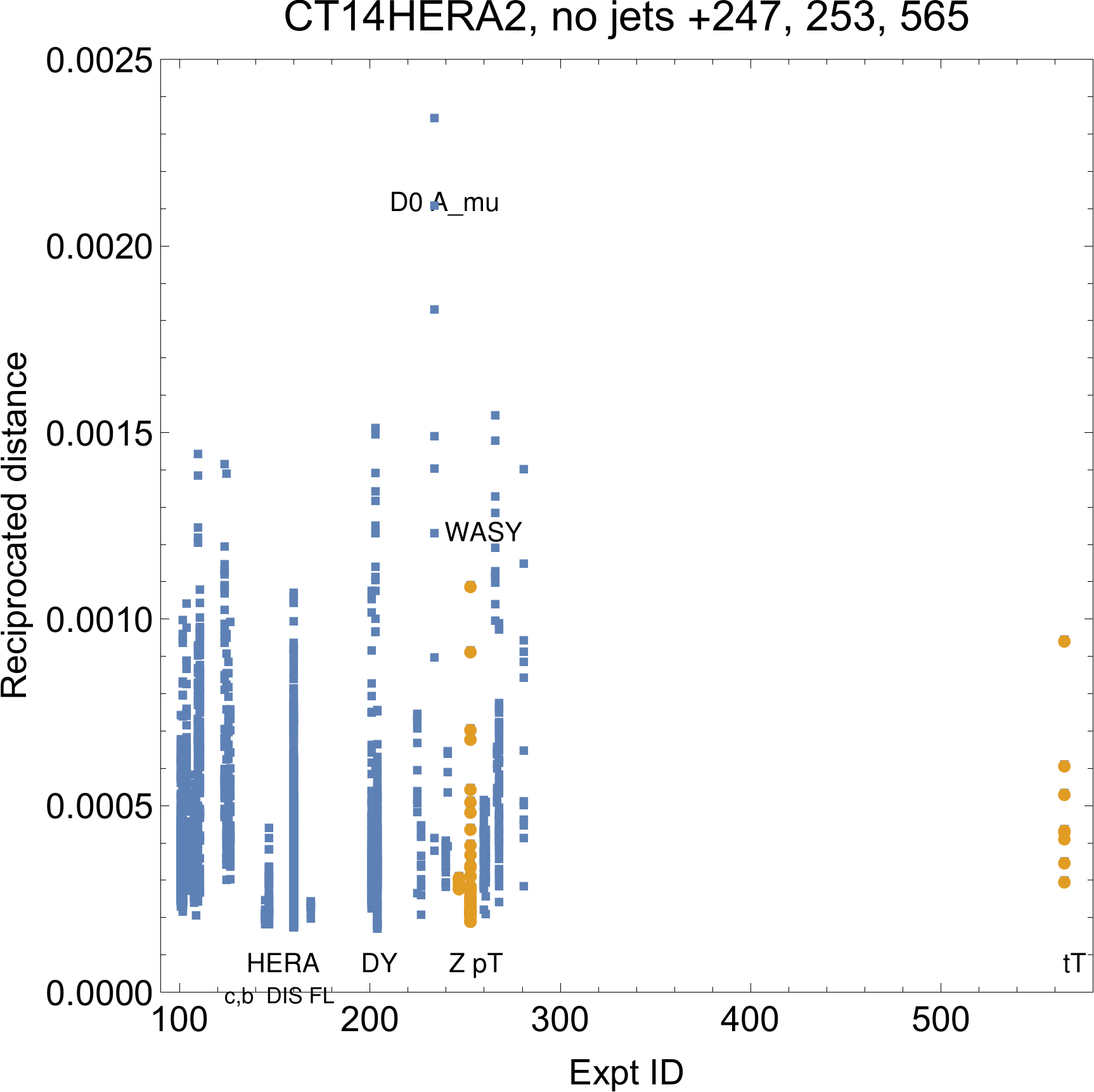}
\caption{ A plot of the reciprocated distances $\mathcal{D}_{i}$ obtained
from the PDFs fitted to the full CT14HERA2 dataset {[}left{]} and
to the CT14HERA2 dataset without jet production experiments {[}right{]}.
The horizontal axis displays numerical experimental CT IDs of the constituent
CTEQ-TEA datasets, for each of which is shown a column of values of
the reciprocated distance. We highlight columns corresponding to Expt.~IDs
ATL7ZpT'14 [247], ATL8ZpT'16 [253], and ATL8ttb-pt'16 [565] as discussed in text. \label{fig:recip}}
\end{figure*}

As a complement to the visualization methods based on PCA and t-SNE
just presented, it is also possible to evaluate another similarity
measure based on the distances between the vectors of the
residual variations. For example,
rather than applying the PCA 
to an ensemble of $\vec{\delta}_{i}$ vectors to perform dimensionality reduction,
we might instead compute over the vector space a pair-wise
\textit{reciprocated distance} measure, which we define as 
\begin{equation}
\mathcal{D}_{i}\ \equiv\ \left(\sum_{j\neq i}^{N_{\mathit{all}}}\frac{1}{|\vec{\delta}_{j}-\vec{\delta}_{i}|}\right)^{-1}\ ,\label{eq:recip}
\end{equation}
and evaluate for the $i$ points in each experimental dataset. We
allow the sum over $j$ in Eq.~(\ref{eq:recip}) to run over all
the data points in the CTEQ-TEA set regardless of experiment
(denoted by $N_{\mathit{all}}$). The distances can be computed either
in the 56-dimensional space or in the reduced dimensionality space.\footnote{Alternative definitions for the reciprocated distance can be also
used, with qualitatively similar conclusions. For example, we could
sum over all experimental data, but excluding those points belonging
to the same experiment as point $i$, and normalizing $\mathcal{D}_{i}$
by $(N_{\mathit{pt}}-N_{\mathit{all}})/N_{\mathit{pt}}$ to compensate for different numbers of
points in the experiment. } We plot the result of applying Eq.~(\ref{eq:recip}) to the 56-dimensional
residual variations of the full CTEQ-TEA dataset computed using two PDF ensembles:
CT14HERA2 fitted to all data in the left panel, and CT14HERA2 fitted
only to the DIS and vector boson production data (excluding jet production
data) in the right panel. Fig. \ref{fig:recip} represents the distribution
of the reciprocated distances over individual experiments of the CTEQ-TEA
dataset. The CT Experiment ID \# is shown on the abscissa, and the $\mathcal{D}_{i}$
values for every point of the experiment are indicated by the scatter
points. 

The advantage of the definition in Eq. (\ref{eq:recip}) is that it
enables a quantitative measure of the degree to which separate experiments
broadly differ in terms of their residual variations, and therefore
provides information analogous to that found in Figs. \ref{fig:PCA-TSNE}\textendash \ref{fig:PCA-CTExperiments}.
For example, by inspection of Eq. (\ref{eq:recip}) it can be seen
that those experimental measurements which are widely separated from
the rest of the CTEQ-TEA dataset in space of $\vec{\delta}_{i}$ vectors
will correspond to comparatively large values of $\mathcal{D}_{i}$,
and experiments that systematically differ from the rest of the total
dataset are thus expected to have especially tall distributions in
the panels of Fig. \ref{fig:recip}. On this basis, it can be seen
that information yielded by W asymmetry measurements (D02Masy'08, CMS7Masy2'14,
D02Easy2'15) are particularly distinct, as well as the combined HERA DIS data
(HERAI+II'15) and fixed-target Drell-Yan measurements, such as E605
(E605'91) and E866 data (E866rat'01 and E866pp'03). Similarly,
direct comparison of the $\mathcal{D}_{i}$ distributions in the panels
of Fig. \ref{fig:recip} allows one to compare constraints with and
without the jet data. We note that the 7 and 8 TeV ATLAS high-$p_{T}$
$Z$ production (ATL7ZpT'14 and ATL8ZpT'16) and $t\bar{t}$ production (ATL8ttb-pt'16)
provide a number of ``remote'' points and hence are potentially
useful in the fits sensitive to the gluon. On the other hand, new
jet production experiments (CMS7jets'14, ATLAS7jets'15, CMS8jets'17) all include large numbers
of points characterized by significant reciprocated distances. 

\section{Quantifying distributions of residual variations\label{sec:QuantifyingDistributionsOfResiduals}}

We have demonstrated that the multi-dimensional distribution of
the $\vec{\delta}_i$ vectors reflects the PDF dependence of individual data points. In this section,
we will focus on numerical metrics to assess the emerging geometrical
picture associated with the $\vec{\delta}_i$ distribution, and to visualize the regions of partonic momentum fractions
$x$ and QCD factorization scales $\mu$ where the experiments impose
strong constraints on a given PDF-dependent observable $X$.
Gradients of $r_{i}$ in a space of Hessian eigenvector PDF parameters
$\vec{a}$ are naturally related to the PDF uncertainty. Recall that
in the Hessian method the PDF uncertainty on $X(\vec{a})$ is found
as 
\begin{equation}
\Delta X(\vec{a})=X(\vec{a})-X(\vec{a}_{0})=\vec{\nabla}X|_{\vec{a}_{0}}\cdot\Delta\vec{a},
\end{equation}
where $\vec{a}_{0}$ is the best-fit combination of PDF parameters,
and $\Delta\vec{a}$ is the maximal displacement along the gradient
that is allowed within the tolerance hypersphere of radius $T$ centered
on the best fit \cite{Pumplin:2001ct,Pumplin:2002vw}. The standard
master formula 
\begin{equation}
\Delta X=\left\vert \vec{\nabla}X\right\vert =\frac{1}{2}\sqrt{\sum_{l=1}^{N}\left(X_{l}^{+}-X_{i}^{-}\right)^{2}}\label{masterDX-1}
\end{equation}
is obtained by representing the components of $\vec{\nabla}X$ by
a finite-difference formula 
\begin{equation}
\frac{\partial X}{\partial a_{i}}=\frac{1}{2}(X_{i}^{+}-X_{i}^{-}),\label{dXdzi-1-1}
\end{equation}
in terms of the values $X_{l}^{\pm}$ for extreme displacements of
$\vec{a}$ within the tolerance hypersphere along the $l$-th direction.

In this setup, a dot product between the gradients provides a convenient
measure of the degree of similarity between PDF dependence of two
quantities \cite{Nadolsky:2008zw}. A dot product $\vec{\nabla}r_{i}\cdot\vec{\nabla f}$
between the gradients of a shifted residual $r_{i}$ and another QCD
variable $f$, such as the PDF at some $\{x,\mu\}$ or a cross section,
can be cast in a number of useful forms.

\subsection{Correlation cosine }
\label{sec:Correlations}

The correlation for the $i^{th}$ $\{x,\mu\}$ point, which we define
following Refs.~\cite{Pumplin:2001ct,Nadolsky:2001yg,Nadolsky:2008zw,Gao:2017yyd}
as 
\begin{equation}
	C_{f}\,\equiv\,\mbox{Corr}[f,r_{i}]=\frac{\vec{\nabla} f\cdot\vec{\nabla} r_{i}}{\Delta f\,\Delta r_{i}},\label{eq:corr}
\end{equation}
can determine whether there \emph{may} exist a predictive relationship
between $f$ and goodness of fit to the $i^{th}$ point. The correlation
function $\mathrm{\mbox{Corr}}[X,Y]$ for the quantities $X,\,Y$
in Eq.~(\ref{eq:corr}) represents the realization in the Hessian
formalism of Pearson's correlation coefficient, which we express as
\begin{align}
\mathrm{\mbox{Corr}}[X,Y] & =\frac{1}{4\Delta X\Delta Y}\sum_{j=1}^{N}(X_{j}^{+}-X_{j}^{-})(Y_{j}^{+}-Y_{j}^{-})\ ,\label{eq:corr-def}
\end{align}
with the sum in these expressions being over the $j$ parameters of
the full PDF model space. Geometrically, $\mbox{Corr}[X,Y]$ represents
the cosine of the angle that determines the eccentricity of an ellipse
satisfying $\chi^{2}(\vec{a})<\chi^{2}(\vec{a}_{0})+T^{2}$ in the
$\{X,Y\}$ plane. This latter point follows from the fact that the
mapping of the tolerance hypersphere onto the $\{X,Y\}$ plane is
an ellipse with an eccentricity that depends on the correlation of
$X$ and $Y,$ which is given in turn by Eq.~(\ref{eq:corr-def})
above.

$\mbox{Corr}[f,r_{i}]$ does not indicate how constraining the residual
is, but it may indicate a predictive relation between $r_{i}$ and
$f$. On the basis of previous work \cite{Nadolsky:2008zw}, we say
that the (anti-)correlation between $X$ and $Y$ is significant roughly
if $\left|\mbox{Corr}[X,Y]\right|\gtrsim0.7$, while smaller (anti-)correlation
values are less robust or predictive. Following this rule-of-thumb,
correlations have been used successfully to identify PDF combinations
that dominate PDF uncertainties of complicated observables, for instance
to show that the gluon uncertainty dominates the total uncertainty
on LHC $W$ and $Z$ production, or that the uncertainty on the ratio
$\sigma_{W}/\sigma_{Z}$ of $W^{\pm}$ and $Z^{0}$ boson cross sections
at the LHC is dominated by the strangeness PDF, rather than $u$ and
$d$ (anti-)quark PDFs \cite{Nadolsky:2008zw}.

\subsection{Sensitivity in the Hessian method}
\label{sec:Sensitivities}
The correlation $C_{f}$ alone does not fully encode the potential
impact of separate or new measurements on improving PDF determinations
in terms of the uncertainty reduction. Rather, we employ $\vec{\nabla} f\cdot\vec{\nabla} r_{i}$
again to define the \textit{sensitivity} $S_{f}$ to $f$ of the $i^{th}$
point in experiment $E$: 
\begin{equation}
	S_{f}\equiv\frac{\vec{\nabla} f\cdot\vec{\nabla} r_{i}}{\Delta f\,\langle r_{0}\rangle_{E}}=\frac{\Delta r_{i}}{\langle r_{0}\rangle_{E}}\,C_{f}\ ,\label{eq:sens}
\end{equation}
where $\Delta r_{i}$ and $\langle r_{0}\rangle_{E}$ are computed
according to Eqs.~(\ref{DelX}) and (\ref{r0E}), respectively. In
other words, $\Delta r_{i}$ again represents the variation of the
residuals across the set of Hessian error PDFs, and we normalize it
to the r.m.s.\ residual for the whole dataset $E$ to reduce the impact
of random fluctuations in the data values $D_{i,\mathit{sh}}$. This definition
has the benefit of encoding not only the correlated relationship of
$f$ with $r_{i}$, but also the comparative size of the experimental
uncertainty with respect to the PDF uncertainty. In consequence, for
example, if new experimental data have reported uncertainties that
are much tighter than the present PDF errors, these data would then
register as high-sensitivity points by the definition in Eq.~(\ref{eq:sens}).

\begin{figure*}
\includegraphics[clip,width=0.48\textwidth]{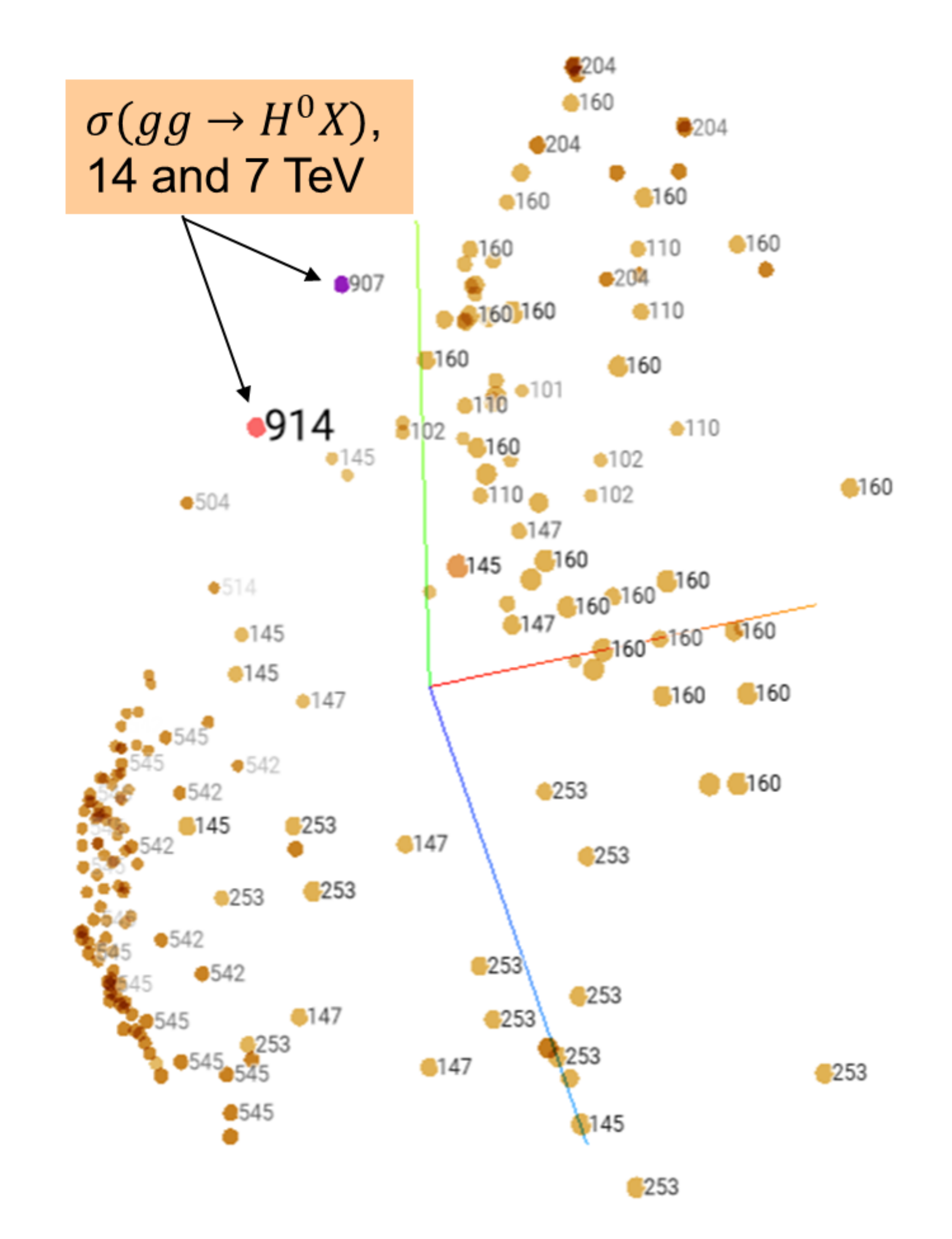}
\quad\quad \includegraphics[clip,width=0.4\textwidth]{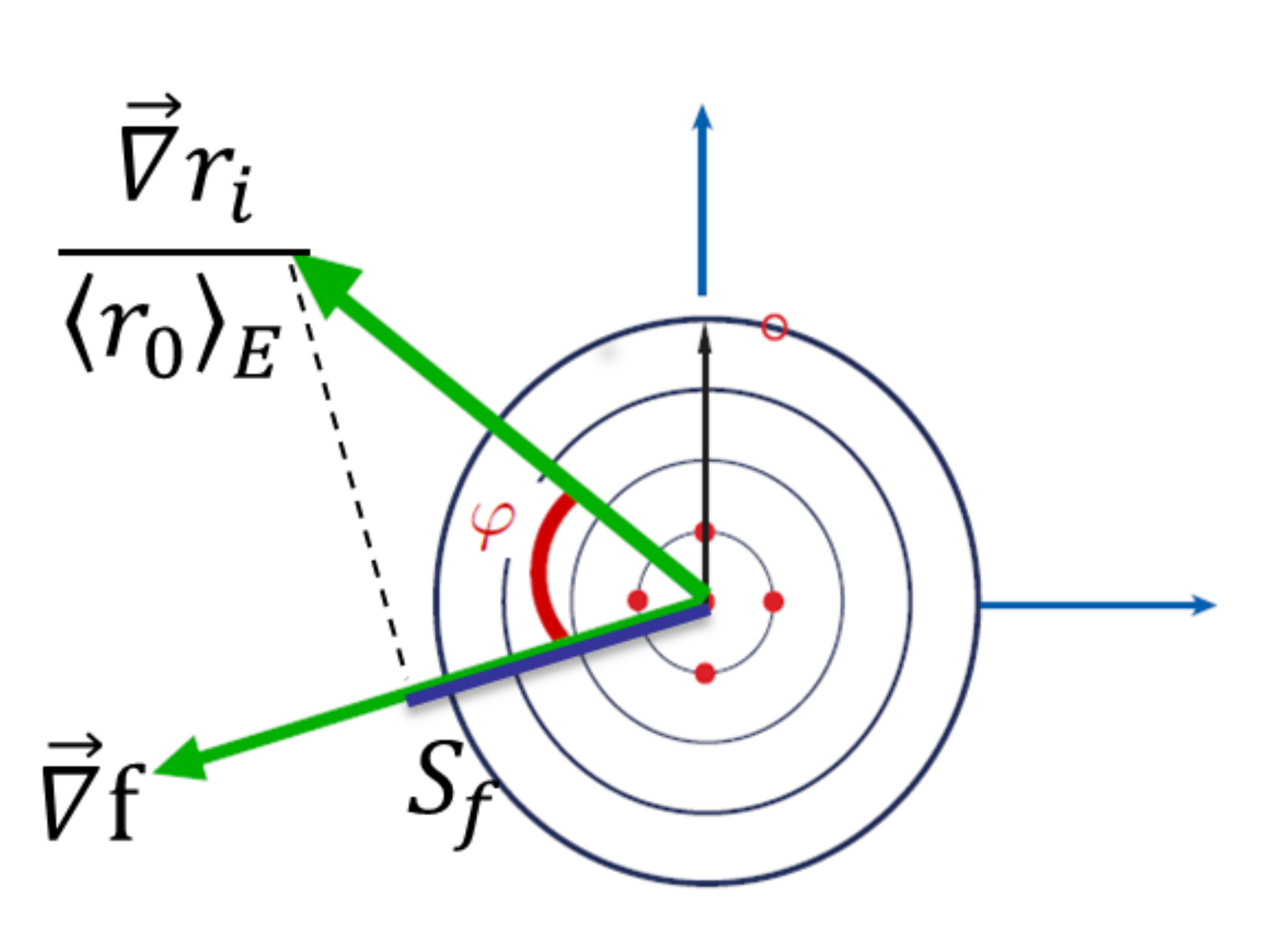}
\caption{Left: A PDF-dependent quantity $f$ defines a direction in
  space of $(2)N$ PDF parameters. The direction is specified by the
  gradient $\vec\nabla f$ in the symmetric convention.
  Here, the Embedding
  Projector \cite{EmbeddingProjector} visualizes the
  vectors $\vec \delta_{907}$ and $\vec \delta_{914}$ for
  NNLO cross sections for Higgs boson production at 7 and 14 TeV, and
  vectors $\vec\delta_i$ for CT14HERA2 NNLO data points  from
  \cite{PDFSenseWebsite} (brown circles), showing only $\vec\delta_i$
  with the smallest angular distances to $\vec\delta_{914}$. These
  points impose the strongest constraints on the PDF dependence of the
  Higgs cross sections in the CT14HERA2 analysis, if they have large enough
  $|\vec \delta_i|$. Again, in the numbering scheme used here, points labeled
  1XX correspond to fixed-target measurements, 2XX to Drell-Yan processes and boson production, and
	5XX to jet and $t\bar{t}$ production as given in Tables~\ref{tab:EXP_1}--\ref{tab:EXP_3}. Right: the sensitivity $S_f$ of the $i$-th data
  residual can be interpreted as the projection of $\vec \delta_i
  \equiv \vec \nabla r_i/\langle r_0\rangle_E$ onto the direction of $\vec\nabla f$.}
\label{fig:sfprojection} 
\end{figure*}

Geometrically, $S_{f}$ represents a projection onto the direction of
the gradient $\vec{\nabla}f$ of the residual variation $\vec{\delta_{i}}$, defined in Sec.~\ref{sec:QuantifyingDistributionsOfResiduals}
using the symmetrized formula for $\delta_{i,l}$ noted in footnote~\ref{fn:sym-deltail}, namely,
\begin{equation}
\delta_{i,l}\equiv\left(r_{i}(\vec{a}_{l}^{+})-r(\vec{a}_{l}^{-})\right)/\left(2\langle r_{0}\rangle_{E}\right)\ .
\end{equation}
Figure~\ref{fig:sfprojection} shows a pictorial illustration of this interpretation. This interpretation
suggests that the total strength of constraints along the direction
of $\vec{\nabla}f$ can be quantified by summing projections $S_{f}$
onto this direction of all individual vectors $\vec{\delta}_{i}$.

As with correlations, only a sufficiently large absolute magnitude
of $\left|S_{f}\right|$ is indicative of a predictive constraint
of the $i^{th}$ point on $f$. Recall that $r_{i}^{2}$ is the contribution
of the $i^{th}$ point to $\chi^{2},$ and that only residuals with
a large enough $\Delta r_{i}$ as compared to the r.m.s.\ residual
$\langle r_{0}\rangle_{E}$ are sensitive to PDF variations. The $S_{f}$
magnitude is of order $\Delta r_{i}/\langle r_{0}\rangle_{E},$ which
suggests an estimate of a minimal value of $S_{f}$ that would be
deemed sensitive according to the respective $\chi^{2}$ contribution.
For the numerical comparisons in this study, we assume that $\left|S_{f}\right|$
must be no less than 0.25 to indicate a predictive constraint, as
the PDF uncertainty of the $i^{th}$ residual contributes no less
than $r_{i}^{2}=$0.0625 to the variation in the global $\chi^{2}$.
The reader can choose a different minimal value in the
\textsc{PDFSense} figures depending
on the desired accuracy. The cumulative sensitivities that we obtain
in later sections are independent of this choice. 

Yet another possible definition, which we list for completeness, is
to further normalize the sensitivity as 
\begin{equation}
	S_{f}^{\prime}\equiv\frac{\vec{\nabla} f\cdot\vec{\nabla} r_{i}}{f_{0}\,\langle r_{0}\rangle_{E}}=\frac{\Delta f}{f_{0}}\,S_{f}\ .\label{eq:sens-prime}
\end{equation}
For instance, if $f$ is the PDF $f(x_{i},\mu_{i})$ or parton luminosity
evaluated at the $\{x_{i},\mu_{i}\}$ points extracted according to
the data, the definition of $S_{f}^{\prime}$ in Eq. (\ref{eq:sens-prime})
de-emphasizes those points where the PDF uncertainty $\Delta f(x_{i},\mu_{i})$
is small compared to the best-fit PDF value $f_{0}(x_{i},\mu_{i})$
\textemdash{} analogously to how $S_{f}$ de-emphasizes (relative
to the correlation $C_{f}$) those data points whose normalized
residual variations $\Delta r_{i}/\langle r_{0}\rangle_{E}$
have already been more tightly constrained.

\subsection{Sensitivity in the Monte-Carlo method}
The above statistical measures are general enough and can be extended
to other representations for the PDF uncertainties, such as the
representation based on Monte-Carlo replica PDFs \cite{Giele:1998gw,Giele:2001mr,Ball:2008by}
of the kind employed, e.g., in the NNPDF framework.  
A family of Monte-Carlo PDFs consists of $N_{\rm rep}$ member PDF sets
$q_a^{(k)}(x,\mu)\equiv \{ q^{(k)} \}$, with $k=1,\ ...,\ N_{\rm rep}$, and
those are used to determine an expectation value
$\langle X\rangle$ for a PDF-dependent quantity $X[\{ q \}]$ such as a
high-energy cross section:
\begin{equation}
\langle X \rangle = \frac{1}{N_{\rm rep}} \sum_{k=1}^{N_{\rm rep}}
X [ \{  q^{(k)} \}]\ .
\label{eq:NNPDF_masterave}
\end{equation}
The resulting Monte-Carlo uncertainty on $X$ can be extracted from the ensemble as
\begin{equation}
\Delta_{\rm MC} X\ =\ \left( \frac{1}{N_{\rm rep}-1}
\sum_{k=1}^{N_{\rm rep}}   
	\left( X [ \{  q^{(k)} \}] 
	-   \langle X \rangle\right)^2 
 \right)^{1/2}\ .
\label{eq:NNPDF_error}
\end{equation}
In consequence of these definitions, the central value of a particular
PDF itself in the NNPDF framework is specified as
\begin{equation}
q_{(0)} \equiv \langle  q \rangle = \frac{1}{N_{\rm rep}}
\sum_{k=1}^{N_{\rm rep}} q^{(k)} \ .
\label{eq:NNPDF_mcav}
\end{equation}
Akin to the Pearson correlation defined in Eq.~(\ref{eq:corr}) of
Sec.~\ref{sec:Correlations}, statistical correlations between two PDF-dependent
quantities $X[\{ q \}]$ and $Y[\{ q \}]$ can be constructed from the PDF
replica language above in terms of ensemble averages \cite{Ball:2008by}: 
\begin{equation}
\mbox{Corr}_{\rm MC} \left[ X, Y \right]
=\frac{\langle X Y \rangle
	- \langle X \rangle \langle Y \rangle}{\Delta_{\rm MC} X \Delta_{\rm MC} Y}\ .
\label{eq:NNPDF_corrPDF}
\end{equation}
Then, using our definitions in Eqs.~(\ref{eq:corr}) and
(\ref{eq:sens}), we immediately construct the realizations of the
correlation and sensitivity for a PDF-dependent quantity $f$
in the Monte-Carlo method:
\begin{eqnarray}
  	C_{f,\ {\rm MC}} &=& \mbox{Corr}_{\rm MC}[f,r_{i}]\ , \label{eq:corrMC}\\
	S_{f,\ {\rm MC}} &=& \frac{\Delta_{\rm MC} r_{i}}{\langle
          r_{0}\rangle_{E}}\,\mbox{Corr}_{\rm MC}\left[f, r_i \right].\label{eq:sensMC}
\end{eqnarray}

\section{Case study: CTEQ-TEA global data \label{sec:CaseCTEQ-TEA}}

\subsection{Maps of correlations and sensitivities}
\begin{figure*}
\includegraphics[clip,width=0.38\textwidth]{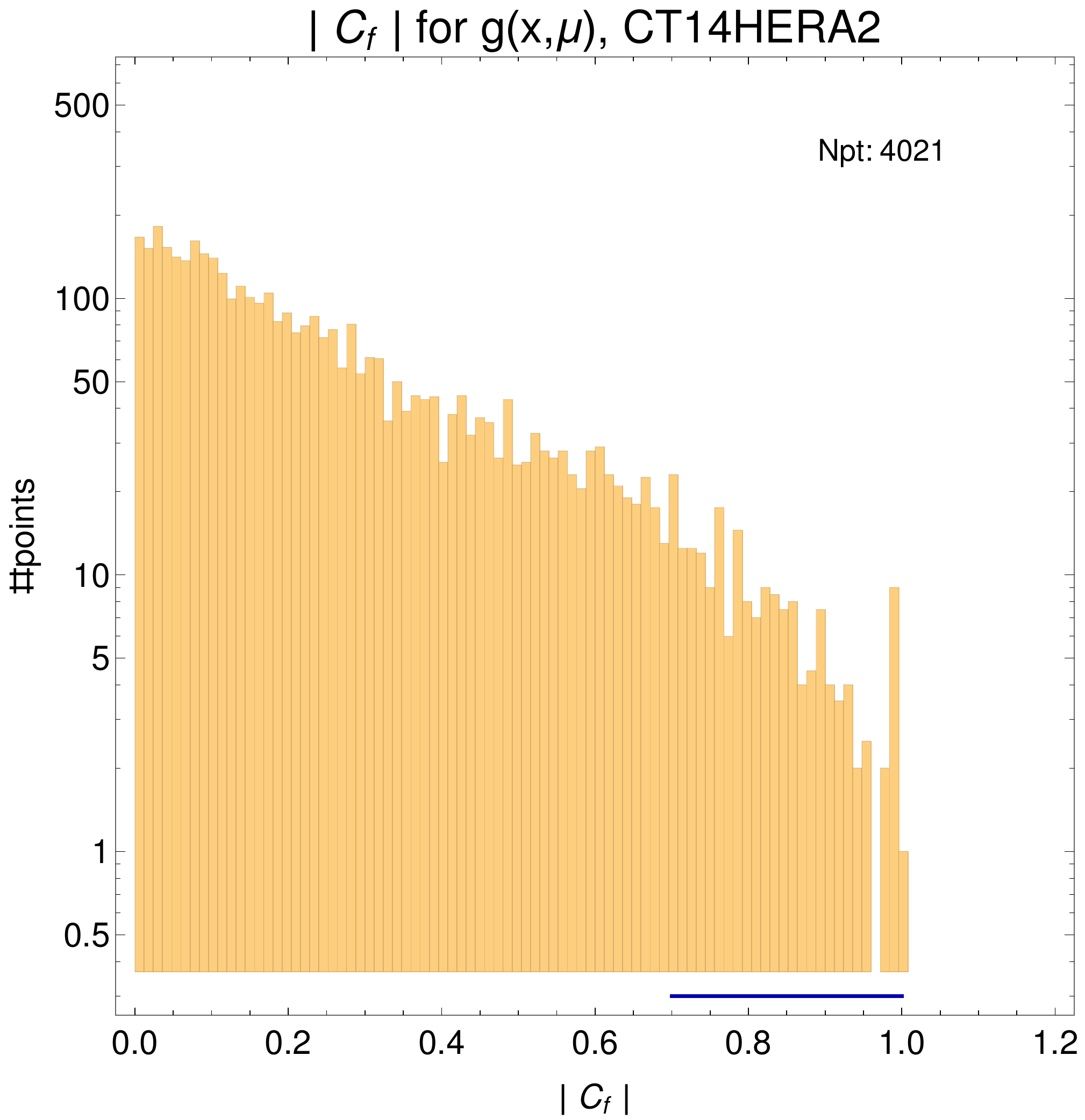}
\includegraphics[clip,width=0.60\textwidth]{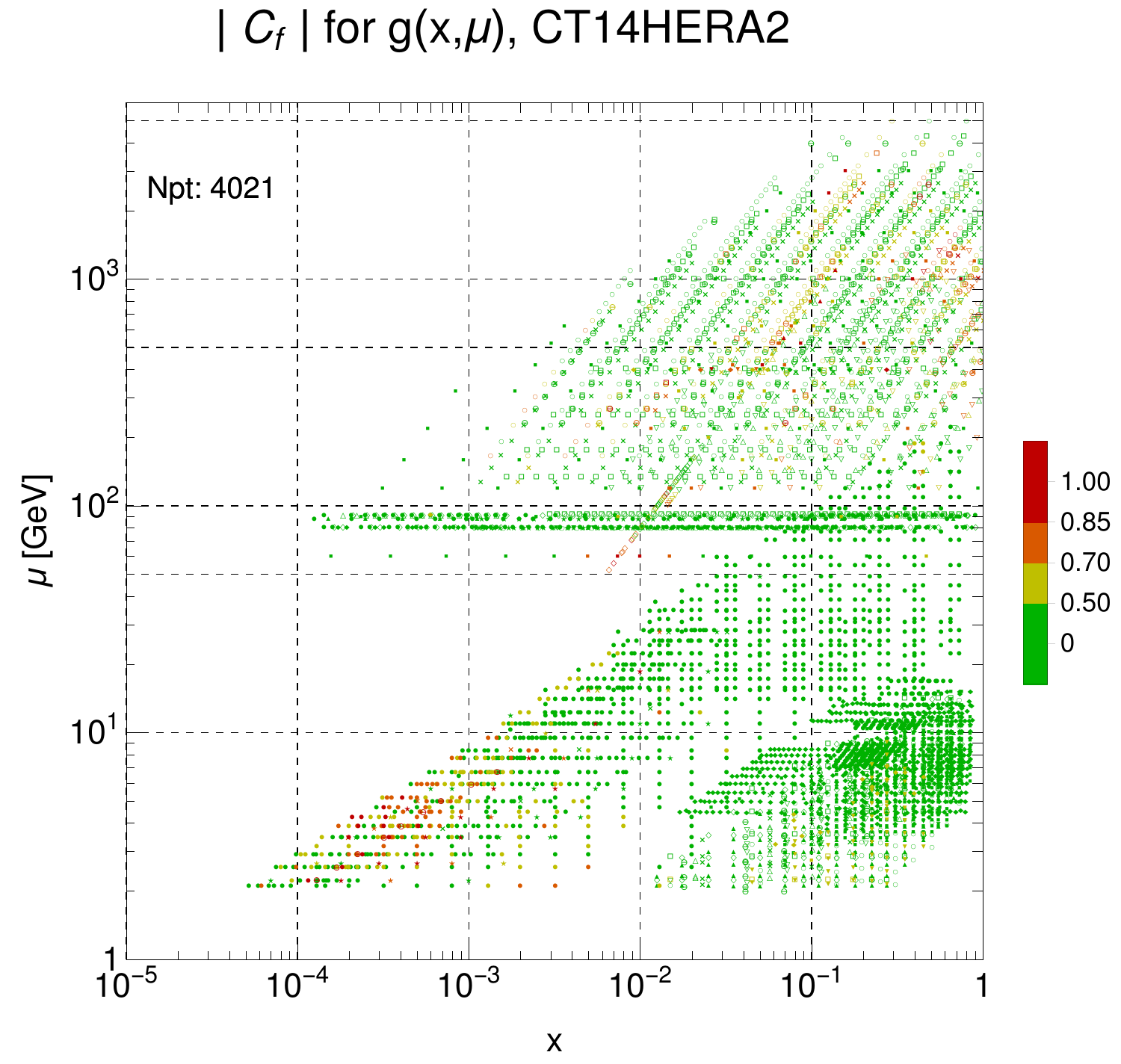}
\\
\includegraphics[clip,width=0.60\textwidth]{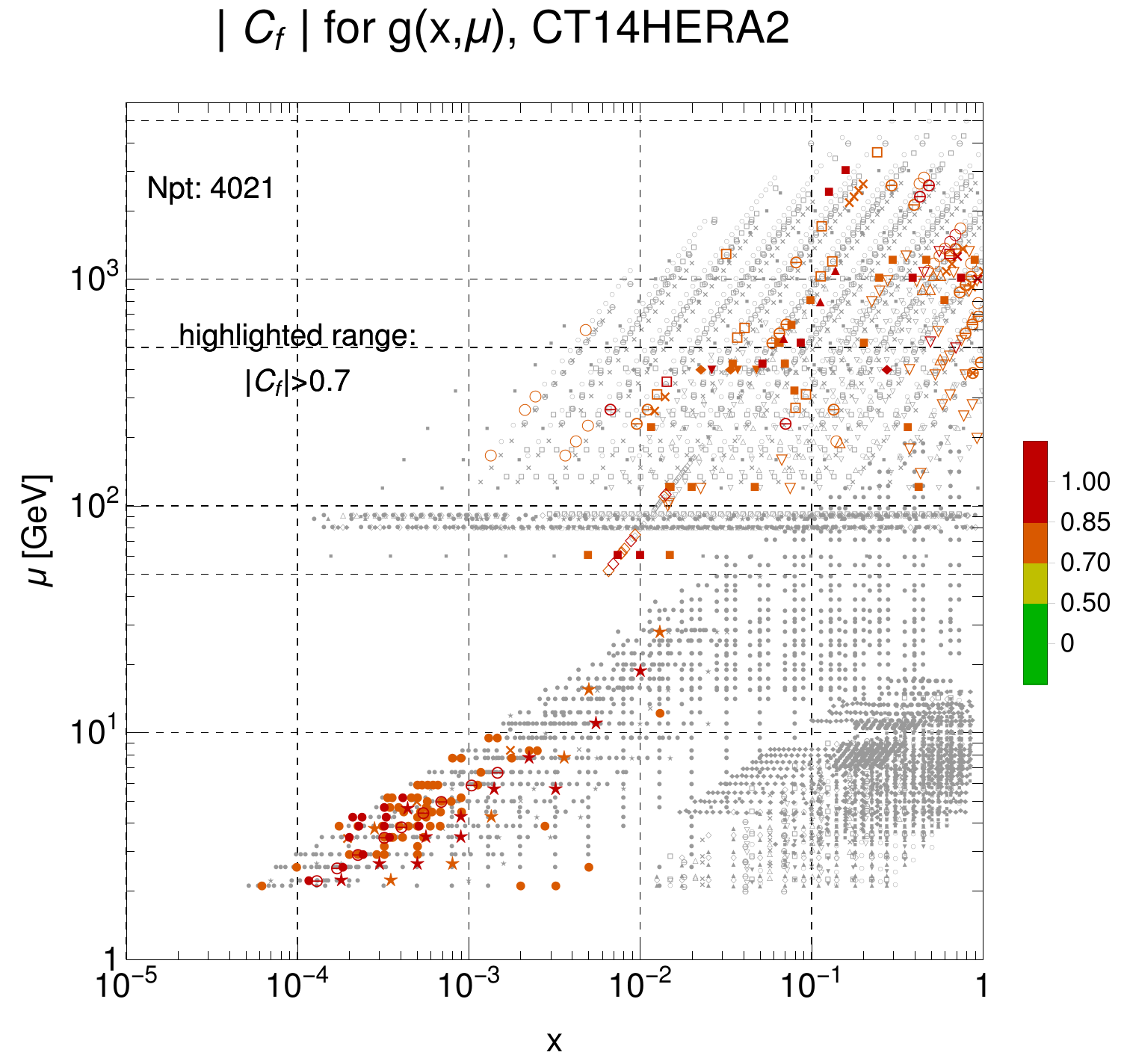}
\caption{Representations of the correlation $|C_{g}|(x_{i},\mu_{i})$ of
the gluon PDF $g(x,\mu)$ with the point-wise residual $r_{i}$ of
the augmented CT14HERA2 analysis. In the first panel, we plot a histogram
showing the distribution of correlations for 4021 physical measurements.
In the second panel we show the 5227-point $\{x_{i},\mu_{i}\}$ map corresponding
to these data within the full dataset, generated as in Appendix~\ref{sec:supp}.
To adjust for the fact that some measurements of rapidity dependent quantities
match to two distinct points in $\{x_{i},\mu_{i}\}$ space using the rules of
Appendix~\ref{sec:supp}, we assign weights of $0.5$ to these complementary
$\{x_{i},\mu_{i}\}$ points in computing the $N_{\mathit{pt}}=4021$-count histogram
at left. The third figure is the same as the second one, but only the
data points satisfying $|C_f|>0.7$ are highlighted.
}
\label{fig:corr-main} 
\end{figure*}
We will now discuss a number of practical examples of using $C_{f}$
or $S_{f}$ to quickly evaluate the impact of various hadronic data
sets upon the knowledge of the PDFs in a fashion that does not require
a full QCD analysis of the type described in Sec.~\ref{sec:PDF-preliminaries}.
For this demonstration, we will continue to study the dataset shown
in Fig.~\ref{fig:data} of the CT14HERA2 analysis~\cite{Hou:2016nqm}
augmented by the candidate LHC data.

We have already noted the extent of this dataset in the $\{x,\mu\}$
plane in Fig.~\ref{fig:data}, where it is decomposed into constituent
experiments labeled according to the conventions in Tables~\ref{tab:EXP_1}-\ref{tab:EXP_3}.
It is instructive to create similar maps in the $\{x,\mu\}$ plane
showing the $C_{f}$ or $S_{f}$ values for each data point.
Such maps are readily produced by the
\textsc{PDFSense} program for a variety of PDF flavors and for user-defined
observables, such as the Higgs cross section. For demonstration we
have collected a large number of these maps at the companion website
\cite{PDFSenseWebsite}. We invite the reader to review these additional
figures while reading the paper
to validate the conclusions that will be summarized below. 

\begin{figure*}
\includegraphics[clip,width=0.38\textwidth]{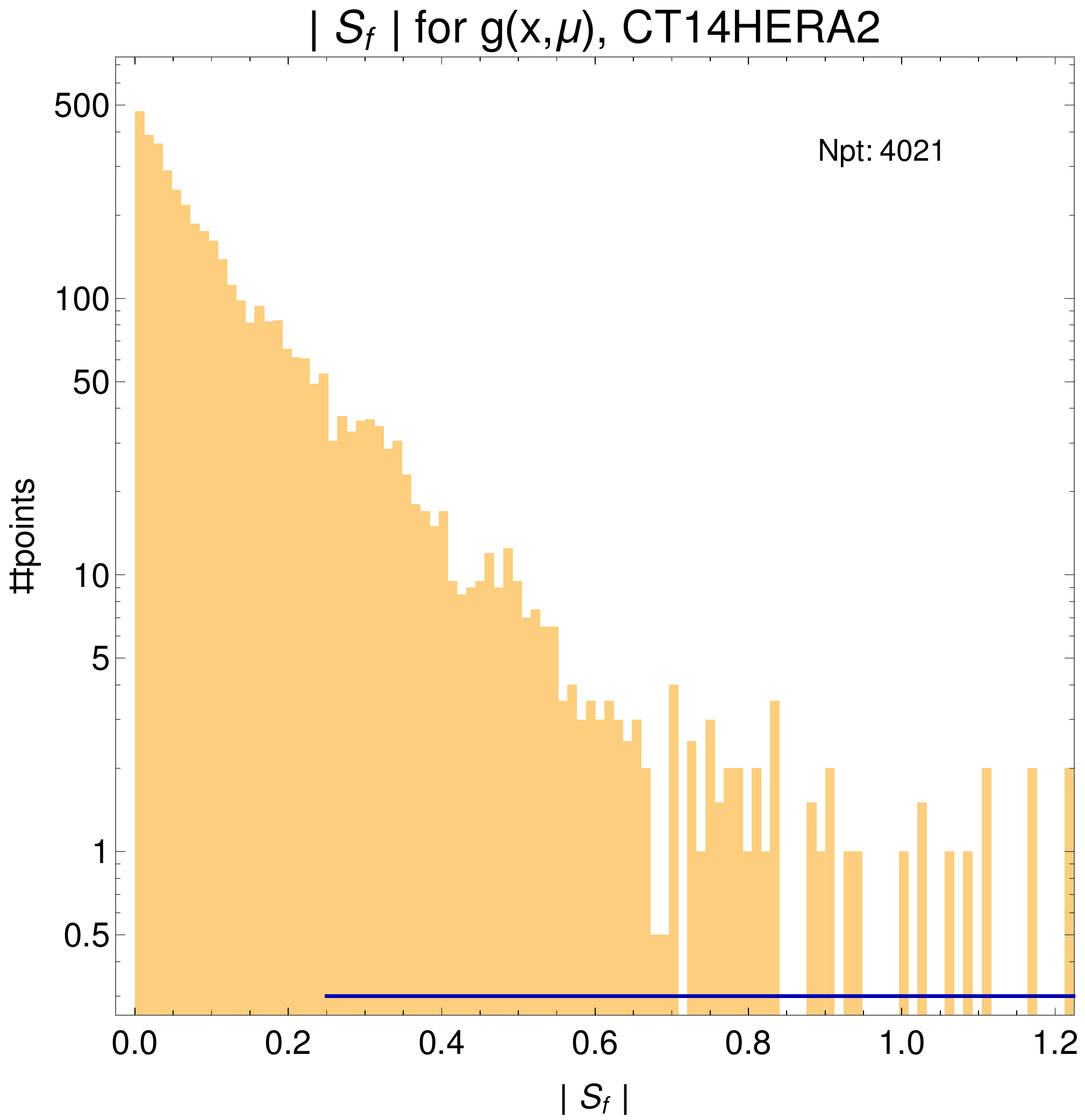}
\includegraphics[clip,width=0.60\textwidth]{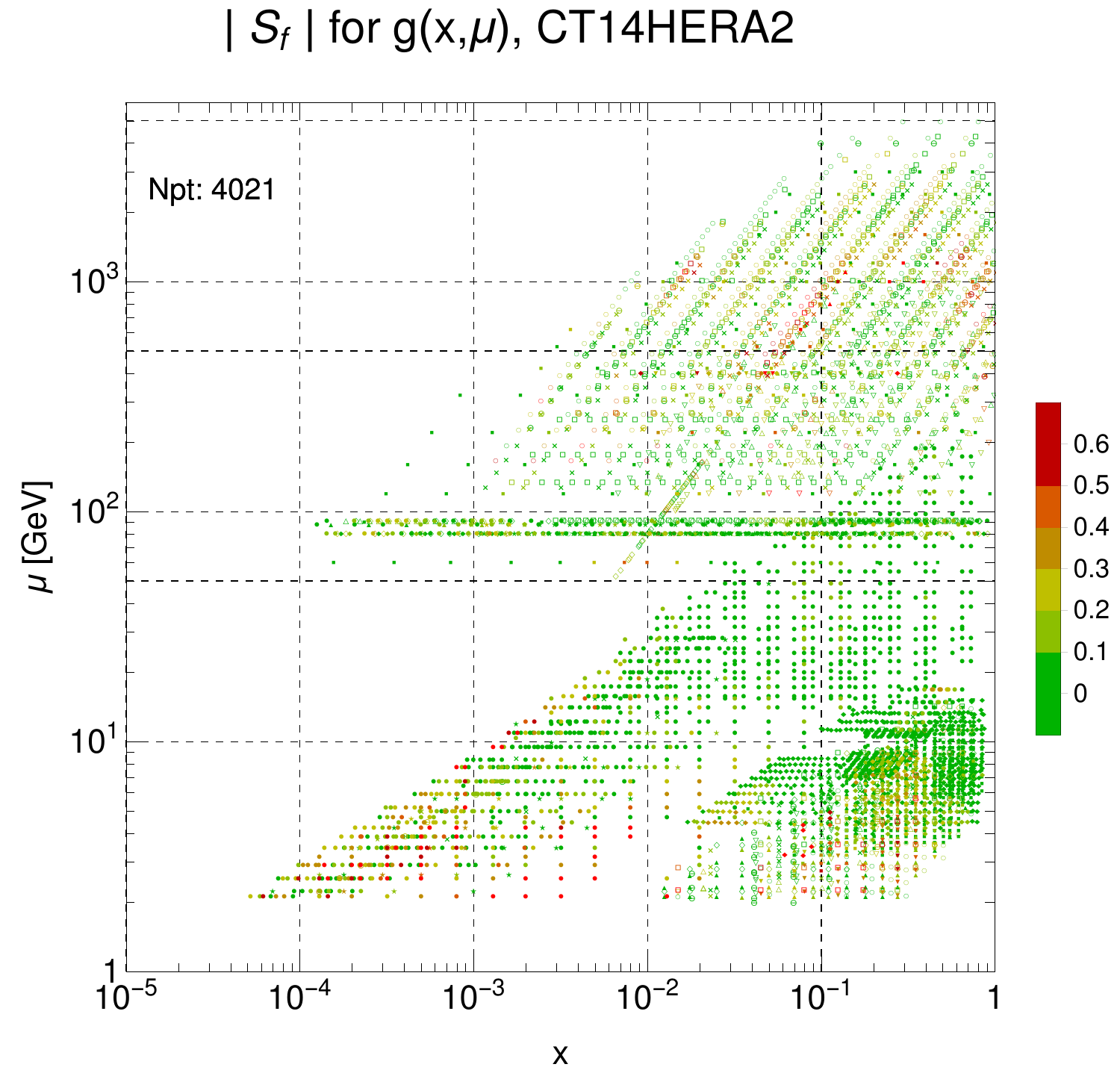}\\
\includegraphics[clip,width=0.60\textwidth]{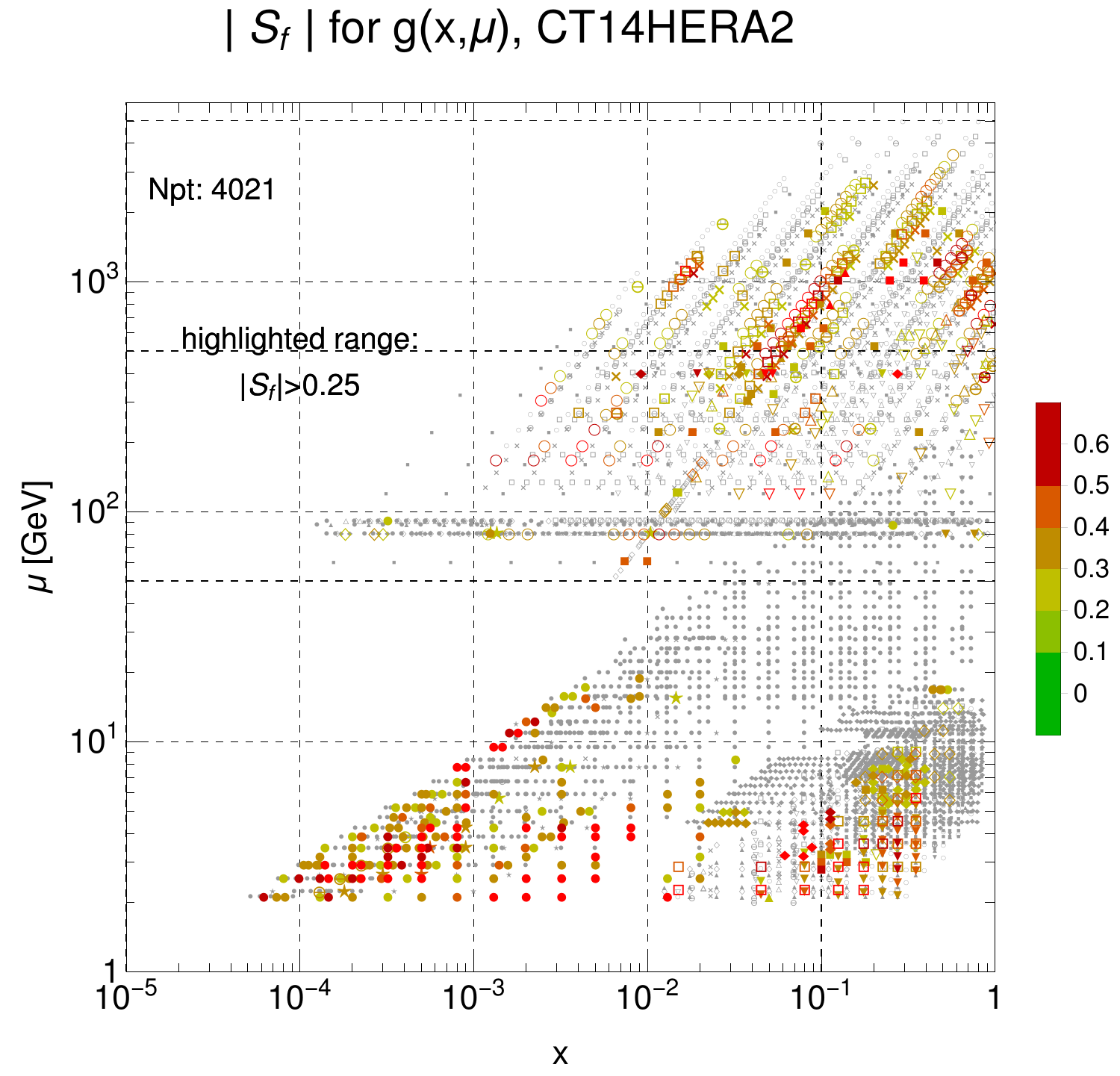}

\caption{Like Fig.~\ref{fig:corr-main}, but for the gluon sensitivity $|S_{g}|(x_{i},\mu_{i})$
	as defined in Eq.~(\ref{eq:sens}). In the third figure, only the
data points satisfying $|S_f|>0.25$ are highlighted.}
\label{fig:sens-main} 
\end{figure*}

Thus, we obtain scatter plots of $C_{f}(x_{i},\mu_{i})$ or $S_{f}(x_{i},\mu_{i})$
for a given QCD observable $f=\sigma$, such as the LHC Higgs production
cross section shown in Fig.~\ref{fig:CorrSensH14}, or with a PDF
$f$ evaluated at the same $\{x_{i},\mu_{i}\}$ determined by the
data points, with examples shown for $g(x_{i},\mu_{i})$ in Figs.~\ref{fig:corr-main}
and \ref{fig:sens-main}. The typical $\{x_{i},\mu_{i}\}$ values
characterizing the data points are found according to Born-level approximations
appropriate for each scattering process included in the CTEQ-TEA dataset,
with the formulas to compute these kinematic matchings summarized
in App.~\ref{sec:supp}. Here and in general, we find it preferable to
consider the absolute
values $|C_f|$ and $|S_f|$ on the grounds that the
signs of $C_f$ and $S_f$ flip when the data points randomly overshoot
or undershoot their theory predictions.

Together with the map in the $\{x,\mu\}$ plane, \textsc{PDFSense
}also returns a histogram of the values for each quantity it plots.
An example is shown for $\left|C_{g}\right|(x_{i},\mu_{i})$ in the
first panel of Fig.~\ref{fig:corr-main}. One would judge that stronger
constraints are in general provided to those PDFs for which the $|C_{f}|$
histogram has many entries comparatively closely to $|C_{f}|\sim1$.
In the first panel of Fig.~\ref{fig:corr-main}, we can see that,
while the distribution peaks at low correlations,
$|C_{g}|\sim0$, the distribution has an extended
tail in the region $0.7\lesssim|C_{g}|\lesssim1$.
This feature shows that, of the 4021 experimental data
points within the augmented CT14HERA2 set in Fig.~\ref{fig:data},
nearly two-hundred --- specifically, 192 --- have especially strong
($|C_{f}|\ge0.7$) correlations (or anti-correlations) with the gluon
PDF. This region of such strong correlations within the histogram
is indicated by the horizontal blue bar that runs along the abscissa.

To identify these points, we plot complementary information in the
second panel of the same figure \textendash{} specifically, a map in
$\{x,\mu\}$ space of each of the data points shown in
Fig.~\ref{fig:data}.
As before, they are colorized according to the magnitude of $|C_{g}|$
following the color palette in the ``rainbow strip'' on the right.
``Cooler'' colors (green/yellow) correspond to weaker correlation
strengths, while ``hotter'' colors (orange/red) represent comparatively
stronger correlations, as indicated. To reveal the data points with
the highest correlations, we reproduce the same figure in the third
panel, but showing in color only the data points satisfying $|C_f|>0.7$.
Thus, we obtain two maps in the
$\{ x,\mu \}$ plane that look similar to the $|C_f|$ map in the left
panel of Fig.~\ref{fig:CorrSensH14}, apart from the differences that
(a) Fig.~\ref{fig:corr-main} shows the correlation $|C_g|$ for
$g(x_i,\mu_i)$ at the same typical values $\{x_i,\mu_i\}$ as in the
data, rather than $|C_{\sigma_{H^0}}|$ for Higgs production cross
  section in Fig.~\ref{fig:CorrSensH14}; and (b)
  Fig.~\ref{fig:CorrSensH14} highlights 310 points with the highest $|C_{\sigma_{H^0}}|$.

The correlations for the LHC Higgs production cross section trace those for
$g(x_i,\mu_i)$, but not entirely, as we will see in a moment.    
Large magnitudes of $|C_{g}|$ in Fig.~\ref{fig:corr-main}
are found for inclusive jet production measurements, especially those
recently obtained by CMS at 8 TeV \cite{Khachatryan:2016mlc} (Expt.~CMS8jets'17,
inverted triangles) with $|C_{g}|(x_{i},\mu_{i})$ as high as
0.85, including at the highest values of $x$ and $\mu$. Beyond these,
a sizable cluster of HERA (HERAI+II'15) data points at the lowest values of
$x$ are also seen to have large correlations with the gluon PDF,
consistent with the common wisdom that HERA DIS constrains the gluon
PDF at small $x$ via DGLAP scaling violations. Under the jet production
cluster, high-$p_{T}$ $Z$ production (ATL7ZpT'14, ATL8ZpT'16) and $t\bar{t}$
production (ATL8ttb-pt'16, ATL8ttb-y\_ave'16, ATL8ttb-mtt'16,
ATL8ttb-y\_ttb'16) at the LHC show a high $|C_{g}|(x_{i},\mu_{i})$
correlation. At the same time, many other measurements, including
fixed-target data at large $x$ and $W$ asymmetry data near $\mu\!\sim\!100$
GeV, have feeble correlations with $g(x_i,\mu_i)$ and would therefore be less
emphasized by an analysis based solely upon the PDF-residual correlations.

We can also consider the analogous plots for the sensitivity
$\left|S_{g}\right|(x_{i},\mu_{i})$ as defined in Eq.~(\ref{eq:sens}), which we plot in Fig.~\ref{fig:sens-main}. In the first panel, we again consider
the histogram, here for the magnitudes of the gluon sensitivity $|S_{g}|(x_{i},\mu_{i})$,
in which the correlations $|C_{g}|$ are now weighted by the relative
size of the PDF uncertainty $\Delta r_{i}$ in the residual. As discussed
in Sec.~\ref{sec:Sensitivities}, this additional weighting
emphasizes those data points for which the PDF-driven fluctuations
in the residuals are comparatively large relatively to experimental
uncertainties. This leads to a redistribution of the data points
shown in the $|C_{g}|$ histogram of Fig.~\ref{fig:corr-main}, with
the result being a considerably longer-tailed histogram for $|S_{g}|$
such that, in this instance, there are 546 raw data points
with larger sensitivities,
$|S_{f}|\ge0.25$, indicated by the horizontal blue bar. Unlike the correlation, $|S_{g}|$ can be arbitrarily large,
depending on the $\Delta r_{i}$ value. It is suppressed at the data points with
large uncertainties or smeared over the regions of data points with correlated systematic
uncertainties.

In the second and third panels, we show the respective $\{x,\mu\}$
maps for $\left|S_{g}\right|$, with color highlighting given either for all points or only
those with high sensitivities $|S_f|>0.25$, respectively. $\left|S_{g}\right|$ places additional
emphasis on the combined HERA dataset (HERAI+II'15) constraining $g(x_{i},\mu_{i})$
at lowest $x$. In contrast to the $|C_{g}|$ plot, we observe increased
sensitivity in the precise fixed-target DIS data from BCDMS (BCDMSp'89, BCDMSd'90) and CCFR
(CCFR-F2'01, CCFR-F3'97), which are sensitive to the gluon via scaling violations
despite only moderate correlation values.
Similarly, we observe heightened sensitivities at
highest $x$ for the LHC (CMS7jets'14, ATLAS7jets'15, CMS8jets'17) and Tevatron (D02jets'08) jet production
data, which have both large correlations with $g(x_{i},\mu_{i})$
and small experimental uncertainties. Sensitivity $\left|S_{g}\right|$
of LHC jet experiments, CMS7jets'14, ATLAS7jets'15, CMS8jets'17, varies in a large range, and
can significantly improve, depending on the implementation of
experimental systematic uncertainties in the analysis, cf.\ the
discussion of the jet data in the next section. 

We also observe enhanced
sensitivity for \emph{individual points} in a large number of experiments,
including CDHSW DIS (CDHSW-F2'91); HERA $F_{L}$ (HERA-FL'11); the Drell-Yan process
(E605'91, E866pp'03); CDF 8 TeV $W$ charge asymmetry (CMS7Masy2'14); HERA charm SIDIS
(HERAc'13); ATLAS high-$p_{T}$ $Z$ production (ATL7ZpT'14, ATL8ZpT'16); and especially
strongly sensitive points in $t\bar{t}$ production (ATL8ttb-pt'16, ATL8ttb-y\_ave'16,
ATL8ttb-mtt'16, ATL8ttb-y\_ttb'16). However, since the latter category includes fewer
points per each experiment, it constrains the gluon less than the high-statistics DIS
and jet production data.

These findings comport with the idea that the gluon PDF remains dominated
by substantial uncertainties at both $x\!\sim\!0$ and in the elastic
limit $x\!\to\!1$, a fact which has driven an intense focus upon production
of hadronic jets, $t\bar{t}$ pairs, and high-$p_{T}$ $Z$ bosons,
which themselves are measured at large center-of-mass energies $\sqrt{s}$
and are expected to be sensitive to the gluon PDF across a wide interval
of $x,$ including $x\!\sim\!0.01$ typical for Higgs boson production
via gluon fusion at the LHC. Turning back to the distributions
of $\left|C_{\sigma_{H}}\right|(x_{i},\mu_{i})$ and $\left|S_{\sigma_{H}}\right|(x_{i},\mu_{i})$
for the Higgs cross section $\sigma_{H}$ at $\sqrt{s}=14$ TeV in
Fig.~\ref{fig:CorrSensH14}, we notice that they largely reflect
the distributions
of $\left|C_{g}\right|(x_{i},\mu_{i})$ and $\left|S_{g}\right|(x_{i},\mu_{i})$
around $x \sim M_H/\sqrt{s}=125/14000=0.009$ and $\mu= M_H=125$ GeV.
We also see some differences: although the average $x$ and $\mu$
are fixed in $\sigma_{H}$, it is nonetheless sensitive to some constraints
at much lower $x$ values as a result of the momentum sum rule.

The reader is welcome to examine the plots of sensitivities and
correlations available on the \textsc{PDFSense} website for a large
collection of PDF flavors and PDF ratios, such as
$d/u$, $\overline{d}/\overline{u}$, and
$\left( s +
\overline{s}\right)\!/\!\left(\overline{u}+\overline{d}\right)$. Sensitivities
for other PDF combinations and hadronic cross sections can be 
computed and plotted in a matter of minutes
using the \textsc{PDFSense} program. We will
now turn to another aspect of this analysis: summarizing the abundant
information contained in the sensitivity plots. For this purpose, we
will introduce
numerical indicators and propose a practical procedure to
rank the experimental data sets according to their
sensitivities to the PDFs or PDF-dependent observables of interest.

\subsection{Experiment rankings according to cumulative sensitivities \label{sec:Experiment-rankings-according}}
Being one-dimensional projections of normalized residual variations
$\vec\delta_i$ on a given direction in the PDF parameter space,
sensitivities can be linearly added to construct a number of useful
estimators. By summing absolute sensitivities $|S_f^{(i)}|$ over the
data points $i$ of a given data set $E$, we find the maximal cumulative
sensitivity of $E$ to the PDF dependence of a QCD observable $f$.

Alternatively, from the examination of multiple $\{x,\mu\}$ maps for $\left|S_{f}\right|$
of various PDF flavors collected on the website \cite{PDFSenseWebsite},
we find that the most precise experiments constrain several flavors
at the same time; most notably, the combined HERA
data. For the purpose of identifying such experiments, we can compute an
overall sensitivity statistic for each experiment $E$ to the parton
distributions $f_a(x_i,\mu_i)$ evaluated at the same kinematic parameters
$\{x_i, \mu_i\}$ as the data.
Furthermore, to obtain one overall ranking, we can add
up sensitivity measures as an unweighted sum over the ``basis PDF'' flavors,
such as the six light flavors ($\overline{d},\,\overline{u},\,g,\,u,\,d,\,s$).
To obtain these measures, we say that an experiment $E$ consisting of
$N_{\mathit{pt}}$ physical measurements can be characterized by its mean sensitivity per raw data point\footnote{
For those circumstances in which an individual measurement, {\it e.g.}, obtained
via the Drell-Yan process, maps to two sensitivity values in $\{x,\mu\}$ space, we
compute the average of these and assign the result to that specific measurement.
}
to a PDF of given flavor $f_a(x,\mu)$:
$\langle|S^E_f|\rangle \equiv (N_{\mathit{pt}})^{-1} \sum_{i=1}^{N_{\mathit{pt}}}\left|S_{f}\right|(x_{i},\mu_{i})$,
from which we derive several additional statistical measures of experimental sensitivity.
For each experiment and flavor we then determine a cumulative sensitivity measure, numerically adjusted to the size of each experimental dataset $E$,
according to $|S^E_{f}| \equiv N_{\mathit{pt}}\, \langle|S^E_f|\rangle$.
In addition, we also track cumulative, flavor-summed sensitivity
measures $\sum_{f}|S^E_{f}|$ and
$\langle\sum_{f}|S^E_f|\rangle$, with $f$ running
over $\overline{d},\,\overline{u},\,g,\,u,\,d,\,s$.

We list the corresponding values of these four types of sensitivities
for each experiment of the CTEQ-TEA dataset in summary tables
in App.~\ref{sec:Tables}
as well as extensive Supplementary Material in App.~\ref{sec:SM}. This is also
detailed for categories of experiments from the CTEQ-TEA dataset.

With the above estimators, we {\it quantify}
and {\it compare} the cumulative sensitivities of each experiment to the basis 6
parton flavors. In fact, based on the various trials that we
performed, we find that the cumulative
sensitivity to the 6 basic flavors is a good measure of the overall
sensitivity to a large range of PDF combinations.
Recall that the $N_f=5$ CT14HERA2 PDFs (with up to 11 independent
parton species) are obtained by DGLAP evolution of the 6
basic parton flavors from the initial scale of order 1 GeV. 
There exist alternative approaches for measuring the importance of a given
experiment in a global fit, for example, by counting the numbers of eigenvector
parameters \cite{Pumplin:2009sc} or eigenvector directions
\cite{Harland-Lang:2014zoa} that the experiment constrains. Those
other methods, however, require access to the full machinery of the global fit,
while the sensitivities allow the reader to rank the experiments
according to much the same information, for a variety of PDF-dependent
observables, with the help of
\textsc{PDFSense}, and at a fraction of computational cost. 

In fact, in a companion study we use the above sensitivity estimators to select the new LHC
experiments for the inclusion in the next generation of the CTEQ-TEA
PDF analysis. Full tables given in App.~\ref{sec:Tables} and in the Supplementary Material of App.~\ref{sec:SM}
provide detailed information about the PDF sensitivities of every
experiment of the CTEQ-TEA data set.
For a non-expert reader, along the full tables, we provide
their simplified versions in Tables~\ref{tab5}-\ref{tab6},
where we rank the experimental sensitivities
according to a reward system described in the caption of Table~\ref{tab5}.
In each table, experiments are listed in descending order according
to the cumulative sensitivity measure $\sum_{f}|S^E_{f}|$ to the
six light-parton flavors. For each PDF flavor, the experiments with
especially high overall flavor-specific sensitivities receive an ``\textbf{A}''
rating (shown in bold), per the convention in the caption of
Table~\ref{tab5}. Successively 
weaker overall sensitivities receive marks of ``B'' and ``C,'' while
those falling below a lower limit $|S^E_{f}|=20$ are left unscored.

We similarly evaluate each experimental dataset based on its point-averaged
sensitivity, in this case scoring according to a complementary scheme
in which the highest score is ``\textbf{1}''.  The short-hand names of
the candidate experiments that were {\it not} included in the
CT14HERA2 NNLO fit, that is, the new LHC experiments, are also shown
in bold to facilitate their recognition in the tables.   

Not only do the sensitivity rankings confirm findings known
by applying other methods, they also provide new insights. 
According to this ranking system in
Tables~\ref{tab5}-\ref{tab6}, we find that the expanded HERA dataset
(HERAI+II'15) tallies the highest overall sensitivity to the PDFs,
with enhanced sensitivity to the distributions of the $u$- and $\bar{u}$-quarks,
as well as that of the gluon. 
On similar footing, but with slightly weaker overall sensitivities, are
a number of other fixed-target measurements, including structure
function measurements from BCDMS for $F^{p,d}_2$ (BCDMSp'89, BCDMSd'90) and CCFR extractions
of $xF^p_3$ (CCFR-F3'97) --- as well as several other DIS datasets.
Among the LHC experiments, the inclusive jet measurements have the
highest cumulative sensitivities, with CMS jets at 8 TeV (CMS8jets'17),
7 TeV (CMS7jets'13, CMS7jets'14), and ATLAS 7 TeV (ATLAS7jets'15)
occupying positions 10, 12/13, and 16 in the total sensitivity
rankings. They demonstrate the strongest sensitivities among the candidate LHC
experiments, and at the same time are not precise enough and fall behind the top
fixed-target DIS and Drell-Yan experiments: BCDMS, CCFR, E605, E866,
and NMC. The two versions CMS7jets'13 and CMS7jets'14 of the CMS 7 TeV jet
data that largely overlap have very
close sensitivities and rankings in
Tables~\ref{tab5}-\ref{tab6}. The set CMS7jets'13 that extends to higher $p_{Tj}$ has a
slightly better overall sensitivity, surpassing the larger data set
CMS7jets'14 that includes the extra data points at $p_{Tj}<100$ GeV or
$|y_{j}|>2.5$, yet cannot beat CMS7jets'13 except for in
the overall sensitivity to the Higgs cross section at 7 TeV. 

Going beyond the rankings based upon overall sensitivities, which are
more closely tied to the impact of an entire experimental dataset
in aggregate, it is useful to consider the point-averaged sensitivity
as well, which quantifies how sensitive each individual point
is. [Some experiments with very high point-averaged sensitivity have a
  small cumulative sensitivity because of a small number of points.]
Based on their high point-averaged sensitivity, CMS $\mu$ asymmetry
measurements at 8 and 7 TeV (CMS8Wasy'16 and CMS7Masy2'14) especially stand
out, despite their small number of individual points, $N_{\mathit{pt}}=11$);
this is especially true again for the gluon, $\overline{d}$-, and
$u$-quark PDFs, for which this set of measurements is particularly
highly rated in Table~\ref{tab5}. Another ``small-size'' data set with
the exceptional point-average sensitivity is the
$\sigma_{pd}/(2\sigma_{pp})$ ratio from the E866 lepton pair
production experiment (E866rat'01). The average sensitivity of this data set to
$\overline u$ and $\overline d$ PDFs is 0.8, making it extremely valuable
for constraining the ratio $\overline{d}/\overline{u}$ at $x\sim 0.1$,
in spite of its small size (15 data points). 

Aside from the quark- and gluon-specific rankings of specific measurements,
we can also assess experiments based upon the constraints they impose
on various interesting flavor combinations and observables as presented
in Table~\ref{tab6}. As was the case with Table~\ref{tab5}, a
considerable amount of information resides in Table~\ref{tab6} of
which we only highlight several notable features here. Among these
features are the sharp sensitivities to the Higgs cross section (\textit{e.g.}, $|S|_{H7}$, $\langle|S_{H7}|\rangle$, \textit{etc.}) found for
Run I$+$II HERA data, as well as the tier-C overall sensitivities
 of the BCDMS $F^{p,d}_2$ and CMS jet
 production measurements, corresponding to Exps.~BCDMSd'90, BCDMSp'89, CMS8jets'17 and CMS7jets'14.
 While their
overall sensitivity is small, the corresponding ATLAS $t\overline{t}$
data also possesses significant point-averaged sensitivity. On the other
hand, measurements of $p_{T}$-dependent $Z$ production (ATL7ZpT'14, ATL8ZpT'16)
appear to have somewhat less pronounced sensitivity to the gluon and other PDF flavor
combinations. The total and mean sensitivities of high-$p_T$ $Z$ boson
production experiment ATL8ZpT'16 at 8 TeV is on par with HERA
charm SIDIS data (HERAc'13) and provides comparable constraints to charm
DIS production, albeit in a different $\{x,\mu\}$ region.

For the light-quark PDF combinations like $u_{v},\,d_{v},\,d/u,$
and $\overline{d}/\overline{u}$, the various DIS datasets \textemdash{}
led by Run II of HERA and CCFR measurements of the proton structure
function \textemdash{} demonstrate the greatest sensitivity. At the
same time, however, Run-2 Tevatron data from D0 on the $\mu$ asymmetry (D02Easy2'15)
and Run-1 CDF measurements for the corresponding $A_e(\eta^e)$ asymmetry (CDF1Wasy'96)
also exhibit substantial point-wise sensitivity as well. We collect
a number of other observations in the conclusion below,
Sec.~\ref{sec:Conclusions}.

\subsection{Estimating the impact of LHC datasets on CTEQ-TEA fits}
\label{sec:CTEQfit}
The presented rankings suggest that including the candidate LHC data
sets will produce mild improvements in the uncertainties of
the CT14 HERA2 PDFs. This projection may appear underwhelming, but
 keep in mind that the CT14HERA2 NNLO analysis
already includes significant experimental constraints, for example,
imposed on the gluon PDF at $x>0.01$ by the Tevatron and LHC jet
experiments, CDF2jets'09, D02jets'08, ATL7jets'12, CMS7jets'13. If all jet experiments are eliminated
from the PDF fit, as illustrated in the Supplementary Material tables of App.~\ref{sec:SM}, the candidate LHC
experiments will be promoted to higher rankings, with the CMS 8 and 7 TeV jet experiments
(CMS8jets'17 and CMS7jets'13/CMS7jets'14) elevated to positions 4 and 7/8 in the overall sensitivity rankings,
respectively. 

Our investigations also find that the sensitivities of
CMS jet experiments may improve considerably if
the current correlated systematic effects are moderately reduced
compared to the published values.
For instance, by
requiring a full correlation of the JEC2 correlation error over all
rapidity bins in the CMS 7 TeV jet data set CMS7jets'14, instead of its
partial decorrelation implemented according to the CMS recommendation
\cite{Khachatryan:2014waa}, we obtain a very strong sensitivity of the
data set CMS7jets'14 to $g$ over the full $\{x,\mu\}$ region; but also strong
sensitivities to $\overline u, \overline d$, and even $\overline s$
PDFs.\footnote{With the fully correlated jet energy correction JEC2 source, the data set CMS7jets'14
  would provide a strong overall constraint on $s(x,\mu)$ comparable to
  one of the NuTeV or neutrino CCFR experimental data sets.}
 The overall sensitivity of the data set CMS7jets'14 in this case is
elevated to the 4th position from the 13th position in the CT14HERA2
NNLO analysis in Tables~\ref{tab5} and \ref{tab6}. Similarly, for the CMS
8 TeV jet data set CMS8jets'17, the sensitivity to the above flavors can
increase under moderate reduction of systematic uncertainties, easily
surpassing the sensitivity of CMS7jets'14 because of the larger number of
points in CMS8jets'17. 
\subsection{Comparing {\sc PDFSense} predictions to post-fit constraints from Lagrange Multiplier scans}
\label{sec:Validation}
\begin{figure}[p]
\hspace*{-0.2cm}\includegraphics[clip,width=0.46\textwidth]{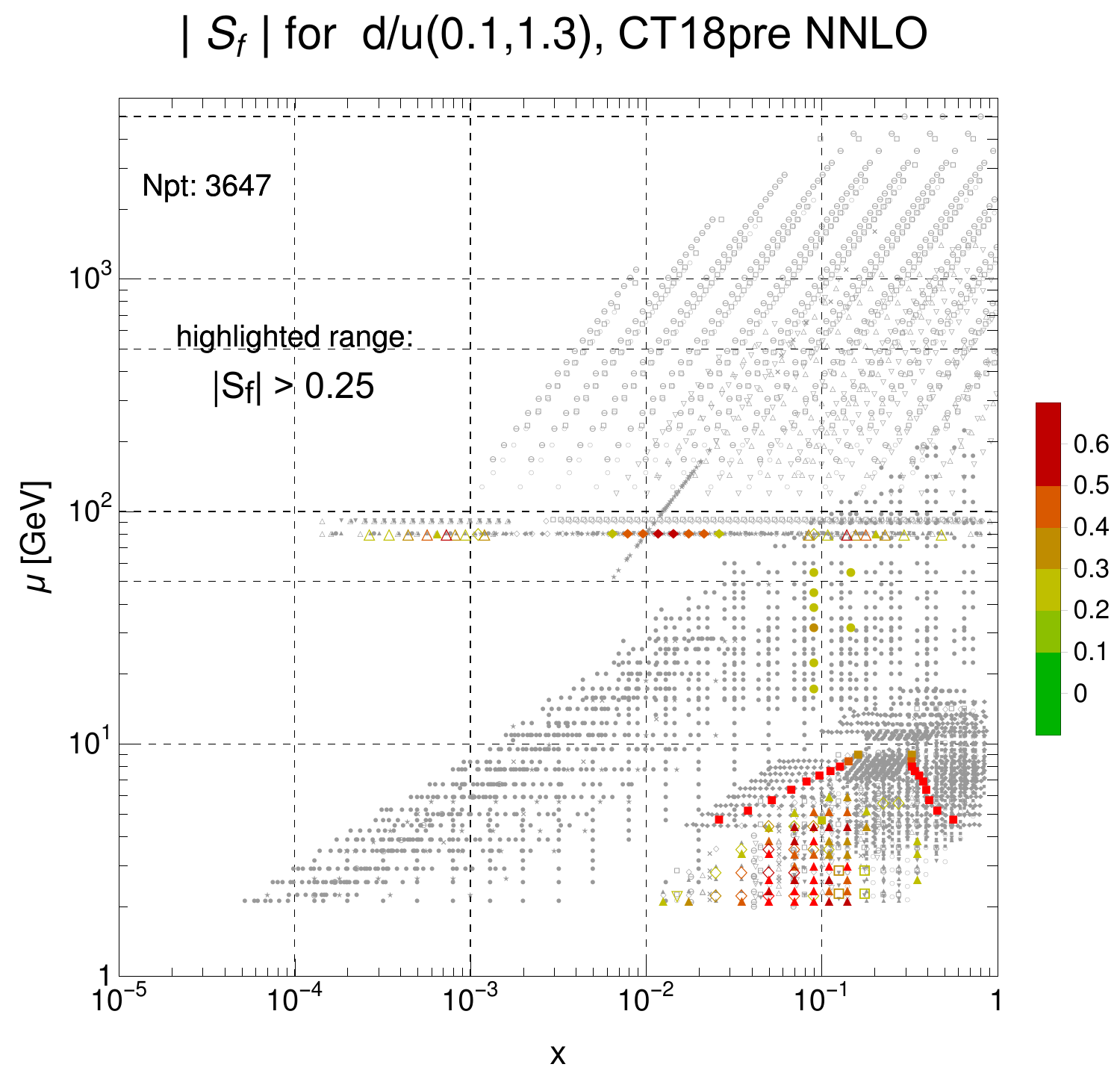} \ \
\includegraphics[clip,width=0.53\textwidth]{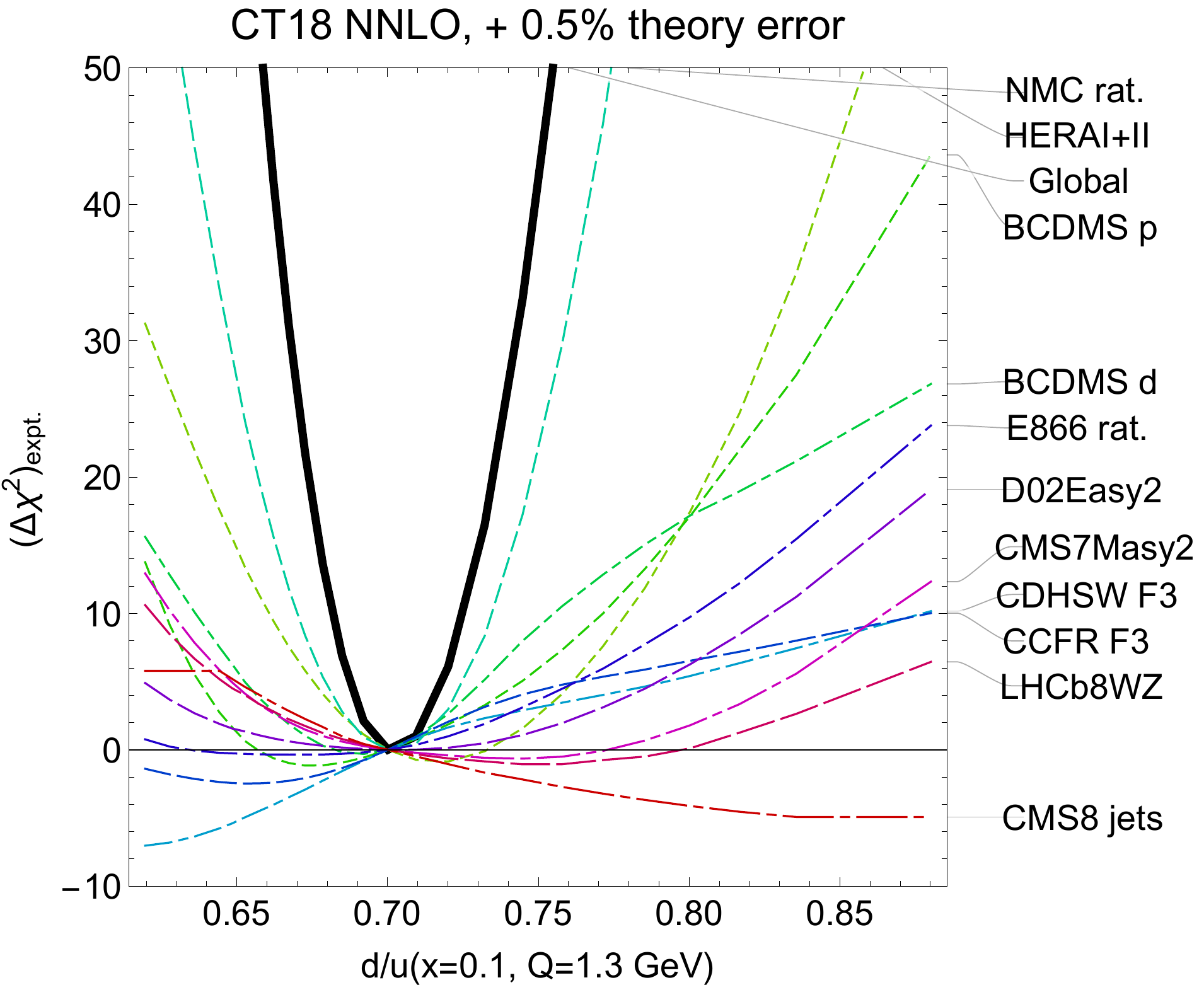}
\caption{
Left: the \textsc{PDFSense}  map for the sensitivity of the
fitted dataset of the CT18pre NNLO analysis to the $d/u$ PDF ratio,
$d/u(x\!=\!0.1,\mu\!=\!1.3\, \mbox{ GeV})$. Right:
Dependence of $\chi^2$ for the individual and all experiments 
of the CT18pre dataset on the value of $d/u(x\!=\!0.1,\mu\!=\!1.3\, \mbox{ GeV})$
obtained with the LM scan technique. The curves show the deviations 
$\Delta \chi^2_\mathrm{expt.}\equiv\chi^2_\mathrm{expt.}(\vec a)-\chi^2_\mathrm{expt.}(\vec a_0)$
from the best-fit values in $\chi^2$ for the indicated experiments, as well as for the totality of
all experiments.
}
\label{fig:valid_du}
\end{figure}

\begin{figure}[p]
\hspace*{-0.2cm}\includegraphics[clip,width=0.46\textwidth]{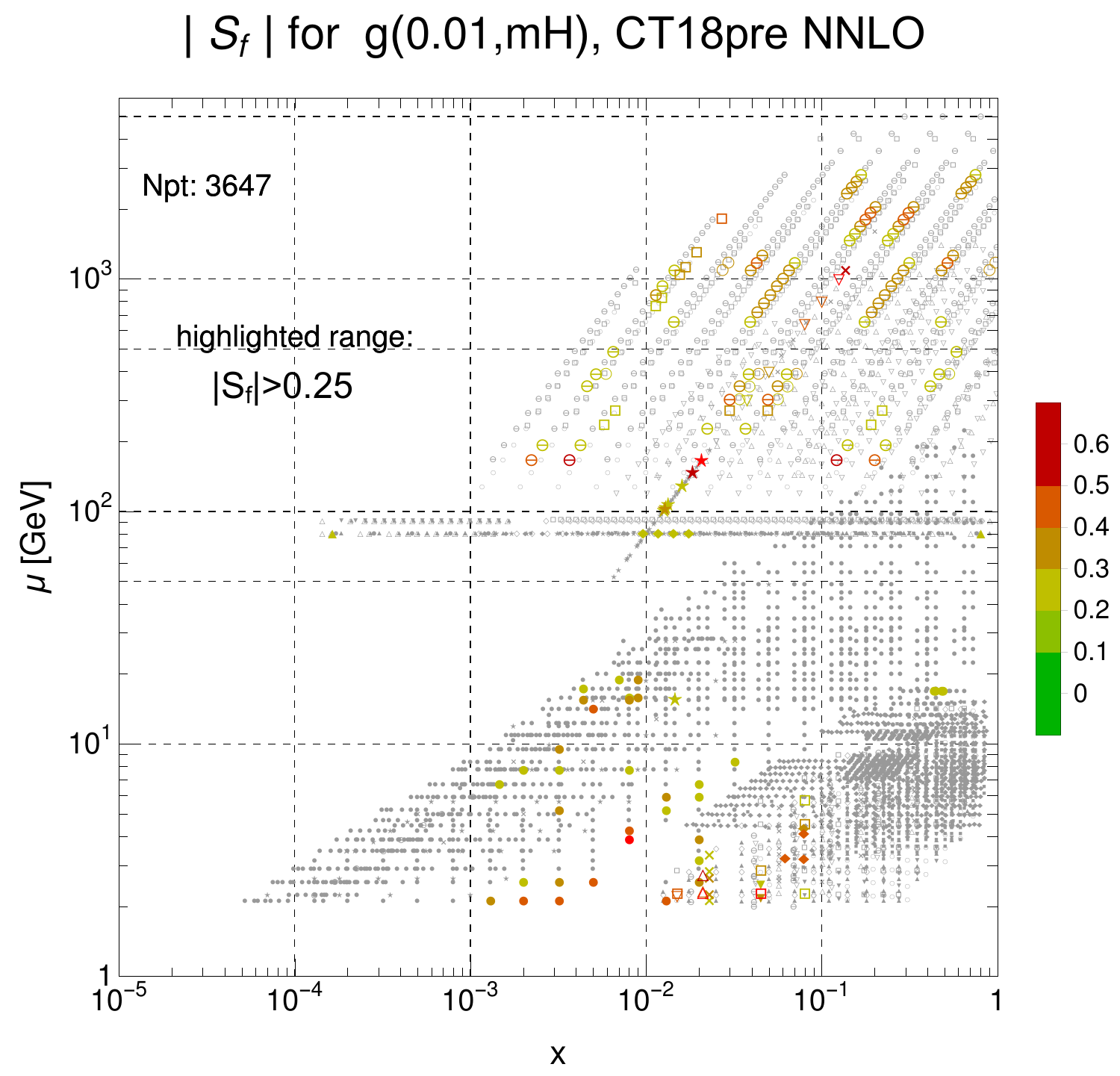} \ \
\includegraphics[clip,width=0.53\textwidth]{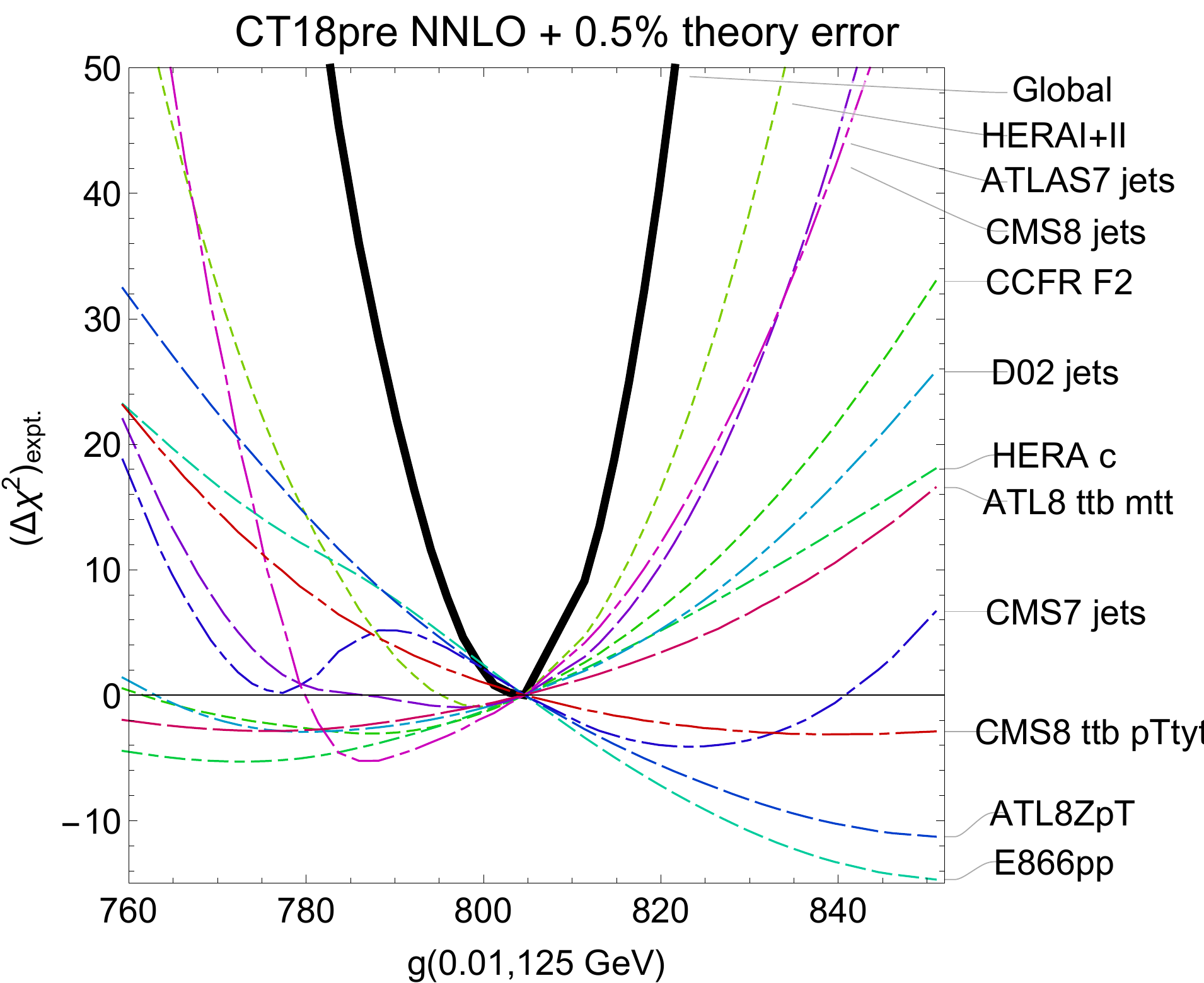}
\caption{
  Like Fig.~\ref{fig:valid_du}, but comparing the \textsc{PDFSense} map (left)
	and LM scan (right) for the gluon PDF $g(x\!=\!0.01,\mu\!=\!m_H)$ in the Higgs boson production region.
}
\label{fig:valid_higgs}
\end{figure}

How do the surveys based on \textsc{PDFSense} compare against the actual
fits? As we noted, the \textsc{PDFSense} method is designed to provide a
fast large-scope estimation of the impact of the existing and future
data sets in conjunction with other tools, such as the \textsc{ePump} \cite{Schmidt:2018hvu}
program for PDF reweighting. It works the best in the quadratic
(Hessian) approximation near the best fit, and when the new experiments
are compatible with the old ones. When detailed understanding of the
experimental constraints is necessary, the \textsc{PDFSense} approach
must be supplemented by other techniques, such as Lagrange
multiplier (LM) scans \cite{Stump:2001gu,Pumplin:2000vx,Brock:2000ud}. 

As an illustration of the scope of the differences between the
\textsc{PDFSense} predictions before and after the fit,
the left panels in Figs.~\ref{fig:valid_du} and \ref{fig:valid_higgs}
show the \textsc{PDFSense} maps for
$d/u(x\!=\!0.1,\mu\!=\!1.3\,\mbox{ GeV})$ and
$g(x\!=\!0.01,\mu\!=\!125 \mbox{ GeV})$ evaluated using a 
preliminary CT18 NNLO fit (designated as ``CT18pre'') that includes 11 new
LHC experimental data sets, namely CMS8jets'17, CMS7jets'14,
ATLAS7jets'15, LHCb8WZ'16, CMS8Wasy'16, LHCb8Zee'15, LHCb7ZWrap'15, ATL8ZpT'16, ATL8ttb-pt'16, ATL8ttb-mtt'16, 
and 8 TeV $t\bar{t}$ production at CMS (`CMS8 ttb pTtyt') \cite{Sirunyan:2017azo} in
  addition to the experiments included in the CT14HERA2 fit. The full
  details of the CT18 fit will be presented in an upcoming
  publication \cite{CT18}. Some modifications were made in the
  methodology adopted in CT18, as compared to CT14HERA2; notably the PDF
  parametrization forms and treatment of NNLO radiative contributions
  have been changed, while some shown curves are also subject to a
  theoretical uncertainty associated with the QCD scale
  choices. In accord with the
  \textsc{PDFSense} predictions based on the CT14HERA2 NNLO PDFs, we find
  that including the above LHC experiments into the fit
  produces only mild differences between the CT18pre and
  CT14HERA2 NNLO PDFs. Consequently the \textsc{PDFSense} $\{x,\mu\}$ maps based on
  CT18pre NNLO PDFs are similar to the CT14HERA2 ones
  \cite{PDFSenseWebsite}. One noticeable difference is that the
  sensitivity of the new experiments decreases after they are included
  in the CT18pre fit, because the new information from the newly
  added experiments suppresses PDF uncertainties of data residuals.  

  In the right panels of Figs.~\ref{fig:valid_du}
  and \ref{fig:valid_higgs}, we illustrate the constraints on the same
  quantities, $d/u(0.1,1.3\mbox{ GeV})$ and $g(0.01,125\mbox{ GeV})$
  in the candidate CT18pre NNLO fit, now obtained with the help of
  LM scans. A LM scan \cite{Stump:2001gu,Pumplin:2000vx,Brock:2000ud}
  is a powerful technique that elicits detailed information about a
  PDF-dependent quantity $X(\vec a)$, such as a PDF or cross section,
  from a constrained global fit in which the value of $X(\vec a)$ is fixed by
  an imposed condition. By minimizing a modified goodness-of-fit function
$\chi^2_{LM}(\lambda,\vec{a})$ that includes a `generalized-force'
term equal to $X(\vec{a})$ with weight $\lambda$, in addition to the
global $\chi_{global}^2$ in Eq.~(\ref{eq:chi2glob}), a LM scan reveals
the parametric relationship between $X(\vec a)$ and
$\chi^2_\mathrm{global}$ or $\chi^2_\mathrm{expt.}$ contributions from
individual experiments, including any non-Gaussian dependence.
In the LM scans at hand, the modified fitted function takes the form
\begin{equation}
	\chi^2_{LM}(\lambda,\vec{a}) = \chi^2_\mathrm{global}(\vec{a}) + \lambda X(\vec{a}),\ 
\end{equation}
and  $X(\vec a)$ are $d/u(x,\mu)$ or $g(x,\mu)$ at a specific location in
$\{x,\mu\}$ space. For the optimal parameter combination
$\vec{a} \equiv \vec{a}_0$ at which $\chi^2_\mathrm{global}(\vec{a})$ is minimized,
we find in Fig.~\ref{fig:valid_du} that $d/u(0.1,1.3\mbox{ GeV}) \approx 0.7$. The LM scan for the
$d/u$ then consists of a series of refits of the parameters
$\vec{a}_k$, as the multiplier parameter $\lambda$ is dialed along a set
of discrete values $\lambda_k$, effectively
pulling $d/u$ away from the value $\sim\! 0.7$ at $\vec{a} = \vec{a}_0$
preferred by the global fit. The right panel of Fig.~\ref{fig:valid_du}
shows the relationship between
$d/u(0.1,1.3\mbox{ GeV})$ and $\chi^2_\mathrm{global}$ that is quantified
this way; and similarly for $g(0.01,125\mbox{ GeV})$.

%

We can also examine how the $\chi^2$ changes for the individual
experiments. Figs.~\ref{fig:valid_du} and \ref{fig:valid_higgs} show
the curves for 11 experiments with the largest variations
$\mathrm{max}(\chi^2)-\mathrm{min}(\chi^2)$ in the shown ranges of $d/u$
and $g$, i.e., the most constraining experiments.
We notice that, while the $\Delta\chi^2$ dependence is
nearly Gaussian for the total $\chi^2$, it is sometimes less so for
the individual experiments.  Some experiments may be inconsistent
when they have a large best-fit $\chi^2(\vec a_0)$ or
prefer an incompatible $X$ value. Figure~\ref{fig:valid_du} is an
example of a good agreement between the experiments, when the individual
$\Delta\chi^2_{expt.}$ curves are approximately quadratic and
minimized at about the same location. Figure~\ref{fig:valid_higgs}
shows more pronounced inconsistencies, notably in the case of the E866pp
and ATL8ZpT curves that prefer a significantly larger
$g(0.01, 125\mbox{ GeV})$ than in the rest of the
experiments.

The LM procedure thus allows a systematic exploration
of the exact constraints from the experiments on $X$
without relying on the Gaussian assumption
that is inherent to the \textsc{PDFSense} method.
Both \textsc{PDFSense} and LM scans successfully identify the experiments with the strongest sensitivity
to $X$, while their specific rankings of such experiments are not
strictly identical and reflect the chosen
ranking prescription and settings of the global fit.
We emphasize that, though informative, the LM scans are computationally intensive, with a typical
30-point scan at NNLO requiring $\sim\!\! 6500$ CPU core-hours on a
high-performance cluster. This is in contrast to the \textsc{PDFSense}
analysis, which can be run for our entire 4021-point dataset on a single CPU core
of a modern workstation in $\sim\!\! 5$ minutes, representing
a $\sim\! 0.8 \times 10^5$ savings in computational cost.

\begin{table}
  \begin{tabular}{|c c| c || c c| c|}
    \hline
	\multicolumn{3}{|c||}{$d/u(x\!=\!0.1,\mu\!=\!1.3\,\mbox{ GeV})$}
        &\multicolumn{3}{c|}{$g(x\!=\!0.01,\mu\!=\!125 \mbox{ GeV})$} \tabularnewline
	\multicolumn{2}{|c|}{\textsc{PDFSense}}  & LM scan & \multicolumn{2}{c|}{\textsc{PDFSense}}  & LM scan \tabularnewline
	\hspace{0.5cm}   CT14HERA2 \ \  & \ \ CT18pre \ \  & \ \  CT18pre \ \ & \ \  CT14HERA2 \ \   & \ \ CT18pre \ \  & \ \ CT18pre \hspace{0.3cm} \tabularnewline
        \hline 
	\hspace{0.5cm}   HERAI+II'15   &   NMCrat'97      &   NMCrat'97     &    HERAI+II'15    &   HERAI+II'15     & HERAI+II'15    \hspace{0.3cm}  \tabularnewline
        \hspace{0.5cm}   BCDMSp'89     &   HERAI+II'15    &   CCFR-F3'97    &    CMS8jets'17    &   CMS8jets'17     & CMS8jets'17    \hspace{0.3cm}  \tabularnewline
        \hspace{0.5cm}   NMCrat'97     &   BCDMSp'89      &   HERAI+II'15   &    CMS7jets'14    &   CMS7jets'14     & ATL8ZpT'16     \hspace{0.3cm}  \tabularnewline
        \hspace{0.5cm}   CCFR-F3'97    &   CCFR-F3'97     &   BCDMSd'90     &    ATLAS7jets'15  &   E866pp'03       & E866pp'03      \hspace{0.3cm}  \tabularnewline
        \hspace{0.5cm}   E866pp'03     &   BCDMSd'90      &   BCDMSp'89     &    E866pp'03      &   ATLAS7jets'15   & ATLAS7jets'15  \hspace{0.3cm}  \tabularnewline
        \hspace{0.5cm}   BCDMSd'90     &   E605'91        &   CDHSW-F3'91   &    BCDMSd'90      &   BCDMSd'90       & CCFR-F2'01     \hspace{0.3cm}  \tabularnewline
        \hspace{0.5cm}   CDHSW-F3'91   &   E866pp'03      &   E866rat'01    &    CCFR-F3'97     &   BCDMSp'89       & D02jets'08     \hspace{0.3cm}  \tabularnewline
        \hspace{0.5cm}   CMS8jets'17   &   E866rat'01     &   CMS7Masy2'14  &    D02jets'08     &   D02jets'08      & HERAc'13       \hspace{0.3cm}  \tabularnewline
        \hspace{0.5cm}   E866rat'01    &   CMS8jets'17    &   NuTeV-nu'06   &    NMCrat'97      &   NMCrat'97       & NuTeV-nub'06   \hspace{0.3cm}  \tabularnewline
        \hspace{0.5cm}   LHCb8WZ'16    &   CDHSW-F3'91    &   CMS8jets'17   &    BCDMSp'89      &   CDHSW-F2'91     & CCFR-F3'97     \hspace{0.3cm}  \tabularnewline
\hline 
\end{tabular}
\caption{
We list the top 10 experiments predicted to drive knowledge of the
$d/u$ PDF ratio and of the gluon distribution in the Higgs region
according to \textsc{PDFSense} and LM scans.
	For both, we list the \textsc{PDFSense} evaluations based both
	on the CT14HERA2 fit and on a preliminary CT18pre fit
         in the first and second columns on either side of the
	double-line partition.
}
\label{tab:valid}
\end{table}

Let us further illustrate these observations by referring again to
Figs.~\ref{fig:valid_du} and \ref{fig:valid_higgs}, as well as to
Table~\ref{tab:valid} that displays the top 10 experiments with the
largest cumulative sensitivity to $d/u(0.1,1.3 \mbox{ GeV})$ and
$g(0.01,\ 125 \mbox{ GeV})$ according to \textsc{ PDFSense} and LM
scans, with either CT14HERA2 or CT18pre PDFs used to
construct the \textsc{PDFSense} rankings. 
In the \textsc{PDFSense} columns, the experiments are ranked in order
of descending cumulative sensitivities $\sum_{i=1}^{N_{pt}} |S_{f}|(x_i,\mu_i)$ according to the 
same prescription as in Sec.~\ref{sec:Experiment-rankings-according}.
For the LM scans, the table shows the experiments that have the
largest variations $\mathrm{max}(\chi^2)-\mathrm{min}(\chi^2)$ in the
range of $X$ corresponding to $\Delta \chi^2_{global}\leq 100$, that
is, within approximately the 90\% probability level interval of the CT18pre NNLO
PDFs. As the residual uncertainties $\Delta r_i$ in the sensitivities $S_f$ are normalized to
the root-mean-squared residuals $\langle r_0\rangle_E$  at the best fit, 
cf. Eq.~(\ref{eq:sens}), we similarly divide
$\mathrm{max}(\chi^2)-\mathrm{min}(\chi^2)$ by the best-fit
$\chi^2(\vec{a}_{0})/N_\mathit{pt}$ of the experiment in the rankings for the LM scans in Table~\ref{tab:valid}. 

From the side-by-side examination of the figures and the table, we can draw a broad conclusion
that both the pre-fit \textsc{PDFSense} and post-fit LM scan
approaches agree in identifying the most constraining
experiments, even though they may result in different orderings of these
experiments. This agreement is especially impressive
in the instance of $d/u(x\!=\!0.1,\mu\!=\!1.3\,\mbox{ GeV})$, when
the rankings agree on 8 out of 10 leading experiments, confirming the
dominance of the NMC $p/d$ ratio, HERAI+II,  CCFR $F_3$, and
BCDMS $p$ and $d$ measurements. For $g(x\!=\!0.01,\mu\!=\!m_H)$, 
for which we see more tension and non-Gaussian behavior in Fig.~\ref{fig:valid_higgs},
both \textsc{PDFSense} and LM scans concur on the crucial role played by the top 5-6 experiments, namely,
HERAI+II, E866pp, and inclusive jet production data from CMS, ATLAS, and
D0 Run-2. The upward pull on $g$ from the incompatible ATL8ZpT data set seen
in Fig.~\ref{fig:valid_higgs} modifies the rankings of the trailing
experiments, such as CMS7 jets or BCDMS. Based upon an extended
battery of LM scans we have performed, including the two examples
presented here, we conclude that the \texttt{PDFSense} surveys perform
as intended.

Lastly, we reiterate that a number of subtleties exists in comparing
the results of LM scans and \textsc{PDFSense} sensitivity plots. Most importantly,
\textsc{PDFSense} is intended by conception as a tool to quantify the
anticipated {\it average} impact of potentially unfitted data
based upon their precision in comparison to the PDF
uncertainties. We discussed simplifying assumptions made in
\textsc{PDFSense} in order to bypass certain complexities of the full
fit and obtain quick estimates.   
LM scans, on the other hand, provide post-fit assessments of the contributions of specific
data to the global $\chi^2$ function, as specific quantities predicted
by the QCD analysis are dialed away from
their optimal values. In the comparisons we made,
the detailed pictures produced by both \textsc{PDFSense} and the LM
scans depend on a variety of theoretical settings like pQCD scale
choices, as well as upon the specific implementation of correlated
experimental uncertainties [from up to $\sim\!\! 100$ different sources in
some experiments] and the parametric forms chosen for the nonperturbative
parametrizations at the starting scale $\mu = Q_0$. The inclusion of additional theory uncertainties
and decorrelation of some experimental correlated errors are
necessitated in a few experiments
by the relatively large $\chi^2$ values that would otherwise be obtained.
All these have some peripheral effect on the specific orderings of experiments shown in Table~\ref{tab:valid}.
Thus, rather than anticipating an exact point-to-point matching
between the \textsc{PDFSense} and LM methods, we instead expect, and
indeed find, the general congruity between the most important
experiments identified by the two approaches illustrated in this section.
\section{Conclusions }
\label{sec:Conclusions}
In the foregoing analysis, we have confronted the modern challenge
of a rapidly growing set of global QCD data with new statistical methodologies
for quantifying and exploring the impact of this information. These
novel methodologies are realized in a new analysis tool \textsc{PDFSense
\cite{PDFSenseWebsite},} which allows the rapid exploration of the
impact of both existing and potential data on PDF determinations,
thus providing a means of weighing the impact of measurements of QCD
processes in a way that allows meaningful conclusions to be drawn
without the cost of a full global analysis. We expect this approach
to guide future PDF fitting efforts by allowing fitters to examine
the world's data \textit{a priori,} so as to concentrate analyses on
the highest impact datasets. In particular, this work builds upon
the existing CT framework with its reliance on the Hessian formalism
and assumed quasi-Gaussianity, but these features do not impact the validity
of our analysis and conclusions.
Our approach provides a means to carry out a detailed study of
data residuals, for which we explored novel visualizations in several
ways, including the PCA, t-SNE, and reciprocated distance approaches
discussed in Sec. \ref{subsec:Manifold-learning}. These techniques
show promise for moving forward by providing useful insights into the
numerical relationships among datasets and experimental processes. 

Crucial to this analysis is the leveraging of both the existing and
proposed statistical measures laid out in Secs.~\ref{sec:Correlations} and \ref{sec:Sensitivities}.
Of these, the flavor-specific sensitivity $S_{f}$ of Eq.~(\ref{eq:sens})
for a data point to the PDF serves as a particularly powerful discriminator,
and we deployed it and the correlation $C_{f}$ of Eq.~(\ref{eq:corr})
to map PDF constraints provided by data over a wide range in $\{x,\mu\}$.
This was facilitated by the fact that the sensitivity and correlation
are readily computable over the extent of the global dataset. The
companion website collects a large number of figures illustrating
the sensitivities to various flavors as a function of $x$ and $\mu$.

To quantify the abundant information contained in the maps of sensitivities,
in Sec.~\ref{sec:Experiment-rankings-according}
we presented statistical estimators to systematically rank and assess subsidiary
datasets within the world's data according to their potential to be influential
in constraining PDFs. We note that one
is allowed some freedom in choosing a specific ranking prescription, but we find our
conclusions to be stable against variations among these possible choices. In this context,
we reaffirmed the unique advantage of DIS and jet production for determination of the PDFs.

Many intriguing physics results can be established using our
sensitivity methods, and the specific results in the previous sections
are only illustrative examples. We stress that these results take
the complementary form of sensitivity tables (for example, Table \ref{tab5})
and $\{x,\mu\}$ plots (such as Fig. \ref{fig:CorrSensH14}), which
respectively offer global categorizations of the experimental landscape and
detailed mappings of the placements of PDF constraints in
$\{x,\mu\}$ space. In totality, the full range of physics insights from this method 
is beyond the scope of the present article, but the interested user
can explore them using our \textsc{PDFSense} package at \cite{PDFSenseWebsite}.
We mention only a representative sample of these to motivate the reader:
\begin{itemize}
\item A wide range of experimental processes possess sensitivity to the
nucleon's quark sea distributions; for example, for the distribution
$\overline{d}(x,\mu)$, the $\sigma_{pd}$ DY measurements of E866
(E866rat'01) exhibit strong sensitivity, but so do DY data from E605 (E605'91)
as well as (at larger $\mu$) information on the $\mu$-production
asymmetry $A_{\mu}(\eta)$ from CMS at 7 TeV (CMS7Masy2'14); at high $x$ and
$\mu$, CMS inclusive jet data (CMS8jets'17, CMS7jets'14) also acquire some sensitivity
to $\bar u$ and $\bar d$.
Still, however, the recent HERA data (HERAI+II'15) registers the greatest
overall sensitivity.
\item Were they taken cumulatively together as a single dataset, CMS jet
  production at 7 and 8 TeV (CMS7jets'14 and CMS8jets'17) would provide a total sensitivity
  $|S^E_s| = 11.9 + 8.11$ to $s(x,\mu)$ that is comparable to
  one of the NuTeV (NuTeV-nu'06) or CCFR (CCFR SI nu'01, CCFR SI nub'01) dimuon SIDIS experiments,
  which have very strong average sensitivity to the strange
  distribution. Still, the strongest constraint is contributed by 
a mix of the DIS measurements, including $\nu\mu\mu$ data from NuTeV
(NuTeV-nu'06), data on $\nu(\overline{\nu})\mu\mu$ processes from SIDIS
at CCFR (CCFR SI nu'01 and CCFR SI nub'01), as well as the inclusive DIS data at lower $x$ from
HERA1+2 (HERAI+II'15) that actually has the strongest cumulative sensitivity. 
Similarly, various vector boson production data sets have a rank-3
point-averaged sensitivity to the strangeness, including  
the $A_{\mu}(\eta^\mu)$ data from D0 (D02Masy'08) and CMS (CMS8Wasy'16, CMS7Masy2'14), as well
ATLAS $W/Z$ production (ATL8DY2D'16, ATL7WZ'12) and high-$p_T$ $Z$ production (ATL8ZpT'16)
cross sections. Although each of the
individual vector boson production data set has a weak cumulative sensitivity
to $s(x,\mu)$ because of a small number of data points,
in totality a group of {\it mutually consistent} LHC experiments on
vector boson production can
provide a competing constraint on $s(x,\mu)$ that confronts the
 low-energy CCFR/NuTeV constraints. 
\item Knowledge of the charm distribution $c(x,\mu)$ is most influenced
by a number of datasets, with HERA (HERAI+II'15) at low $x$ especially important.
Fixed target measurements, particularly those of CDHSW on the proton's
$F_{2}^{p}$ structure function (CDHSW-F2'91) have strong sensitivity at slightly
higher $x\!\sim\!10^{-1}$, while a wide range of jet measurements, including
7 TeV data from ATLAS (ATLAS7jets'15) and CMS (CMS7jets'14), and 8 TeV CMS (CMS8jets'17) points
are also sensitive. This pattern of sensitive
measurements broadly follows the corresponding plot for $|S_{g}|(x_{i},\mu_{i})$
{[}as well as $|S_{b}|(x_{i},\mu_{i})${]} due to the dominance of
boson fusion graphs in heavy quark production. The datasets of
importance we identify are broadly consistent with the conclusions
of the recent CT14 analysis \cite{Hou:2017khm} of the nucleon's
intrinsic charm \cite{Hobbs:2013bia}.
\item One can also study the correlations and sensitivities for various
derived PDF combinations. For instance, for the $\overline{d}/\overline{u}$
ratio representing deviations from flavor symmetry in the nucleon
sea, the E866 experiment (E866rat'01) shows exceptional point-averaged
sensitivity, $\langle|S_{\bar{d}/\bar{u}}|\rangle=1.67$ such that
its ``C'' ranking for its overall sensitivity to $\bar{d}/\bar{u}$
places it in the company of only a few other DIS and DY experiments, despite
their much larger number of measurements, $N_{\mathit{pt}}=15$. At somewhat lower $x\gtrsim0.01$,
NMC data on the structure function ratio $F_{2}^{d}/F_{2}^{p}$ (NMCrat'97)
show sensitivity in the range $0.8<|S_{\overline{d}/\overline{u}}|<2$.
At still lower $x$, the CMS 8 and 7 TeV $A_{\mu}$ points (CMS8Wasy'16, CMS7Masy2'14) and
$W/Z$ data from LHCb (LHCb8WZ'16) show strong pull, corresponding to
point-averaged rankings of ``2,'' ``{\bf 1},'' and ``2,'' respectively.
\item We also consider the PDF ratio $d/u(x,\mu)$, which often serves as
a discriminant among various nucleon structure models, especially at
high $x$. For $x>0.1$ an amalgam of fixed-target experiments, including
the NMC $F_{2}^{d}/F_{2}^{p}$ data (NMCrat'97) particularly, but also $F_{2}^{p}$
measurements from BCDMS (BCDMSp'89) and CCFR (CCFR-F2'01) as well as $xF_{3}^{p}$ data
from CCFR drive the current status.
At higher $\mu$, however, the LHCb $W/Z$ data (LHCb8WZ'16) and $A_{e}(\eta)$
measurements from Run-2 of D0 (D02Easy2'15) also constrain the high $x$
behavior of $d/u$ together with $A_{\mu}(\eta)$ points
from CMS at 7 TeV (CMS7Masy2'14).
\item
More generally, we note that, among the new LHC experiments to be considered for future
global fits, the datasets for inclusive jet production are expected to have the greatest
impact, followed by a group of vector boson production experiments at ATLAS, CMS, and LHCb.
We find that the constraints from jet production at the LHC depend significantly on
the treatment of experimental systematic uncertainties --- especially the correlated systematic errors.
It is conceivable that, with the full implementation of NNLO theoretical cross sections and
modest reduction in the experimental systematic uncertainties, the constraints from the LHC jet
production will catch up in strength to the effect of adding a large fixed-target DIS dataset, such as
BCDMS $F^p_2$ (BCDMSp'89). Meanwhile, the magnitude of the constraint on the gluon PDF from high-$p_T$ $Z$
production (ATL8ZpT'16) is comparable to those from the combined HERA SIDIS charm dataset (HERAc'13) or
inclusive jet production from CDF Run-2 (CDF2jets'09); that is, the high-$p_T$ $Z$ data are significant
in the event that the jet datasets are not included, in overall consistency with the findings in
Ref.~\cite{Boughezal:2017nla}. The smaller ATLAS $t\overline{t}$ production data sets (ATL8ttb-pt'16,
ATL8ttb-y\_ave'16, ATL8ttb-mtt'16, ATL8ttb-y\_ttb'16) have strong point-by-point sensitivity to the gluon,
but will have a more diminished role when combined with other, larger data sets. HERA DIS (HERAI+II'15),
BCDMS $F_2^d$ (BCDMSd'90), and CMS inclusive jets at 8 TeV (CMS8jets'17) render the strongest overall
constraints on the Higgs production cross section at the LHC according to the rankings in Table~\ref{tab6}.
\end{itemize}
Quantifying correlations and sensitivities thus provides a comprehensive means of
evaluating the ability of a global dataset to constrain our knowledge
of nucleon structure. It must be emphasized,
however, that this analysis is not a substitute for actually performing
a QCD global analysis, which remains the single most robust means
of determining the nucleon PDFs themselves. Rather, the method presented
in the paper is a guiding tool to both supplement and direct fits
by gauging the potential for improving PDFs with the incorporation of new
datasets.

The essential ingredients of this study are the PDF-residual correlation
and sensitivity $|C_{f}|$ and $|S_{f}|$, with the latter representing
an extension of the correlation used elsewhere in the modern PDF literature.
These definitions are robust enough that we can exhaustively score
the data points in an arbitrary global dataset to construct and map
the resulting distributions, as shown in Figs.~\ref{fig:corr-main}
and \ref{fig:sens-main}. Accordingly, we found it possible to impose
cuts on these distributions to identify points of especially strong
correlation ($|C_{f}|>0.7$) or sensitivity ($|S_{f}|>0.25$); we
stress that these cuts are chosen as approximate indicators, and any
user can adjust them freely. On the other hand, the distributions
themselves, as shown in the second panels of Figs.~\ref{fig:corr-main} and \ref{fig:sens-main},
are not subject to such cut choices.
Although the conclusions of this analysis are resistant to alterations
in the basic approach, it is worth noting that other formats are possible
for evaluating experimental sensitivities and performing the rankings of
measurements. For example, one might use somewhat different matchings than
those outlined in App.~\ref{sec:supp} to extract $\{x,\mu\}$ points from the
experimental data, but we expect the resulting impact on
the overall picture to be minor. Similarly, while the ordering inside
ranking tables like Table~\ref{tab5} was decided according to the total
sensitivity to serve our specific goal of identifying the most valuable experiments
for the CTEQ-TEA fit, for other purposes one might produce alternative tables ranked
according to point-averaged sensitivities, or sensitivities to
specific flavors. Such alternate conventions would also yield important
information, and \textsc{PDFSense} allows the user to do this. It should
be stressed that these elections for the form of our presentation
can always be recovered from the more fundamental information ---
the numerical values of the sensitivities detailed in the Supplementary
Material of App.~\ref{sec:SM}. 

While we have demonstrated these techniques in the context of the
CT14 family of global fits, they are of sufficient generality that
one could readily repeat our analysis using alternative PDF sets.
For the sake of testing this point and validating our predictions
for the most decisive experiments in the CTEQ-TEA dataset, we performed
a preliminary fit including the CT14HERA2 and the candidate LHC experiments
(`CT18pre'), and directly compared \textsc{PDFSense} predictions against
Lagrange multiplier scans quantifying the constraints these fitted measurements
imposed on select quantities. This provided a demonstration of the robustness
of our sensitivity-based analysis, which identified the same sets of high-impact
measurements {\it before fitting}.
The results of this study can be expected to vary somewhat depending
on the specifics of the PDF sets used to compute $|C_{f}|$ and $|S_{f}|$,
but we see this as an advantage of \textsc{PDFSense}. One
could imagine exploiting them to undertake a systematic analysis of the impact of various theoretical
assumptions implemented in competing global fits (\textit{e.g.}, the
choice of input PDF parametrization or the status of the perturbative
QCD treatment implemented in various processes). The sensitivity $S_f$
can be constructed either from the Hessian or Monte-Carlo PDF uncertainties,
as prescribed by Eqs.~(\ref{eq:sens}) and (\ref{eq:sensMC}), while the shifted
residuals that are crucial to our analysis can be recovered from any type of
covariance matrix, as argued in relation to Eq.~(\ref{eq:res-cov}).
In the same spirit but on
the side of the data, \textsc{PDFSense} empowers the user to evaluate
the combined impact of multiple experimental datasets \textemdash{}
for example, to evaluate the extent to which the impact of a proposed
experiment might be diminished by the constraints already imposed
by existing measurements. These various functions collectively suggest
a number of possible avenues to use the presented approach and the \textsc{PDFSense}
tool to advance PDF knowledge in the coming
years. 

\subsection*{Acknowledgments}

We thank our CTEQ-TEA colleagues, Davison Soper, and Madeline Hamilton
for support and insightful discussions, and
appreciate helpful clarifications concerning the LHC
experimental data sets from Alexander Glazov, Uta Klein,
Bogdan Malescu, and Klaus Rabbertz. We also thank German Valencia,
Ursula Laa, and Dianne Cook for helpful discussions related to data visualizations based
on the PCA and t-SNE methods. This work was supported in part by the
U.S.~Department of Energy under Grant No.~DE-SC0010129 and by the
National Natural Science Foundation of China under the Grant No.~11465018.
T.J.~Hobbs acknowledges support from an EIC Center Fellowship.
The work of J.G. is sponsored by Shanghai Pujiang Program.

\appendix

\section{Approximate kinematical variables\label{sec:supp}}

In this section, we describe in detail our method for identifying
the values of $\{x_{i},\mu_{i}\}$ that correspond to experimental
data.

For each experimental data point $i$, we can establish an approximate
relation between the kinematical quantities for that data point, and
unobserved quantities specifying the PDFs: the partonic momentum fraction
$x$ and QCD factorization scale $\mu$. For example, in DIS, $x$
and $\mu$ are approximately equal to Bjorken $x_{B}$ and momentum
transfer $Q$ according to the Born-level kinematic relation. Although
this relation is violated by higher-order radiative contributions,
it will approximately hold in most scattering events. The same overall
logic can be followed to relate the kinematical quantities in every
process of the CTEQ-TEA global set to the \emph{approximate} unobserved
quantities $x$ and $\mu$ in the PDFs. These relations vary by process
and are used to assign approximate pairs $\{x_{i},\mu_{i}\}$ for
each data point.\footnote{It should be pointed out that, while there are 5227 $\{x,\mu\}$ points
generated by the 4021 physical measurements in the
default CTEQ-TEA dataset of this study, occasionally there are instances
in which $|C_{f}|$ and $|S_{f}|$ cannot be meaningfully computed
for select flavors. For example, since the bottom quark PDF $b(x,\mu)$
has no sensible definition below its partonic threshold (\textit{i.e.},
for $\mu<m_{b}=4.75\,\mbox{ GeV}$), it is not possible to evaluate
$|S_{b}|$ for data points extracted at $\mu$ scales below the $b$-quark
mass. Similarly, there are situations when the extracted parton fraction
$x_{i}\approx1$, such that some PDF flavors $f(x_{i},\mu_{i})\approx0$,
and the Hessian procedures described in this paper do not yield a
well-defined correlation or sensitivity. In these cases, we simply
redact the associated $\{x_i,\mu_i\}$ points.}

Specifically, for DIS, which primarily measures the differential cross
sections of the form $d^{2}\sigma/(dx_{B}dQ^{2})$, we simply take
\begin{equation}
\mu_{i}\approx\left.Q\right|_{i},\ x_{i}\approx\left.x_{B}\right|_{i}
\end{equation}
as mentioned above, where the kinematical variables inside ``$\left.\right|_{i}$''
are evaluated at their experimentally measured values for the $i^{th}$
data point. The above approximate relations hold even when (N)NLO
radiative contributions are included.

For one-particle-inclusive particle production in hadron-hadron scattering
of the form $AB\rightarrow CX$ , we plot two $x$ values if the rapidity
$y_{C}$ is known:
\begin{equation}
\mu_{i}\approx\left.Q\right|_{i},\ x_{i}^{\pm}\approx\left.\frac{Q}{\sqrt{s}}\,\exp(\pm y_{C})\right|_{i}.\label{eq:rap-match}
\end{equation}
We set $y_{C}=0$ if the rapidity is integrated away. We point out
that for processes of this type, Eq. (\ref{eq:rap-match}) implies
that a measurement in a single rapidity bin can in fact probe two
distinct values of $x$; for this and other potential reasons, the
number of raw data points in such an experiment ($N_{\mathit{pt}}$) should
not be expected to match the number of extracted $\{x,\mu\}$ points in the figures.

In vector boson production, $AB\rightarrow(\gamma^{*},Z\rightarrow\ell\bar{\ell})X$
or $AB\rightarrow(W\rightarrow\ell\nu_{\ell})X$, we set $Q=m_{\ell\bar{\ell}}$
(invariant mass of the lepton pair), and $y_{C}=y_{\ell}$ if a single-lepton
rapidity is provided or $y_{C}=y_{\ell\bar{\ell}}$ if the lepton-pair
rapidity is provided. If the rapidity $y_{\ell}$ of the lepton is
known, yet $y_{\ell\bar{\ell}}$ of the pair is unknown, we use the
fact that $y_{\ell}\sim y_{\ell\bar{\ell}}\pm1$ for most events because
of the shape of the decay leptonic tensor. Thus, the momentum fractions
$x_{i}^{\pm}$ can still be estimated as $x_{i}^{\pm}\approx\left.(Q/\sqrt{s})\exp(\pm y)\right|_{i}$,
where $y\sim y_{\ell}$ (up to an error of less than 1 unit)

In single-inclusive jet production, $AB\rightarrow j+X$, we set $Q=2p_{Tj},$
$y_{C}=y_{j}.$

In single-inclusive $t\bar{t}$ pair production, $AB\rightarrow t\bar{t}X,$
we set $Q=m_{t\bar{t}},$ $y=y_{t\bar{t}}$ if known, or 0 otherwise.

In single-inclusive top (anti-)quark production, $AB\rightarrow (\overline{t})tX,$
we take $Q=2p_{Tt}$, $y=0$ for $d\sigma/dp_{T_{t}}$ (as in Expt.~ATL8ttb-pt'16).
On the other hand, for $d\sigma/d\langle y_{t}\rangle$ or $d\sigma/dy_{t\overline{t}}$,
in which the $t\bar{t}$ invariant mass is integrated out (Expts. ATL8ttb-y\_ave'16
and ATL8ttb-y\_ttb'16), we take an average mass scale $\mu_{i}=400$~GeV that is
slightly above the observed peak of $d\sigma/dm_{t\bar{t}}$ at $m_{t\bar{t}}\approx2m_{t}$.

Lastly, for the $d\sigma/dp_{T}^{Z}$ measurements from $AB\rightarrow(\gamma^{*},Z\rightarrow\ell\bar{\ell})X$
in Expts.~ATL7ZpT'14 and ATL8ZpT'16, we take $Q=\sqrt{(p_{T}^{Z})^{2}+(M_{Z})^{2}}$,
$y_{C}=y_{Z}$. {[}Here $Q$ denotes the boson's transverse mass,
not the invariant mass.{]}

\section{Tabulated results \label{sec:Tables}}

In Tables~\ref{tab:EXP_1}\textendash \ref{tab:EXP_3} we provide
a detailed key for the individual experiments mapped in Fig.~\ref{fig:data},
including the physical process, number of points, and luminosities,
where available. We group these tables broadly according to subprocess
\textemdash{} Table~\ref{tab:EXP_1} corresponds to DIS experiments,
while Tables~\ref{tab:EXP_2} and \ref{tab:EXP_3} collect various
measurements for the hadroproduction of, \textit{e.g.}, gauge boson,
jet, and $t\bar{t}$ pairs \textemdash{} and thus provide a translation
key for the experimental short-hand names given in Fig.~\ref{fig:data}.

In Tables~\ref{tab5} and \ref{tab6}, we collect the flavor-specific
($|S^E_{f}|$) and overall ($\sum_{f}|S^E_{f}|$)
sensitivities for the experimental datasets contained in this analysis.
In Table~\ref{tab5} we list the total and point-averaged sensitivities
for each main flavor ($\bar{d},\bar{u},g,u,d,s$), while Table~\ref{tab6}
gives the corresponding information for a number of quantities derived
from these, as explained in the associated captions.

%

\begin{table}
\begin{tabular}{|l|c|lr|c|}
\hline 
\textbf{Experiment name}  &  \textbf{CT\ ID\#}  & \textbf{Dataset details}  &  & $N_{\mathit{pt}}$ \tabularnewline
\hline                         
 BCDMSp'89      &  101  & BCDMS $F_{2}^{p}$  & \cite{Benvenuti:1989rh}  & 337 \tabularnewline
\hline                       
 BCDMSd'90      &  102  & BCDMS $F_{2}^{d}$  & \cite{Benvenuti:1989fm}  & 250 \tabularnewline
\hline                       
 NMCrat'97     &  104  & NMC $F_{2}^{d}/F_{2}^{p}$  & \cite{Arneodo:1996qe}  & 123 \tabularnewline
\hline                       
 CDHSW-F2'91     &  108  & CDHSW $F_{2}^{p}$  & \cite{Berge:1989hr}  & 85 \tabularnewline
\hline                       
 CDHSW-F3'91     &  109  & CDHSW $F_{3}^{p}$  & \cite{Berge:1989hr}  & 96 \tabularnewline
\hline                       
 CCFR-F2'01      &  110  & CCFR $F_{2}^{p}$  & \cite{Yang:2000ju}  & 69 \tabularnewline
\hline                       
 CCFR-F3'97      &  111  & CCFR $xF_{3}^{p}$  & \cite{Seligman:1997mc}  & 86 \tabularnewline
\hline                     
 NuTeV-nu'06    &  124  & NuTeV $\nu\mu\mu$ SIDIS  & \cite{Mason:2006qa}  & 38 \tabularnewline
\hline                     
 NuTeV-nub'06    &  125  & NuTeV $\bar{\nu}\mu\mu$ SIDIS  & \cite{Mason:2006qa}  & 33 \tabularnewline
\hline                     
 CCFR SI nu'01   &  126  & CCFR $\nu\mu\mu$ SIDIS  & \cite{Goncharov:2001qe}  & 40 \tabularnewline
\hline                     
 CCFR SI nub'01  &  127  & CCFR $\bar{\nu}\mu\mu$ SIDIS  & \cite{Goncharov:2001qe}  & 38 \tabularnewline
\hline                     
 HERAb'06       &  145  & H1 $\sigma_{r}^{b}$ ($57.4\mbox{ pb}^{-1}$)  & \cite{Aktas:2004az}\cite{Aktas:2005iw}  & 10 \tabularnewline
\hline                     
 HERAc'13       &  147  & Combined HERA charm production ($1.504\mbox{ fb}^{-1}$)  & \cite{Abramowicz:1900rp}  & 47 \tabularnewline
\hline                     
 HERAI+II'15     &  160  & HERA1+2 Combined NC and CC DIS ($1\mbox{ fb}^{-1}$)  & \cite{Abramowicz:2015mha}  & 1120 \tabularnewline
\hline                     
 HERA-FL'11      &  169  & H1 $F_{L}$ ($121.6\mbox{ pb}^{-1}$)  & \cite{Collaboration:2010ry}  & 9 \tabularnewline
\hline 
\end{tabular}\caption{Experimental datasets considered as part of CT14HERA2 and included
in this analysis: deep-inelastic scattering. We point out that the numbering scheme (CT ID\#)
included in this and subsequent tables follows the standard CTEQ labeling system
with, {\it e.g.}, Expt.~IDs of the form 1XX representing DIS experiments, {\it etc.}
The HERA combined data set HERAI+II'15 consists of both neutral-current (NC) and charge-current
(CC) scattering events.	\label{tab:EXP_1} }
\end{table}

\begin{table}
\begin{tabular}{|l|c|lr|c|}
\hline 
\textbf{Experiment name}  &  \textbf{CT\ ID\#}  & \textbf{Dataset details}  &  & $N_{\mathit{pt}}$ \tabularnewline
\hline 
\hline 
 E605'91      &   201   & E605 DY  & \cite{Moreno:1990sf}  & 119 \tabularnewline
\hline                     
 E866rat'01 &   203   & E866 DY, $\sigma_{pd}/(2\sigma_{pp})$  & \cite{Towell:2001nh}  & 15 \tabularnewline
\hline                     
 E866pp'03    &   204   & E866 DY, $Q^{3}d^{2}\sigma_{pp}/(dQdx_{F})$  & \cite{Webb:2003ps}  & 184 \tabularnewline
\hline                     
 CDF1Wasy'96 &   225   & CDF Run-1 $A_{e}(\eta^{e})$ ($110\mbox{ pb}^{-1}$)  & \cite{Abe:1996us}  & 11 \tabularnewline
\hline                     
 CDF2Wasy'05 &   227   & CDF Run-2 $A_{e}(\eta^{e})$ ($170\mbox{ pb}^{-1}$)  & \cite{Acosta:2005ud}  & 11 \tabularnewline
\hline                     
 D02Masy'08   &   234   & D$\emptyset$~ Run-2 $A_{\mu}(\eta^{\mu})$ ($0.3\mbox{ fb}^{-1}$)  & \cite{Abazov:2007pm}  & 9 \tabularnewline
\hline                     
 LHCb7WZ'12   &   240   & LHCb 7 TeV $W/Z$ muon forward-$\eta$ Xsec ($35\mbox{ pb}^{-1}$)  & \cite{Aaij:2012vn}  & 14 \tabularnewline
\hline                     
 LHCb7Wasy'12 &   241   & LHCb 7 TeV $W$ $A_{\mu}(\eta^{\mu})$ ($35\mbox{ pb}^{-1}$)  & \cite{Aaij:2012vn}  & 5 \tabularnewline
\hline                     
 ZyD02'08     &   260   & D$\emptyset$~ Run-2 $Z$ $d\sigma/dy_{Z}$ ($0.4\mbox{ fb}^{-1}$)  & \cite{Abazov:2006gs}  & 28 \tabularnewline
\hline                     
 ZyCDF2'10    &   261   & CDF Run-2 $Z$ $d\sigma/dy_{Z}$ ($2.1\mbox{ fb}^{-1}$)  & \cite{Aaltonen:2010zza}  & 29 \tabularnewline
\hline                     
 CMS7Masy2'14 &   266   & CMS 7 TeV $A_{\mu}(\eta)$ ($4.7\mbox{ fb}^{-1}$)  & \cite{Chatrchyan:2013mza}  & 11 \tabularnewline
\hline                     
 CMS7Easy'12 &   267   & CMS 7 TeV $A_{e}(\eta)$ ($0.840\mbox{ fb}^{-1}$)  & \cite{Chatrchyan:2012xt}  & 11 \tabularnewline
\hline                     
 ATL7WZ'12    &   268   & ATLAS 7 TeV $W/Z$ Xsec, $A_{\mu}(\eta)$ ($35\mbox{ pb}^{-1}$)  & \cite{Aad:2011dm}  & 41 \tabularnewline
\hline                     
 D02Easy2'15  &   281   & D$\emptyset$~ Run-2 $A_{e}(\eta)$ ($9.7\mbox{ fb}^{-1}$)  & \cite{D0:2014kma}  & 13 \tabularnewline
\hline                     
 CDF2jets'09 &   504   & CDF Run-2 incl. jet ($d^2\sigma/dp_{T}^{j}dy_{j}$) ($1.13\mbox{ fb}^{-1}$)  & \cite{Aaltonen:2008eq}  & 72 \tabularnewline
\hline                     
 D02jets'08  &   514   & D$\emptyset$~ Run-2 incl. jet ($d^2\sigma/dp_{T}^{j}dy_{j}$) ($0.7\mbox{ fb}^{-1}$)  & \cite{Abazov:2008ae}  & 110 \tabularnewline
\hline                     
 ATL7jets'12 &   535   & ATLAS 7 TeV incl. jet ($d^2\sigma/dp_{T}^{j}dy_{j}$) ($35\mbox{ pb}^{-1}$)  & \cite{Aad:2011fc}  & 90 \tabularnewline
\hline                     
 CMS7jets'13 &   538   & CMS 7 TeV incl. jet ($d^2\sigma/dp_{T}^{j}dy_{j}$) ($5\mbox{ fb}^{-1}$)  & \cite{Chatrchyan:2012bja}  & 133 \tabularnewline
\hline 
\end{tabular}\caption{Same as Table~\ref{tab:EXP_1}, showing experimental datasets for
production of vector bosons, single-inclusive jets, and $t\bar{t}$
pairs. \label{tab:EXP_2} }
\end{table}

\begin{table}
\begin{tabular}{|l|c|lr|c|}
\hline 
\textbf{Experiment name}  &  \textbf{CT\ ID\#}  & \textbf{Dataset details}  &  & $N_{\mathit{pt}}$ \tabularnewline
\hline 
\hline 
	\textbf{LHCb7ZWrap'15}      &  \textbf{245}   & LHCb 7 TeV Z/W muon forward-$\eta$ Xsec ($1.0\mbox{ fb}^{-1}$)  & \cite{Aaij:2015gna}  & 33 \tabularnewline
\hline                                                    
	\textbf{LHCb8Zee'15}        &  \textbf{246}   & LHCb 8 TeV Z electron forward-$\eta$ $d\sigma/dy_{Z}$ ($2.0\mbox{ fb}^{-1}$)  & \cite{Aaij:2015vua}  & 17 \tabularnewline
\hline                                                    
	\textbf{ATL7ZpT'14}         &  \textbf{247}   & ATLAS 7 TeV $d\sigma/dp_{T}^{Z}$ ($4.7\mbox{ fb}^{-1}$)  & \cite{Aad:2014xaa}  & 8 \tabularnewline
\hline                                                    
	\textbf{CMS8Wasy'16}        &  \textbf{249}   & CMS 8 TeV W muon, Xsec, $A_{\mu}(\eta^{\mu})$ ($18.8\mbox{ fb}^{-1}$)  & \cite{Khachatryan:2016pev}  & 33 \tabularnewline
\hline                                                    
	\textbf{LHCb8WZ'16}         &  \textbf{250}   & LHCb 8 TeV W/Z muon, Xsec, $A_{\mu}(\eta^{\mu})$ ($2.0\mbox{ fb}^{-1}$)  & \cite{Aaij:2015zlq}  & 42 \tabularnewline
\hline                                                    
	\textbf{ATL8DY2D'16}       &  \textbf{252}   & ATLAS 8 TeV Z ($d^{2}\sigma/d|y|_{ll}dm_{ll}$) ($20.3\mbox{ fb}^{-1}$)  & \cite{Aad:2016zzw}  & 48 \tabularnewline
\hline                                                    
	\textbf{ATL8ZpT'16}         &  \textbf{253}   & ATLAS 8 TeV ($d^{2}\sigma/dp_{T}^{Z}dm_{ll}$) ($20.3\mbox{ fb}^{-1}$)  & \cite{Aad:2015auj}  & 45 \tabularnewline
\hline                                                    
	\textbf{CMS7jets'14}       &  \textbf{542}   & CMS 7 TeV incl. jet, R=0.7, ($d^2\sigma/dp_{T}^{j}dy_{j}$) ($5\mbox{ fb}^{-1}$)  & \cite{Chatrchyan:2014gia}  & 158 \tabularnewline
\hline                                                    
	\textbf{ATLAS7jets'15}     &  \textbf{544}   & ATLAS 7 TeV incl. jet, R=0.6, ($d^2\sigma/dp_{T}^{j}dy_{j}$) ($4.5\mbox{ fb}^{-1}$)  & \cite{Aad:2014vwa}  & 140 \tabularnewline
\hline                                                    
	\textbf{CMS8jets'17}       &  \textbf{545}   & CMS 8 TeV incl. jet, R=0.7, ($d^2\sigma/dp_{T}^{j}dy_{j}$) ($19.7\mbox{ fb}^{-1}$)  & \cite{Khachatryan:2016mlc}  & 185 \tabularnewline
\hline                                                    
	\textbf{ATL8ttb-pt'16}     &  \textbf{565}   & ATLAS 8 TeV $t\overline{t}\:d\sigma/dp_{T}^{t}$ ($20.3\mbox{ fb}^{-1}$)  & \cite{Aad:2015mbv}  & 8 \tabularnewline
\hline                                                  
	\textbf{ATL8ttb-y\_ave'16} &  \textbf{566}   & ATLAS 8 TeV $t\overline{t}\:d\sigma/dy_{<t/\overline{t}>}$ ($20.3\mbox{ fb}^{-1}$)  & \cite{Aad:2015mbv}  & 5 \tabularnewline
\hline                                                  
	\textbf{ATL8ttb-mtt'16}    &  \textbf{567}   & ATLAS 8 TeV $t\overline{t}\:d\sigma/dm_{t\overline{t}}$ ($20.3\mbox{ fb}^{-1}$)  & \cite{Aad:2015mbv}  & 7 \tabularnewline
\hline                                                  
	\textbf{ATL8ttb-y\_ttb'16} &  \textbf{568}   & ATLAS 8 TeV $t\overline{t}\:d\sigma/dy_{t\overline{t}}$ ($20.3\mbox{ fb}^{-1}$)  & \cite{Aad:2015mbv}  & 5 \tabularnewline
\hline 
\end{tabular}\caption{Same as Table~\ref{tab:EXP_1}, showing experimental datasets for
production of vector bosons, single-inclusive jets, and $t\bar{t}$
pairs that were not incorporated in the CT14HERA2 fit but included
in our augmented CTEQ-TEA set. \label{tab:EXP_3} }
\end{table}



\begin{table}
\hspace*{-0.5cm}\begin{tabular}{c|c|c|cc|cc|cc|cc|cc|cc|cc}
\hline 
 &  &  & \multicolumn{14}{c}{Rankings, {\bf CT14 HERA2 NNLO PDFs}}\tabularnewline
No.  & Expt. & $N_{\mathit{pt}}$  & $\sum_{f}|S^E_{f}|$  & $\langle\sum_{f}|S^E_{f}|\rangle$  & $|S^E_{\bar{d}}|$  & $\langle|S^E_{\bar{d}}|\rangle$  & $|S^E_{\bar{u}}|$  & $\langle|S^E_{\bar{u}}|\rangle$  & $|S^E_{g}|$  & $\langle|S^E_{g}|\rangle$  & $|S^E_{u}|$  & $\langle|S^E_{u}|\rangle$  & $|S^E_{d}|$  & $\langle|S^E_{d}|\rangle$  & $|S^E_{s}|$  & $\langle|S^E_{s}|\rangle$ \tabularnewline
\hline 
 &  &  &  &  &  &  &  &  &  &  &  &  &  &  &  & \tabularnewline
1 & HERAI+II'15 & 1120. & 620. & 0.0922 & \text{B} &   & \textbf{A} & 3 & \textbf{A} & 3 & \textbf{A} & 3 & \text{B} &   & \text{C} &   \tabularnewline
2 & CCFR-F3'97 & 86 & 218. & 0.423 & \text{C} & \textbf{1} & \text{C} & \textbf{1} &   & 3 & \text{B} & \textbf{1} & \text{C} & 2 &   &   \tabularnewline
3 & BCDMSp'89 & 337 & 184. & 0.0908 &   &   & \text{C} &   & \text{C} &   & \text{B} & 3 & \text{C} &   &   &   \tabularnewline
4 & NMCrat'97 & 123 & 169. & 0.229 & \text{C} & 2 &   &   &   &   & \text{C} & 2 & \text{B} & 2 &   &   \tabularnewline
5 & BCDMSd'90 & 250 & 141. & 0.0939 & \text{C} &   &   &   & \text{C} & 3 & \text{C} & 3 & \text{C} & 3 &   &   \tabularnewline
6 & CDHSW-F3'91 & 96 & 115. & 0.199 & \text{C} & 2 & \text{C} & 2 &   & 3 & \text{C} & 2 & \text{C} & 3 &   &   \tabularnewline
7 & E605'91 & 119 & 113. & 0.158 & \text{C} & 2 & \text{C} & 2 &   &   &   & 3 &   &   &   &   \tabularnewline
8 & E866pp'03 & 184 & 103. & 0.0935 &   & 3 & \text{C} & 3 &   &   & \text{C} & 3 &   &   &   &   \tabularnewline
9 & CCFR-F2'01 & 69 & 89.1 & 0.215 &   & 3 &   & 3 & \text{C} & 2 &   & 3 &   & 2 &   & 3 \tabularnewline
10 & \textbf{CMS8jets'17} & 185 & 87.6 & 0.0789 &   &   &   &   & \text{C} & 3 &   &   &   &   &   &   \tabularnewline
11 & CDHSW-F2'91 & 85 & 82.4 & 0.162 &   & 3 &   & 3 &   & 3 &   & 3 & \text{C} & 3 &   &   \tabularnewline
12 & CMS7jets'13 & 133 & 63.8 & 0.0799 &   &   &   &   & \text{C} & 3 &   &   &   &   &   &   \tabularnewline
13 & NuTeV-nu'06 & 38 & 58.9 & 0.259 &   & 3 &   & 3 &   &   &   & 3 &   & 3 & \text{C} & \textbf{1} \tabularnewline
14 & \textbf{CMS7jets'14} & 158 & 57.5 & 0.0606 &   &   &   &   & \text{C} & 3 &   &   &   &   &   &   \tabularnewline
15 & CCFR SI nub'01 & 38 & 49.4 & 0.217 &   & 3 &   & 3 &   &   &   & 3 &   & 3 & \text{C} & \textbf{1} \tabularnewline
16 & \textbf{ATLAS7jets'15} & 140 & 48.2 & 0.0574 &   &   &   &   &   & 3 &   &   &   &   &   &   \tabularnewline
17 & CCFR SI nu'01 & 40 & 48. & 0.2 &   & 3 &   & 3 &   &   &   & 3 &   & 3 & \text{C} & \textbf{1} \tabularnewline
18 & \textbf{LHCb8WZ'16} & 42 & 41.4 & 0.164 &   & 3 &   & 3 &   & 3 &   & 3 &   & 2 &   &   \tabularnewline
19 & ATL7WZ'12 & 41 & 39.6 & 0.161 &   & 3 &   & 3 &   &   &   & 3 &   & 3 &   & 3 \tabularnewline
20 & \textbf{CMS8Wasy'16} & 33 & 39.2 & 0.198 &   & 2 &   & 3 &   &   &   & 3 &   & 2 &   & 3 \tabularnewline
21 & D02jets'08 & 110 & 37.5 & 0.0568 &   &   &   &   &   & 3 &   &   &   &   &   &   \tabularnewline
22 & NuTeV-nub'06 & 33 & 36.7 & 0.185 &   & 3 &   & 3 &   &   &   & 3 &   & 3 &   & 2 \tabularnewline
23 & \textbf{ATL8DY2D'16} & 48 & 34.7 & 0.121 &   & 3 &   & 3 &   &   &   & 3 &   &   &   & 3 \tabularnewline
24 & E866rat'01 & 15 & 33.3 & 0.37 &   & \textbf{1} &   & \textbf{1} &   &   &   & 3 &   & 2 &   &   \tabularnewline
25 & ATL7jets'12 & 90 & 30.4 & 0.0563 &   &   &   &   &   & 3 &   &   &   &   &   &   \tabularnewline
26 & \textbf{LHCb7ZWrap'15} & 33 & 30.2 & 0.152 &   & 3 &   & 3 &   & 3 &   & 3 &   & 3 &   &   \tabularnewline
27 & CMS7Masy2'14 & 11 & 29.4 & 0.446 &   & \textbf{1} &   & 2 &   & 2 &   & 2 &   & \textbf{1} &   & 3 \tabularnewline
28 & CDF2jets'09 & 72 & 21.5 & 0.0497 &   &   &   &   &   & 3 &   &   &   &   &   &   \tabularnewline
29 & \textbf{ATL8ZpT'16} & 45 & 17.2 & 0.0638 &   &   &   &   &   & 3 &   &   &   &   &   & 3 \tabularnewline
30 & HERAc'13 & 47 & 15.1 & 0.0537 &   &   &   &   &   & 3 &   &   &   &   &   &   \tabularnewline
31 & D02Masy'08 & 9 & 15. & 0.278 &   & 3 &   & 3 &   &   &   & 2 &   & 2 &   & 2 \tabularnewline
32 & CMS7Easy'12 & 11 & 14.3 & 0.216 &   & 2 &   & 3 &   & 3 &   & 3 &   & 2 &   &   \tabularnewline
33 & D02Easy2'15 & 13 & 14. & 0.18 &   & 3 &   & 3 &   &   &   & 3 &   & 2 &   &   \tabularnewline
34 & ZyD02'08 & 28 & 11.6 & 0.0693 &   &   &   &   &   &   &   & 3 &   & 3 &   &   \tabularnewline
35 & ZyCDF2'10 & 29 & 11.2 & 0.0647 &   &   &   &   &   &   &   & 3 &   &   &   &   \tabularnewline
36 & CDF1Wasy'96 & 11 & 8.83 & 0.134 &   & 3 &   & 3 &   &   &   & 3 &   & 2 &   &   \tabularnewline
37 & LHCb7WZ'12 & 14 & 7.27 & 0.0866 &   & 3 &   &   &   &   &   &   &   & 3 &   &   \tabularnewline
38 & \textbf{LHCb8Zee'15} & 17 & 7.1 & 0.0696 &   &   &   &   &   &   &   & 3 &   &   &   &   \tabularnewline
39 & \textbf{ATL8ttb-pt'16} & 8 & 6.2 & 0.129 &   & 3 &   & 3 &   & 2 &   &   &   &   &   &   \tabularnewline
40 & LHCb7Wasy'12 & 5 & 6.11 & 0.204 &   & 2 &   & 3 &   &   &   & 3 &   & 2 &   & 3 \tabularnewline
41 & \textbf{ATL7ZpT'14} & 8 & 5.84 & 0.122 &   & 3 &   & 3 &   & 3 &   & 3 &   & 3 &   &   \tabularnewline
42 & HERA-FL'11 & 9 & 3.99 & 0.0739 &   &   &   &   &   & 2 &   &   &   &   &   &   \tabularnewline
43 & \textbf{ATL8ttb-mtt'16} & 7 & 3.81 & 0.0907 &   &   &   &   &   & 2 &   &   &   &   &   &   \tabularnewline
44 & CDF2Wasy'05 & 11 & 3.7 & 0.056 &   &   &   &   &   &   &   &   &   & 3 &   &   \tabularnewline
45 & \textbf{ATL8ttb-y\_ttb'16} & 5 & 3.37 & 0.112 &   &   &   &   &   & 2 &   &   &   &   &   &   \tabularnewline
46 & \textbf{ATL8ttb-y\_ave'16} & 5 & 3.2 & 0.107 &   &   &   &   &   & 2 &   &   &   &   &   &   \tabularnewline
47 & HERAb'06 & 10 & 1.14 & 0.0191 &   &   &   &   &   &   &   &   &   &   &   &   \tabularnewline
\hline 
\end{tabular}
\caption{For each experiment $E$ we have defined its flavor-specific sensitivity $|S^E_{f}|$ and its
point-averaged counterpart $\langle|S^E_{f}|\rangle$ in Sec.~\ref{sec:Experiment-rankings-according}.
Using these quantities, we tabulate the total overall (\textit{i.e.},
flavor-summed) sensitivity and a flavor-dependent sensitivity for
the various experiments in our dataset, ordering the table in descending
magnitude for the overall sensitivity. Thus, row $1$ for the
combined HERA Run I $+$ Run 2 dataset has the greatest overall sensitivity,
while row $47$ for the H1 $\sigma_{r}^{b}$ reduced cross section
has the least overall sensitivity according to that metric. For each
flavor, we award particularly sensitive experiments a rank $\mathrm{{\bf A},B,C}$
or ${\bf 1*},{\bf 1},2,3$ based on their total and point-averaged
sensitivities, respectively. These ranks are decided using the criteria:
$C\protect\iff|S^E_{f}|\in[20,50]$, $B\protect\iff|S^E_{f}|\in[50,100]$,
and ${\bf A}\protect\iff|S^E_{f}|>100$ according to the total sensitivities
for each flavor; and, analogously, $3\protect\iff\langle|S^E_{f}|\rangle\in[0.1,0.25]$,
$2\protect\iff\langle|S^E_{f}|\rangle\in[0.25,0.5]$, ${\bf 1}\protect\iff\langle|S^E_{f}|\rangle\in[0.5,1]$,
and ${\bf 1*}\protect\iff\langle|S^E_{f}|\rangle>1$ according to the
point-averaged sensitivities. Experiments with sensitivities
falling below the lowest ranks (that is, with $|S^E_{f}|<20$ or $\langle|S^E_{f}|\rangle<0.1$)
are not awarded a rank for that category/flavor. Note that we sum over
the light quark $+$ gluon flavors to compute $\langle\sum_{f}|S^E_{f}|\rangle$
within this and subsequent tables. Also, new experimental datasets not originally
included in CT14HERA2 are indicated by \textbf{bold} Expt.~names in the
second column.}
\label{tab5} 
\end{table}

\begin{table}
\hspace*{-1.0cm}\begin{tabular}{c|c|cc|cc|cc|cc|cc|cc|cc}
\hline 
 &  & \multicolumn{14}{c}{Rankings, {\bf CT14 HERA2 NNLO PDFs}}\tabularnewline
No.  & Expt.  & $|S^E_{u_{v}}|$  & $\langle|S^E_{u_{v}}|\rangle$  & $|S^E_{d_{v}}|$  & $\langle|S^E_{d_{v}}|\rangle$  & $|S^E_{\bar{d}/\bar{u}}|$  & $\langle|S^E_{\bar{d}/\bar{u}}|\rangle$  & $|S^E_{d/u}|$  & $\langle|S^E_{d/u}|\rangle$  & $|S^E_{H7}|$  & $\langle|S^E_{H7}|\rangle$  & $|S^E_{H8}|$  & $\langle|S^E_{H8}|\rangle$  & $|S^E_{H14}|$  & $\langle|S^E_{H14}|\rangle$ \tabularnewline
\hline 
 &  &  &  &  &  &  &  &  &  &  &  &  &  &  & \tabularnewline
1 & HERAI+II'15  & \text{B} &   & \text{C} &   & \text{C} &   & \text{B} &   & \text{B} &   & \text{B} &   & \text{B} &   \tabularnewline
2 & CCFR-F3'97  & \text{B} & \textbf{1} & \text{B} & \textbf{1} &   &   & \text{C} & 2 &   & 3 &   & 3 &   & 3 \tabularnewline
3 & BCDMSp'89  & \text{B} & 3 & \text{C} &   & \text{C} &   & \text{C} & 3 & \text{C} &   &   &   &   &   \tabularnewline
4 & NMCrat'97  & \text{C} & 2 & \text{C} & 3 & \text{C} & 2 & \text{B} & \textbf{1} &   &   &   &   &   &   \tabularnewline
5 & BCDMSd'90  & \text{C} &   & \text{C} & 3 &   &   & \text{C} &   & \text{C} &   & \text{C} &   &   &   \tabularnewline
6 & CDHSW-F3'91  & \text{C} & 2 & \text{C} & 2 &   &   &   & 3 &   &   &   &   &   &   \tabularnewline
7 & E605'91  & \text{C} & 3 & \text{C} & 3 &   &   &   &   &   &   &   &   &   &   \tabularnewline
8 & E866pp'03  & \text{C} & 3 &   &   &   &   &   &   &   &   &   &   &   &   \tabularnewline
9 & CCFR-F2'01  &   & 3 &   & 3 &   & 3 &   & 3 &   & 3 &   & 3 &   & 3 \tabularnewline
10 & \textbf{CMS8jets'17}  &   &   &   &   &   &   &   &   &   & 3 & \text{C} & 3 & \text{C} & 3 \tabularnewline
11 & CDHSW-F2'91  &   & 3 &   & 3 &   &   &   & 3 &   & 3 &   & 3 &   &   \tabularnewline
12 & CMS7jets'13  &   &   &   &   &   &   &   &   &   & 3 &   & 3 &   & 3 \tabularnewline
13 & NuTeV-nu'06  &   &   &   &   &   &   &   &   &   &   &   &   &   &   \tabularnewline
14 & \textbf{CMS7jets'14}  &   &   &   &   &   &   &   &   &   & 3 &   & 3 &   & 3 \tabularnewline
15 & CCFR SI nub'01  &   &   &   &   &   &   &   &   &   &   &   &   &   &   \tabularnewline
16 & \textbf{ATLAS7jets'15}  &   &   &   &   &   &   &   &   &   &   &   &   &   &   \tabularnewline
17 & CCFR SI nu'01  &   &   &   &   &   &   &   &   &   &   &   &   &   &   \tabularnewline
18 & \textbf{LHCb8WZ'16}  &   & 3 &   & 3 &   & 2 &   & 2 &   & 3 &   & 3 &   &   \tabularnewline
19 & ATL7WZ'12  &   & 3 &   &   &   & 3 &   & 3 &   &   &   &   &   &   \tabularnewline
20 & \textbf{CMS8Wasy'16}  &   & 3 &   & 3 &   & 2 &   & 2 &   &   &   &   &   &   \tabularnewline
21 & D02jets'08  &   &   &   &   &   &   &   &   &   &   &   &   &   &   \tabularnewline
22 & NuTeV-nub'06  &   &   &   &   &   &   &   &   &   &   &   &   &   &   \tabularnewline
23 & \textbf{ATL8DY2D'16}  &   & 3 &   &   &   & 3 &   & 3 &   &   &   &   &   &   \tabularnewline
24 & E866rat'01  &   & 2 &   & 2 & \text{C} & \textbf{1*} &   & 2 &   & 3 &   & 3 &   &   \tabularnewline
25 & ATL7jets'12  &   &   &   &   &   &   &   &   &   & 3 &   & 3 &   & 3 \tabularnewline
26 & \textbf{LHCb7ZWrap'15}  &   & 3 &   & 3 &   & 2 &   & 2 &   & 3 &   & 3 &   &   \tabularnewline
27 & CMS7Masy2'14  &   & 2 &   & 2 &   & \textbf{1} &   & \textbf{1} &   & 3 &   & 3 &   & 3 \tabularnewline
28 & CDF2jets'09  &   &   &   &   &   &   &   &   &   &   &   &   &   &   \tabularnewline
29 & \textbf{ATL8ZpT'16}  &   &   &   &   &   &   &   &   &   &   &   &   &   & 3 \tabularnewline
30 & HERAc'13  &   &   &   &   &   &   &   &   &   & 3 &   & 3 &   & 3 \tabularnewline
31 & D02Masy'08  &   & 2 &   & 2 &   & 2 &   & 2 &   &   &   &   &   & 3 \tabularnewline
32 & CMS7Easy'12  &   & 3 &   & 3 &   & 2 &   & 2 &   &   &   &   &   &   \tabularnewline
33 & D02Easy2'15  &   & 3 &   & 2 &   & 3 &   & 2 &   &   &   &   &   &   \tabularnewline
34 & ZyD02'08  &   & 3 &   &   &   &   &   &   &   &   &   &   &   &   \tabularnewline
35 & ZyCDF2'10  &   & 3 &   &   &   &   &   &   &   &   &   &   &   &   \tabularnewline
36 & CDF1Wasy'96  &   & 3 &   & 2 &   & 3 &   & 2 &   &   &   &   &   &   \tabularnewline
37 & LHCb7WZ'12  &   &   &   &   &   & 3 &   & 3 &   &   &   &   &   &   \tabularnewline
38 & \textbf{LHCb8Zee'15}  &   &   &   &   &   &   &   &   &   &   &   &   &   &   \tabularnewline
39 & \textbf{ATL8ttb-pt'16}  &   & 3 &   &   &   &   &   &   &   & 2 &   & 2 &   & 2 \tabularnewline
40 & LHCb7Wasy'12  &   & 3 &   & 3 &   & 2 &   & 2 &   & 3 &   & 3 &   & 3 \tabularnewline
41 & \textbf{ATL7ZpT'14}  &   &   &   &   &   &   &   & 3 &   & 3 &   & 3 &   & 3 \tabularnewline
42 & HERA-FL'11  &   &   &   &   &   &   &   &   &   & 3 &   & 3 &   &   \tabularnewline
43 & \textbf{ATL8ttb-mtt'16}  &   &   &   &   &   &   &   &   &   & 3 &   & 3 &   & 3 \tabularnewline
44 & CDF2Wasy'05  &   &   &   & 3 &   &   &   & 3 &   &   &   &   &   &   \tabularnewline
45 & \textbf{ATL8ttb-y\_ttb'16}  &   &   &   &   &   &   &   &   &   & 2 &   & 2 &   & 3 \tabularnewline
46 & \textbf{ATL8ttb-y\_ave'16}  &   &   &   &   &   &   &   &   &   & 2 &   & 2 &   & 3 \tabularnewline
47 & HERAb'06  &   &   &   &   &   &   &   &   &   &   &   &   &   &   \tabularnewline
\hline 
\end{tabular}\caption{ A horizontal continuation of the information in Table~\ref{tab5},
containing the flavor-dependent total and mean sensitivities
of a number of derived quantities, as opposed to the individual flavors
given in Table~\ref{tab5}. Going across, the total
and mean sensitivities are tabulated for valence distributions of
the $u$ and $d$ quarks, the partonic flavor ratios $\bar{d}/\bar{u}$
and $d/u$, and the Higgs production cross section $\sigma_{pp\to H^{0}X}$
at $7$, $8$, and $14$ TeV, respectively. The ranking criteria,
ordering, and other conventions are again as described in Table~\ref{tab5}. }
\label{tab6} 
\end{table}

\section{Supplementary Material \label{sec:SM}}
As Supplementary Material, we enclose in this Appendix a series of additional tables
that further illustrate the details of our sensitivity analysis. These include
a detailed breakdown of the various CTEQ-TEA experiments according to physical
process (Table~\ref{tab:IDsOfProcessesType}) and associated sensitivity rankings, both for individual PDF
flavors (Table~\ref{tab7}) and for various derived quantities (Table~\ref{tab8}).
In addition, in Tables~\ref{tab:tab5vals} and \ref{tab:tab6vals}, we give numerical values of sensitivities
corresponding to the rankings shown in Tables~\ref{tab5} and \ref{tab6}. In
Tables~\ref{tab:tab7vals}  and \ref{tab:tab8vals}, numerical values of sensitivities corresponding to 
Tables~\ref{tab7} and \ref{tab8} are also given. Lastly, in Tables~\ref{tab5nojet} and \ref{tab6nojet},
sensitivity ranking tables of the CTEQ-TEA dataset based upon a companion fit
that excluded jet data are given, and corresponding numerical values are shown
in Tables~\ref{tab:tab5nojetvals} and \ref{tab:tab6nojetvals}.
\begin{table}
\begin{tabular}{|c|p{15cm}|}
\hline 
Process  & Experiment Names\tabularnewline
\hline 
\hline 
\text{DIS Old}  & BCDMSp'89, BCDMSd'90, NMCrat'97, CDHSW-F2'91, CDHSW-F3'91, CCFR-F2'01, CCFR-F3'97, NuTeV-nu'06, NuTeV-nub'06, CCFR SI nu'01, CCFR SI nub'01, HERAb'06, HERAc'13, HERAI+II'15,
169 \tabularnewline
\hline 
\text{DISCC Old}  & CDHSW-F2'91, CDHSW-F3'91, CCFR-F2'01, CCFR-F3'97, NuTeV-nu'06, NuTeV-nub'06, CCFR SI nu'01, CCFR SI nub'01 \tabularnewline
\hline 
	\text{JP New}  & \textbf{CMS7jets'14}, \textbf{ATLAS7jets'15}, \textbf{CMS8jets'17} \tabularnewline
\hline 
\text{DISNCCC}  & HERAI+II'15 \tabularnewline
\hline 
\text{VBPZ Old}  & E605'91, E866rat'01, E866pp'03, ZyD02'08, ZyCDF2'10 \tabularnewline
\hline 
\text{DISNC Old}  & BCDMSp'89, BCDMSd'90, NMCrat'97, HERAb'06, HERAc'13, HERA-FL'11 \tabularnewline
\hline 
\text{JP Old}  & CDF2jets'09, D02jets'08, ATL7jets'12, CMS7jets'13 \tabularnewline
\hline 
\text{VBPW Old}  & CDF1Wasy'96, CDF2Wasy'05, D02Masy'08, LHCb7Wasy'12, CMS7Masy2'14, CMS7Easy'12, D02Easy2'15 \tabularnewline
\hline 
	\text{VBPWZ New}  & \textbf{LHCb7ZWrap'15}, \textbf{LHCb8WZ'16} \tabularnewline
\hline 
	\text{VBPZ New}  & \textbf{LHCb8Zee'15}, \textbf{ATL8DY2D'16} \tabularnewline
\hline 
\text{VBPWZ Old}  & LHCb7WZ'12, ATL7WZ'12 \tabularnewline
\hline 
	\text{VBPW New}  & \textbf{CMS8Wasy'16} \tabularnewline
\hline 
	\text{VBPZpT}  & \textbf{ATL7ZpT'14}, \textbf{ATL8ZpT'16} \tabularnewline
\hline 
	$t\overline{t}$  & \textbf{ATL8ttb-pt'16}, \textbf{ATL8ttb-y\_ave'16}, \textbf{ATL8ttb-mtt'16}, \textbf{ATL8ttb-y\_ttb'16} \tabularnewline
\hline 
\end{tabular}\caption{The experimental IDs of the datasets making up the process types considered in this analysis;
we identify these various processes by abbreviated labels: charge
current DIS (DISCC), neutral current DIS (DISNC), NC/CC DIS (DISNCCC), and all DIS; Vector
Boson Production (VBP) of the $W$ (VBPW), $Z$ (VBPZ), and W/Z processes (VBPWZ); $p_{T}^{W/Z}$ of $Z$ (VBPZpT);
jet production (JP) and $t\overline{t}$. ``Old'' sets were in CT14HERA2, but the ``New''
only in CTEQ-TEA.
\label{tab:IDsOfProcessesType}}
\end{table}

\begin{table}
\begin{tabular}{c|c|c|cc|cc|cc|cc|cc|cc|cc}
\hline 
 &  &  & \multicolumn{14}{c}{Rankings, {\bf CT14 HERA2 NNLO PDFs}}\tabularnewline
No.  & Process  & $N_{\mathit{pt}}$  & $\sum_{f}|S^E_{f}|$  & $\langle\sum_{f}|S^E_{f}|\rangle$  & $|S^E_{\bar{d}}|$  & $\langle|S^E_{\bar{d}}|\rangle$  & $|S^E_{\bar{u}}|$  & $\langle|S^E_{\bar{u}}|\rangle$  & $|S^E_{g}|$  & $\langle|S^E_{g}|\rangle$  & $|S^E_{u}|$  & $\langle|S^E_{u}|\rangle$  & $|S^E_{d}|$  & $\langle|S^E_{d}|\rangle$  & $|S^E_{s}|$  & $\langle|S^E_{s}|\rangle$ \tabularnewline
\hline 
 &  &  &  &  &  &  &  &  &  &  &  &  &  &  &  & \tabularnewline
1 & \text{DIS Old} & 2381. & \text{1.83E3} & 0.128 & \textbf{A} & 3 & \textbf{A} & 3 & \textbf{A} & 3 & \textbf{A} & 3 & \textbf{A} & 3 & \textbf{A} &   \tabularnewline
2 & \text{DISCC Old} & 485 & 698. & 0.24 & \textbf{A} & 2 & \textbf{A} & 3 & \text{B} & 3 & \textbf{A} & 2 & \textbf{A} & 2 & \textbf{A} & 3 \tabularnewline
3 & \text{DISNCCC} & 1120. & 620. & 0.0922 & \text{B} &   & \textbf{A} & 3 & \textbf{A} & 3 & \textbf{A} & 3 & \text{B} &   & \text{C} &   \tabularnewline
4 & \text{DISNC Old} & 776 & 514. & 0.11 & \text{B} & 3 & \text{B} &   & \text{B} &   & \textbf{A} & 3 & \textbf{A} & 3 & \text{C} &   \tabularnewline
5 & \text{VBPZ Old} & 375 & 272. & 0.121 & \text{B} & 3 & \text{B} & 3 & \text{C} &   & \text{C} & 3 & \text{C} &   &   &   \tabularnewline
\end{tabular}\caption{Similar to Table~\ref{tab5}, yet we tabulate the various types of
processes rather than the separate experiments in our global CTEQ-TEA dataset.
The process labels are explained in the caption of Table~\ref{tab:IDsOfProcessesType},
which summarized the constituent experiments contributing to each process type.
\label{tab7} }
\end{table}

\begin{table}
\begin{tabular}{c|c|cc|cc|cc|cc|cc|cc|cc}
\hline 
 &  & \multicolumn{14}{c}{Rankings, {\bf CT14 HERA2 NNLO PDFs}}\tabularnewline
No.  & Process  & $|S^E_{u_{v}}|$  & $\langle|S^E_{u_{v}}|\rangle$  & $|S^E_{d_{v}}|$  & $\langle|S^E_{d_{v}}|\rangle$  & $|S^E_{\bar{d}/\bar{u}}|$  & $\langle|S^E_{\bar{d}/\bar{u}}|\rangle$  & $|S^E_{d/u}|$  & $\langle|S^E_{d/u}|\rangle$  & $|S^E_{H7}|$  & $\langle|S^E_{H7}|\rangle$  & $|S^E_{H8}|$  & $\langle|S^E_{H8}|\rangle$  & $|S^E_{H14}|$  & $\langle|S^E_{H14}|\rangle$ \tabularnewline
\hline 
 &  &  &  &  &  &  &  &  &  &  &  &  &  &  & \tabularnewline
1 & \text{DIS Old}  & \textbf{A} & 3 & \textbf{A} & 3 & \textbf{A} &   & \textbf{A} & 3 & \textbf{A} &   & \textbf{A} &   & \textbf{A} &   \tabularnewline
2 & \text{DISCC Old}  & \textbf{A} & 2 & \textbf{A} & 2 & \text{C} &   & \text{B} & 3 & \text{B} & 3 & \text{B} & 3 & \text{C} &   \tabularnewline
3 & \text{DISNCCC}  & \text{B} &   & \text{C} &   & \text{C} &   & \text{B} &   & \text{B} &   & \text{B} &   & \text{B} &   \tabularnewline
4 & \text{DISNC Old}  & \textbf{A} & 3 & \text{B} &   & \text{B} & 3 & \textbf{A} & 3 & \text{B} &   & \text{B} &   & \text{C} &   \tabularnewline
5 & \text{VBPZ Old}  & \text{B} & 3 & \text{B} & 3 & \text{B} & 3 & \text{C} &   &   &   &   &   &   &   \tabularnewline
\end{tabular}\caption{Continuation of Table \ref{tab7}, listing the PDF combinations and Higgs production cross sections similarly to Table~\ref{tab6} of the main paper.
\label{tab8}}
\end{table}

\begin{table}
\hspace*{-1cm}\begin{tabular}{c|c|c|cc|cc|cc|cc|cc|cc|cc}
\hline 
 &  &  & \multicolumn{14}{c}{Values, {\bf CT14 HERA2 NNLO PDFs}}\tabularnewline
No.  & Expt. & $N_{\mathit{pt}}$  & $\sum_{f}|S^E_{f}|$  & $\langle\sum_{f}|S^E_{f}|\rangle$  & $|S^E_{\bar{d}}|$  & $\langle|S^E_{\bar{d}}|\rangle$  & $|S^E_{\bar{u}}|$  & $\langle|S^E_{\bar{u}}|\rangle$  & $|S^E_{g}|$  & $\langle|S^E_{g}|\rangle$  & $|S^E_{u}|$  & $\langle|S^E_{u}|\rangle$  & $|S^E_{d}|$  & $\langle|S^E_{d}|\rangle$  & $|S^E_{s}|$  & $\langle|S^E_{s}|\rangle$ \tabularnewline
\hline 
 &  &  &  &  &  &  &  &  &  &  &  &  &  &  &  & \tabularnewline
1 & HERAI+II'15 & 1120. & 620. & 0.0922 & 78.3 & 0.0699 & 124. & 0.111 & 143. & 0.128 & 146. & 0.13 & 83. & 0.0741 & 45. & 0.0402 \tabularnewline
2 & CCFR-F3'97 & 86 & 218. & 0.423 & 46.2 & 0.537 & 49.8 & 0.579 & 16.8 & 0.195 & 63.7 & 0.741 & 34.7 & 0.404 & 6.85 & 0.0797 \tabularnewline
3 & BCDMSp'89 & 337 & 184. & 0.0908 & 16.6 & 0.0492 & 31.1 & 0.0922 & 25.8 & 0.0765 & 81.2 & 0.241 & 20.3 & 0.0601 & 8.77 & 0.026 \tabularnewline
4 & NMCrat'97 & 123 & 169. & 0.229 & 44.4 & 0.361 & 9.87 & 0.0802 & 9.7 & 0.0789 & 36.4 & 0.296 & 61.4 & 0.499 & 7.44 & 0.0605 \tabularnewline
5 & BCDMSd'90 & 250 & 141. & 0.0939 & 24.3 & 0.0971 & 17.7 & 0.0707 & 26.5 & 0.106 & 26.9 & 0.108 & 36.2 & 0.145 & 9.19 & 0.0368 \tabularnewline
\hline 
\end{tabular}\caption{Here we separately collect the numerical values of the total and point-averaged
	sensitivities used to determine the rankings in Table~\ref{tab5} of the main paper.
\label{tab:tab5vals}}
\end{table}

\begin{table}
\hspace*{-1cm}\begin{tabular}{c|c|cc|cc|cc|cc|cc|cc|cc}
\hline 
 &  & \multicolumn{14}{c}{Values, {\bf CT14 HERA2 NNLO PDFs}}\tabularnewline
No.  & Expt.  & $|S^E_{u_{v}}|$  & $\langle|S^E_{u_{v}}|\rangle$  & $|S^E_{d_{v}}|$  & $\langle|S^E_{d_{v}}|\rangle$  & $|S^E_{\bar{d}/\bar{u}}|$  & $\langle|S^E_{\bar{d}/\bar{u}}|\rangle$  & $|S^E_{d/u}|$  & $\langle|S^E_{d/u}|\rangle$  & $|S^E_{H7}|$  & $\langle|S^E_{H7}|\rangle$  & $|S^E_{H8}|$  & $\langle|S^E_{H8}|\rangle$  & $|S^E_{H14}|$  & $\langle|S^E_{H14}|\rangle$ \tabularnewline
\hline 
 &  &  &  &  &  &  &  &  &  &  &  &  &  &  & \tabularnewline
1 & HERAI+II'15  & 64.6 & 0.0577 & 36.7 & 0.0328 & 39.7 & 0.0354 & 63.7 & 0.0569 & 76.2 & 0.0681 & 74.3 & 0.0663 & 78.6 & 0.0701 \tabularnewline
2 & CCFR-F3'97  & 77.5 & 0.901 & 65.4 & 0.761 & 5.87 & 0.0682 & 25.5 & 0.297 & 11.3 & 0.131 & 11.8 & 0.137 & 13.3 & 0.154 \tabularnewline
3 & BCDMSp'89  & 70.8 & 0.21 & 21.3 & 0.0632 & 20.2 & 0.0598 & 40.4 & 0.12 & 21.4 & 0.0634 & 19.3 & 0.0573 & 11.7 & 0.0349 \tabularnewline
4 & NMCrat'97  & 34.6 & 0.281 & 25.6 & 0.208 & 48.7 & 0.396 & 76.6 & 0.623 & 9.47 & 0.077 & 9.94 & 0.0808 & 10.5 & 0.0852 \tabularnewline
5 & BCDMSd'90  & 20.3 & 0.0813 & 27. & 0.108 & 12.6 & 0.0503 & 22.5 & 0.09 & 23.9 & 0.0958 & 23.1 & 0.0922 & 18.1 & 0.0724 \tabularnewline
\hline 
\end{tabular}\caption{Sensitivity values for Table~\ref{tab6} of the main paper.
\label{tab:tab6vals}}
\end{table}

\begin{table}
{\small{}{}{}}%
\hspace*{-1.5cm}\begin{tabular}{c|c|c|cc|cc|cc|cc|cc|cc|cc}
\hline 
 &  &  & \multicolumn{14}{c}{Values, {\bf CT14 HERA2 NNLO PDFs}}\tabularnewline
No.  & Process & $N_{\mathit{pt}}$  & $\sum_{f}|S^E_{f}|$  & $\langle\sum_{f}|S^E_{f}|\rangle$  & $|S^E_{\bar{d}}|$  & $\langle|S^E_{\bar{d}}|\rangle$  & $|S^E_{\bar{u}}|$  & $\langle|S^E_{\bar{u}}|\rangle$  & $|S^E_{g}|$  & $\langle|S^E_{g}|\rangle$  & $|S^E_{u}|$  & $\langle|S^E_{u}|\rangle$  & $|S^E_{d}|$  & $\langle|S^E_{d}|\rangle$  & $|S^E_{s}|$  & $\langle|S^E_{s}|\rangle$ \tabularnewline
\hline 
 &  &  &  &  &  &  &  &  &  &  &  &  &  &  &  & \tabularnewline
1 & \text{DIS Old} & 2381. & \text{1.83E3} & 0.128 & 287. & 0.121 & 304. & 0.128 & 292. & 0.122 & 441. & 0.185 & 326. & 0.137 & 181. & 0.0759 \tabularnewline
2 & \text{DISCC Old} & 485 & 698. & 0.24 & 122. & 0.251 & 119. & 0.245 & 77. & 0.159 & 148. & 0.306 & 123. & 0.254 & 109. & 0.224 \tabularnewline
3 & \text{DISNCCC} & 1120. & 620. & 0.0922 & 78.3 & 0.0699 & 124. & 0.111 & 143. & 0.128 & 146. & 0.13 & 83. & 0.0741 & 45. & 0.0402 \tabularnewline
4 & \text{DISNC Old} & 776 & 514. & 0.11 & 87.5 & 0.113 & 60.8 & 0.0783 & 71.3 & 0.0919 & 147. & 0.189 & 120. & 0.155 & 27.2 & 0.0351 \tabularnewline
5 & \text{VBPZ Old} & 375 & 272. & 0.121 & 76.6 & 0.204 & 82.2 & 0.219 & 21.7 & 0.0579 & 46.7 & 0.125 & 27. & 0.0721 & 17.6 & 0.047 \tabularnewline
\end{tabular}\caption{Sensitivity values for Table \ref{tab7}. \label{tab:tab7vals}}
\end{table}

\begin{table}
\hspace*{-1.5cm}\begin{tabular}{c|c|cc|cc|cc|cc|cc|cc|cc}
\hline
 &  & \multicolumn{14}{c}{Values, {\bf CT14 HERA2 NNLO PDFs}}\tabularnewline
No.  & Process & $|S^E_{u_{v}}|$  & $\langle|S^E_{u_{v}}|\rangle$  & $|S^E_{d_{v}}|$  & $\langle|S^E_{d_{v}}|\rangle$  & $|S^E_{\bar{d}/\bar{u}}|$  & $\langle|S^E_{\bar{d}/\bar{u}}|\rangle$  & $|S^E_{d/u}|$  & $\langle|S^E_{d/u}|\rangle$  & $|S^E_{H7}|$  & $\langle|S^E_{H7}|\rangle$  & $|S^E_{H8}|$  & $\langle|S^E_{H8}|\rangle$  & $|S^E_{H14}|$  & $\langle|S^E_{H14}|\rangle$ \tabularnewline
\hline 
 &  &  &  &  &  &  &  &  &  &  &  &  &  &  & \tabularnewline
1 & \text{DIS Old}  & 333. & 0.14 & 240. & 0.101 & 151. & 0.0632 & 279. & 0.117 & 194. & 0.0814 & 186. & 0.0782 & 169. & 0.0711 \tabularnewline
2 & \text{DISCC Old}  & 142. & 0.292 & 128. & 0.264 & 28.5 & 0.0588 & 74.9 & 0.154 & 55.6 & 0.115 & 52.7 & 0.109 & 44.7 & 0.0922 \tabularnewline
3 & \text{DISNCCC}  & 64.6 & 0.0577 & 36.7 & 0.0328 & 39.7 & 0.0354 & 63.7 & 0.0569 & 76.2 & 0.0681 & 74.3 & 0.0663 & 78.6 & 0.0701 \tabularnewline
4 & \text{DISNC Old}  & 127. & 0.163 & 74.6 & 0.0961 & 82.4 & 0.106 & 141. & 0.181 & 61.9 & 0.0798 & 59.2 & 0.0763 & 45.9 & 0.0592 \tabularnewline
5 & \text{VBPZ Old}  & 64.7 & 0.173 & 50.8 & 0.135 & 57.1 & 0.152 & 24.9 & 0.0664 & 15.6 & 0.0417 & 16.5 & 0.0439 & 18.4 & 0.0491 \tabularnewline
\end{tabular}\caption{Sensitivity values for Table \ref{tab8}. \label{tab:tab8vals}}
\end{table}

\begin{table}
\hspace*{-1cm}\begin{tabular}{c|c|c|cc|cc|cc|cc|cc|cc|cc}
\hline 
 &  &  & \multicolumn{14}{c}{Rankings, {\bf CT14 HERA2 NNLO PDFs fitted without jet data}}\tabularnewline
No.  & Expt. & $N_{\mathit{pt}}$  & $\sum_{f}|S^E_{f}|$  & $\langle\sum_{f}|S^E_{f}|\rangle$  & $|S^E_{\bar{d}}|$  & $\langle|S^E_{\bar{d}}|\rangle$  & $|S^E_{\bar{u}}|$  & $\langle|S^E_{\bar{u}}|\rangle$  & $|S^E_{g}|$  & $\langle|S^E_{g}|\rangle$  & $|S^E_{u}|$  & $\langle|S^E_{u}|\rangle$  & $|S^E_{d}|$  & $\langle|S^E_{d}|\rangle$  & $|S^E_{s}|$  & $\langle|S^E_{s}|\rangle$ \tabularnewline
\hline 
 &  &  &  &  &  &  &  &  &  &  &  &  &  &  &  & \tabularnewline
1 & HERAI+II'15 & 1120. & 599. & 0.0892 & \text{B} &   & \textbf{A} & 3 & \textbf{A} & 3 & \textbf{A} & 3 & \text{B} &   & \text{B} &   \tabularnewline
2 & BCDMSp'89 & 337 & 230. & 0.114 & \text{C} &   & \text{C} & 3 & \text{C} & 3 & \text{B} & 3 & \text{C} &   & \text{C} &   \tabularnewline
3 & CCFR-F3'97 & 86 & 223. & 0.433 & \text{B} & \textbf{1} & \text{B} & \textbf{1} &   & 3 & \text{B} & \textbf{1} & \text{C} & 2 &   & 3 \tabularnewline
4 & \textbf{CMS8jets'17} & 185 & 205. & 0.185 & \text{C} & 3 & \text{C} & 3 & \text{B} & 2 &   &   &   &   & \text{C} & 3 \tabularnewline
5 & NMCrat'97 & 123 & 168. & 0.228 & \text{C} & 2 &   & 3 &   &   & \text{C} & 2 & \text{B} & 2 &   &   \tabularnewline
\hline 
\end{tabular}
	\caption{Experiment rankings as in Table~\ref{tab5} of the main paper, for the PDFs obtained using the CT14HERA2 NNLO data sets (Tables II, III of the main paper), 
                 with the exclusion of Tevatron and LHC jet production experiments. The {\bf bold} font indicates those experiments which were {\it not} fitted as constraints
                 leading to the PDF parametrization {\it CT14 HERA2 NNLO PDFs fitted without jet data} used to evaluate the sensitivities in this table. 
}
\label{tab5nojet}
\end{table}

\begin{table}
\hspace*{-1cm}
\begin{tabular}{c|c|cc|cc|cc|cc|cc|cc|cc}
\hline 
 &  & \multicolumn{14}{c}{Rankings, {\bf CT14 HERA2 NNLO PDFs fitted without jet data}}\tabularnewline
No.  & Expt. & $|S^E_{u_{v}}|$  & $\langle|S^E_{u_{v}}|\rangle$  & $|S^E_{d_{v}}|$  & $\langle|S^E_{d_{v}}|\rangle$  & $|S^E_{\bar{d}/\bar{u}}|$  & $\langle|S^E_{\bar{d}/\bar{u}}|\rangle$  & $|S^E_{d/u}|$  & $\langle|S^E_{d/u}|\rangle$  & $|S^E_{H7}|$  & $\langle|S^E_{H7}|\rangle$  & $|S^E_{H8}|$  & $\langle|S^E_{H8}|\rangle$  & $|S^E_{H14}|$  & $\langle|S^E_{H14}|\rangle$ \tabularnewline
\hline 
 &  &  &  &  &  &  &  &  &  &  &  &  &  &  & \tabularnewline
1 & HERAI+II'15  & \text{B} &   & \text{C} &   & \text{C} &   & \text{B} &   & \text{B} &   & \text{B} &   & \textbf{A} &   \tabularnewline
2 & BCDMSp'89  & \text{B} & 3 & \text{C} &   & \text{C} &   & \text{C} & 3 & \text{C} &   & \text{C} &   & \text{C} &   \tabularnewline
3 & CCFR-F3'97  & \text{B} & \textbf{1} & \text{B} & \textbf{1} &   & 3 & \text{C} & 2 &   & 3 &   & 3 &   & 3 \tabularnewline
4 & \textbf{CMS8jets'17}  &   &   &   &   &   & 3 &   &   & \text{C} & 3 & \text{C} & 3 & \text{C} & 3 \tabularnewline
5 & NMCrat'97  & \text{C} & 2 & \text{C} & 3 & \text{C} & 2 & \text{B} & \textbf{1} &   &   &   &   &   &   \tabularnewline
\hline 
\end{tabular}
	\caption{Continuation of Table~\ref{tab5nojet} for the PDFs obtained without imposing constraints from jet production experiments. The PDF combinations and Higgs production cross sections are the same as in Table~\ref{tab6} of the main paper. }
\label{tab6nojet}
 \end{table}

\begin{table}
\hspace*{-1cm}\begin{tabular}{c|c|c|cc|cc|cc|cc|cc|cc|cc}
\hline 
 &  &  & \multicolumn{14}{c}{Values, {\bf CT14 HERA2 NNLO PDFs fitted without jet data}}\tabularnewline
No.  & Expt. & $N_{\mathit{pt}}$  & $\sum_{f}|S^E_{f}|$  & $\langle\sum_{f}|S^E_{f}|\rangle$  & $|S^E_{\bar{d}}|$  & $\langle|S^E_{\bar{d}}|\rangle$  & $|S^E_{\bar{u}}|$  & $\langle|S^E_{\bar{u}}|\rangle$  & $|S^E_{g}|$  & $\langle|S^E_{g}|\rangle$  & $|S^E_{u}|$  & $\langle|S^E_{u}|\rangle$  & $|S^E_{d}|$  & $\langle|S^E_{d}|\rangle$  & $|S^E_{s}|$  & $\langle|S^E_{s}|\rangle$ \tabularnewline
\hline 
 &  &  &  &  &  &  &  &  &  &  &  &  &  &  &  & \tabularnewline
1 & HERAI+II'15 & 1120. & 599. & 0.0892 & 73.8 & 0.0659 & 122. & 0.109 & 143. & 0.128 & 136. & 0.121 & 73.8 & 0.0659 & 51.1 & 0.0457 \tabularnewline
2 & BCDMSp'89 & 337 & 230. & 0.114 & 25. & 0.0742 & 38.8 & 0.115 & 45.6 & 0.135 & 76.9 & 0.228 & 22.3 & 0.0662 & 21. & 0.0622 \tabularnewline
3 & CCFR-F3'97 & 86 & 223. & 0.433 & 50.8 & 0.591 & 51.3 & 0.597 & 19.5 & 0.227 & 61.3 & 0.713 & 30.2 & 0.352 & 10.1 & 0.117 \tabularnewline
4 & \textbf{CMS8jets'17} & 185 & 205. & 0.185 & 30.2 & 0.163 & 39.1 & 0.211 & 72.3 & 0.391 & 15.2 & 0.0824 & 14.5 & 0.0784 & 33.6 & 0.181 \tabularnewline
5 & NMCrat'97 & 123 & 168. & 0.228 & 45.7 & 0.371 & 12.9 & 0.105 & 6.69 & 0.0544 & 37.8 & 0.308 & 58.6 & 0.476 & 6.77 & 0.0551 \tabularnewline
\hline 
\end{tabular}
\caption{Sensitivity values for Table \ref{tab5nojet}.}
\label{tab:tab5nojetvals}
\end{table}

\begin{table}
\hspace*{-1cm}\begin{tabular}{c|c|cc|cc|cc|cc|cc|cc|cc}
\hline 
 &  & \multicolumn{14}{c}{Values, {\bf CT14 HERA2 NNLO PDFs fitted without jet data}}\tabularnewline
No.  & Expt. & $|S^E_{u_{v}}|$  & $\langle|S^E_{u_{v}}|\rangle$  & $|S^E_{d_{v}}|$  & $\langle|S^E_{d_{v}}|\rangle$  & $|S^E_{\bar{d}/\bar{u}}|$  & $\langle|S^E_{\bar{d}/\bar{u}}|\rangle$  & $|S^E_{d/u}|$  & $\langle|S^E_{d/u}|\rangle$  & $|S^E_{H7}|$  & $\langle|S^E_{H7}|\rangle$  & $|S^E_{H8}|$  & $\langle|S^E_{H8}|\rangle$  & $|S^E_{H14}|$  & $\langle|S^E_{H14}|\rangle$ \tabularnewline
\hline 
 &  &  &  &  &  &  &  &  &  &  &  &  &  &  & \tabularnewline
1 & HERAI+II'15  & 56.6 & 0.0505 & 36. & 0.0321 & 44.9 & 0.0401 & 58.8 & 0.0525 & 90.6 & 0.0809 & 95.4 & 0.0852 & 111. & 0.0993 \tabularnewline
2 & BCDMSp'89  & 66.3 & 0.197 & 26.4 & 0.0784 & 22.5 & 0.0667 & 38. & 0.113 & 32.9 & 0.0975 & 31.4 & 0.0932 & 25.2 & 0.0748 \tabularnewline
3 & CCFR-F3'97  & 77.1 & 0.896 & 67.2 & 0.781 & 10.4 & 0.121 & 31.6 & 0.367 & 10.1 & 0.117 & 11.3 & 0.131 & 14.4 & 0.167 \tabularnewline
4 & \textbf{CMS8jets'17}  & 17.8 & 0.0962 & 14.9 & 0.0806 & 18.8 & 0.102 & 10.3 & 0.0559 & 39.9 & 0.216 & 40.5 & 0.219 & 41.4 & 0.224 \tabularnewline
5 & NMCrat'97  & 36.9 & 0.3 & 27.8 & 0.226 & 49. & 0.398 & 76. & 0.618 & 8.66 & 0.0704 & 8.53 & 0.0693 & 8.13 & 0.0661 \tabularnewline
\hline 
\end{tabular}\caption{Sensitivity values for Table \ref{tab6nojet}. \label{tab:tab6nojetvals}}
\end{table}

%

\clearpage
\newpage{}

\bibliographystyle{apsrev}
\bibliography{sensitivity}

\end{document}